\documentclass[useAMS,usenatbib,onecolumn]{mn2e}

\usepackage[latin1]{inputenc}  
\usepackage[T1]{fontenc}       
\usepackage[dvips]{graphicx}
\usepackage{amsmath}
\usepackage{amssymb}
\usepackage{indentfirst}

\title[Model for the prompt and high latitude emission in GRBs]{Realistic
analytic model for the prompt and high latitude emission in GRBs}

\author[F. Genet and J.Granot]{F. Genet$^{1}$\thanks{E-mail:
f.genet@herts.ac.uk;} and J.Granot$^{1}$\\
$^{1}$ Center for Astrophysics Research, University of Hertfordshire, UK.}
\begin{document}

\date{Submitted December 2008.}

\pagerange{\pageref{firstpage}--\pageref{lastpage}} \pubyear{2009}

\maketitle

\label{firstpage}

\begin{abstract}

Most gamma-ray bursts (GRBs) observed by the {\it Swift} satellite
show an early steep decay phase (SDP) in their X-ray lightcurve, which
is usually a smooth continuation of the prompt gamma-ray emission,
strongly suggesting that it is its tail. However, the mechanism behind
it is still not clear. The most popular model for this SDP is High 
Latitude Emission (HLE), in which after the prompt emission from a 
(quasi-) spherical shell stops photons from increasingly large angles 
relative to the line of sight still reach the observer, with a smaller 
Doppler factor. This results in a simple relation between the temporal 
and spectral indexes, $\alpha = 2 + \beta$ where 
$F_\nu \propto t^{-\alpha}\nu^{-\beta}$. While HLE is
expected in many models for the prompt GRB emission, such as the
popular internal shocks model, there are models in which it is not
expected, such as sporadic magnetic reconnection events. Therefore,
testing whether the SDP is consistent with HLE can help distinguish
between different prompt emission models. In order to adequately
address this question in a careful quantitative
manner we develop a realistic self-consistent model for the
prompt emission and its HLE tail, which can be used for combined
temporal and spectral fits to GRB data that would provide strict tests
for the HLE model. We model the prompt emission as the sum of its
individual pulses with
their HLE tails, where each pulse arises from an ultra-relativistic
uniform thin spherical shell that emits isotropically in its own rest
frame over a finite range of radii. Analytic expressions for the observed 
flux density are
obtained for the internal shock case with a Band function 
emission spectrum. We find that the observed instantaneous spectrum is
also a Band function. Our model naturally produces, at least
qualitatively, the observed spectral softening and steepening of the
flux decay as the peak photon energy sweeps across the observed energy
range. The observed flux during
the SDP is initially dominated by the tail of
the last pulse, but the tails of one or more earlier pulses can become
dominant later on. A simple criterion is given for the dominant
pulse at late times. The relation $\alpha = 2+\beta$ holds also as 
$\beta$ and $\alpha$ change in time. Modeling several overlapping pulses 
as a single wider pulse would over-predict the emission tail.
\end{abstract}


\section{Introduction}

Before the launch of the {\it Swift} satellite (Gehrels et al. 2004), Gamma
Ray burst (GRB) X-ray afterglows were detected at least several hours
after the burst (Soffitta et al. 2004 and references therein). They
typically displayed a power law decay $\sim t^{-1} - t^{-1.5}$ around
their detection time (De Pasquale et al. 2006). {\it Swift}'s ability to 
rapidly and autonomously
slew when the Burst Alert Telescope (BAT, observing in the energy
range $15-350\;$keV; Barthelmy et al. 2005) instrument detects a GRB
enables it to point its other instruments - the X-Ray Telescope (XRT,
observing in the energy range $0.2-10\;$keV; Burrows et al. 2005a)
and UV/Optical Telescope (UVOT, observing at wavelengths
$170-650\;$nm, i.e. from the optical to the near UV; Roming et
al. 2005) - toward the GRB within tens of seconds from the GRB trigger
time.The XRT thus filled the
observational gap between the end of the prompt emission and the
beginning of the pre-{\it Swift} afterglow observations several hours
later. It revealed a complex behaviour usually consisting
of three phases, followed by most GRBs, and referred to
as a canonical light curve (Nousek et al. 2006), consisting of three
distinct power-law segments where $F_\nu \propto t^{-\alpha}$: an
initial (at $t < t_{\rm break,1}$, with $300\;{\rm s} \lesssim t_{\rm
break,1} \lesssim 500\;{\rm s}$) very steep decay with time $t$ (with
a power-law index $3 \lesssim \alpha_1 \lesssim 5$; see also
Bartherlmy et al. 2005; Tagliaferri et al. 2005); a subsequent (at
$t_{\rm break,1} < t < t_{\rm break,2}$, with $10^3\;{\rm s} \lesssim
t_{\rm break,2} \lesssim 10^4\;{\rm s}$) very shallow decay ($0.2
\lesssim \alpha_2 \lesssim 0.8$); and a final steepening (at $t >
t_{\rm break,2}$) to the familiar pre-{\it Swift} power-law behaviour
($1 \lesssim \alpha_3 \lesssim 1.5$). In many cases there are X-ray
flares superimposed on this underlying smooth component (typically
during the first two phases, at $t < t_{\rm break,2}$; Burrows et
al. 2005b; Falcone et al. 2006; Krimm et al. 2007), and in some cases
there is a later (at $t_j > t_{\rm break,2}$) further steepening due to
a jet.

The third decay phase ($F_\nu \propto t^{-\alpha_3}$) is the afterglow
emission that was observed before {\it Swift} and is well explained by
synchrotron radiation from the forward shock that is driven into the
external medium as the GRB ejecta are decelerated, where the energy in
the afterglow shock is constant in time (no significant energy gains
or losses). The plateau (or shallow decay) phase can be explained
either by pre-{\it Swift} models or by later models that have been
developed especially for this purpose (Nousek et al. 2006; Panaitescu
et al. 2006; Granot 2007).  It could be due to energy injection,
either by a tail of decreasing Lorentz factors at the end of the
ejection phase (Rees \& M\'eszaros 1998; Sari \& M\'eszaros 2000;
Ramirez-Ruiz, Merloni \& Rees 2001; Granot \& Kumar 2006) or by a
relativistic wind produced by a long lasting source activity (Rees
\& M\'eszaros 2000; McFadyen et al. 2001; Lee \& Ramirez-Ruiz
2002; Dai 2004; Ramirez-Ruiz 2004), by an increasing efficiency of
X-ray afterglow emission due to time dependence of the shock
microphysics parameters (Granot, K\"onigl \& Piran 2006), by a viewing
angle slightly outside the region of prominent afterglow emission
(Eichler \& Granot 2006), by a contribution from the reverse-shock
(Genet, Daigne \& Mochkovitch 2007) or by a two component jet
 model (Peng, K\"onigl \& Granot 2005; Granot, K\"onigl \& Piran 2006).

The steep decay phase is observed in most bursts, and is in the great
majority of cases a smooth continuation of the prompt emission, both
temporally and spectrally (O'Brien et al. 2006). This strongly
suggests that it is the tail of the prompt emission. Several
explanations for this phase have been suggested in the context of
previously existing models (Tagliaferri et al. 2005; Nousek et
al. 2006), such as emission from the hot cocoon in the collapsar model
(M\'eszaros \& Rees 2001, Ramirez-Ruiz et al. 2002). The most popular
model, by far, is High Latitude Emission (HLE) originally referred to
as emission from a ``naked'' GRB (Kumar and Panaitescu 2000a). In this
model the prompt GRB emission is from a (quasi-) spherical shell, and
after it turns off at some radius then photons keep reaching the
observer from increasingly larger angles relative to the line of sight,
due to the the added path length caused by the curvature of the
emitting region. Such late arriving photons experience a smaller
Doppler factor. This leads to a simple relation between the temporal
and spectral indexes, $\alpha = 2+\beta$ where $F_{\nu}(t) \propto
t^{-\alpha} \nu^{-\beta}$, that holds at late times when $t-t_0 \gg
\Delta t$, where $t_0$ and $\Delta t$ are the start time and width of
the pulse, respectively. The steep decay phase also shows a softening
of the spectrum with time (see Zhang et al. 2007 and references
therein).

The consistency of the steep decay phase with HLE has been studied by
several authors (Nousek et al. 2006; Liang et al. 2006; Butler \&
Kocevski 2007; Zhang et al. 2007; Qin 2008). However, some simplifying
assumptions were usually made, which may affect the comparison between
this model and the observations. One such assumption is the choice of
the reference time $t_0$ for the steep decay, especially when the
prompt emission consists of several pulses.
Liang et al. (2006) find that when assuming the HLE relation $\alpha =
2+\beta$ and fitting for $t_0$ its derived value is consistent with
the onset of the last pulse of the prompt emission (or of the
individual spike or flare whose tail is being fit). Zhang et
al. (2007) find that the HLE cannot explain the steep decays
accompanied by a spectral softening, but can explain the cases with no
observed spectral evolution. Barniol Duran and Kumar (2008) find that
only $20\%$ of their sample is consistent with HLE.
Butler \& Kocevski (2007) find that for a (physically motivated) time
independent soft X-ray absorption (fixed N$_{\rm H}$) the spectrum
during the steep decay phase, is much better fit by an intrinsic Band
function spectrum (Band et al. 1993) rather than by a power-law, and
that the peak photon energy shifts to lower energies with time. Qin
(2008) finds that such a behavior can, at least qualitatively, be
produced for a delta function emission in radius with a Band function
spectrum. It is therefore still a largely open question whether the
temporal and spectral properties of the steep decay are consistent
with HLE.  Moreover, it appear that a physically motivated model for
the prompt emission with realistic assumptions about the emission
(e.g. over a finite range of radii with a Band function emissivity) is
needed in order to address this question in a more quantitative and
fully self consistent manner.

The nature of the prompt GRB emission is what ultimately determines
the properties of its tail. HLE is expected only in models where the
prompt emission is from a quasi-spherical shell and turns off rather
abruptly at some finite radius (or lab frame time). The best example
for this type of model is internal shocks (Rees \& M\'eszaros 1994;
Sari \& Piran 1997) where variability in the Lorentz factor of the
relativistic GRB outflow causes faster shells of ejecta to collide
with slower sells resulting in shocks going into the shell over a
finite range of radii (typically $\Delta R \sim R$).  On the other
hands, there are models in which HLE is not expected, such as in the
case of isolated sporadic magnetic reconnection events within a
Poynting flux dominated outflow (e.g. Lyutikov \& Blandford 2003) in
which each spike in the GRB light curve is from a distinct small and
well localized region. Therefore, testing whether the steep decay
phase is consistent with HLE would help distinguish between these two
types (or classes) of prompt GRB models. This can be an important step
toward identifying the basic underlying mechanism for the prompt
emission, which is still one the the most striking open questions in
GRB research more than four decades after the discovery of GRBs.

In order to address this question, we develop a model for the prompt
and its HLE tail that is physically motivated, realistic, and easy to
use (fully analytic in its simplest version) in global joint fits (to
all of the available data at all times and photon energies) of the
prompt GRB and its SDP tail. Such global fits can provide a stringent
and fully self-consistent test of HLE model for the SDP in GRBs. 

The prompt emission is modeled as the sum of a finite number of
pulses. Each pulse corresponds to a spike in the prompt GRB light
curve and has its own HLE tail. An individual spike is modeled as
arising from a thin uniform spherical relativistic shell that emits
isotropically in its own rest frame over a finite range of radii,
while the observed flux is calculated by integrating over the equal
arrival time surface (Granot, Piran \& Sari 1999; Granot 2005; Granot,
Cohen-Tanugi \& do Couto e Silva 2008) of photons to a distant
observer. Our model is particularly suitable for internal shocks,
which we focus on in this paper. For the emitted spectrum we consider
the phenomenological Band function, which provides a good fit the the
prompt emission spectrum of the vast majority of GRBs. We point out
that our model can also be used for X-ray flares, which appear to have
temporal and spectral properties similar to the spikes of the prompt
GRB emission. The main text provides the most useful results in an
easy to use form, while the full derivations of these results are
provided in appendixes in order to help understand their origin and
make it easier to extend or generalize our model. We stress here
that our main aim is not necessarily to uniquely determine all of the
model parameters, which may be subject to various degeneracies and may
prove hard when fitting to real data, but instead to test whether our
model can provide a good fit to the data for any set of physical
parameters. While such a good fit would still not prove that the HLE
must be at work, it would definitely support HLE as a viable and
arguably most plausible model.  Our model for an individual pulse is
described in \S~\ref{sec_dyn}, and results for the flux in the case
for internal shocks with a Band function spectrum are given in
\S~\ref{sec_res_flux}. The dependence of a single pulse on the model
parameters is then investigated in
\S~\ref{sec_emission_single}, while \S~\ref{sec_prompt} discusses
how to combine several pulses in order obtain to the total prompt
emission and its tail. Both are intended to help the reader when
using our model to fit data, which is one of the main aims of our
paper. Our conclusions are discussed in
\S~\ref{sec_concl}. This paper describes in detail our theoretical
model and its main properties, and stresses some important caveats
that one should keep in mind when using it to fit data in order to test
the HLE model. In subsequent work we intend to confront it with {\it
Swift} BAT+XRT data.

\section{Description of the Model} \label{sec_dyn}

\subsection{The Basic Physical Model}

We consider an ultra-relativistic ($\Gamma \gg 1$) thin (of width $\ll
R/\Gamma^2$) spherical expanding shell that emits over a range of
radii $R_0 \leqslant R \leqslant R_f \equiv R_0+\Delta R$. The
emission turns on at radius $R_0$ and turns off at radius $R_f
\ge R_0$. The Lorentz factor of the emitting shell is assumed to scale
as a power-law with radius, $\Gamma^2 = \Gamma_0^2 (R/R_0)^{-m}$ where
$\Gamma_0 \equiv \Gamma(R_0)$. The emission is assumed to be isotropic
in the comoving frame of the shell, and uniform over the shell,
i.e. the comoving spectral luminosity depends only on the radius of
the shell, $L'_{\nu'} = L'_{\nu'}(R)$. As the main purpose of this work 
is to check the consistency of the tail of the prompt emission 
with HLE, we need to model the
prompt emission. We therefore use for the emission spectrum the
phenomenological Band function (Band et al. 1993) spectrum that
provides a good fit to the observed prompt emission spectrum of the
vast majority of GRBs. In the following we mainly consider emission 
over a finite range of radii, $R_0 \leq R \leq R_f = R_0 + \Delta R$. 
The comoving luminosity is then:
\begin{eqnarray}\label{eq_bandfunction_lumin}
L'_{\nu'} = L'_0 \left(\frac{R}{R_0}\right)^a
S\left(\frac{\nu'}{\nu'_p}\right)\ , \qquad \qquad
S(x) = e^{1+b_1} \left\{ \begin{array}{ll} x^{b_1} e^{-(1+b_1)x} & x
\leqslant x_b\ ,\\
x^{b_2} x_b^{b_1-b_2} e^{-(b_1-b_2)} & x \geqslant x_b\ ,
\end{array} \right.
\end{eqnarray}
where $\nu'_p(R)\equiv \nu'_0(R/R_0)^d$ 
is the frequency where $\nu' L'_{\nu'}(R)$ peaks, with 
$\nu'_0 \equiv \nu'_p(R_0)$; $x_b = (b_1-b_2)/(1+b_1)$, while $b_1$ and
$b_2$ are the high and low energy slopes of the spectrum. For $b_1 >
-1 > b_2$ the Band function has a peak in the $\nu F_\nu$ spectrum, at
$x = 1$, and therefore since $S(x)$ is normalized such that $S(x) =
xS(x) = 1$ at $x = 1$, it will not affect normalization of $\nu
F_{\nu}$ at its peak. The two functional forms used in the band
function are matched at $\nu'_b = x_b \nu'_p$. The peak luminosity 
$L'_{\nu'_p}$ evolve as a power-law with radius, 
$L'_{\nu'_p} = L'_0 (R/R_0)^a$ where $L'_0 = L'_{\nu'_p}(R_0)$ is a 
normalization factor.

Throughout the paper, primed
quantities are quantities measured in the comoving frame (i.e. the
local rest frame of the emitting shell), unprimed quantities are
measured either in the source rest frame (the lab frame, i.e. the
cosmological frame of the GRB; this includes $\Gamma$, $R$, $\theta$
and $t$) or the observer frame (this refers to observed quantities,
such as $F_\nu$, $\nu$ and $T$).

\subsection{Calculating the Observed Flux}

\begin{figure}
\begin{center}
\includegraphics[viewport=3cm 7.4cm 20cm 15.5cm, clip, width=1\textwidth,
height=0.5\textwidth]{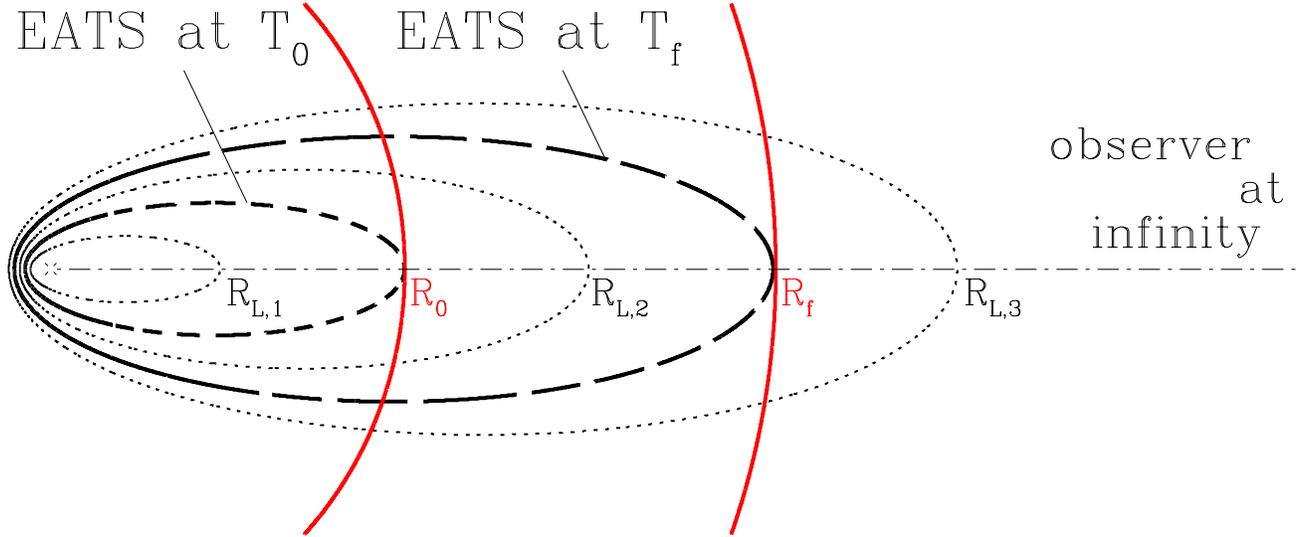}
\caption[EATS]{
\emph{Illustration of Equal Arrival Time Surfaces (EATS). The particular EATS
shown here are for a coasting shell ($m=0$), and are ellipsoids (Rees
1966) with an ellipticity $\epsilon = \beta$ and a semi-major to
semi-minor axis ration of $\Gamma$ (which for display purposes is only
3 here). The red solid lines correspond to the inner ($R_0$) and outer
($R_f$) radii of the emission region. We call $R_L(T)$ the outermost 
radius of the EATS at observed time $T$. Shown are the EATS for the
limiting cases corresponding to $R_L = R_0$ ($T = T_{\rm ej}+T_0$;
thick short-dashed line) and $R_L = R_f$ ($T = T_{\rm ej}+T_f$; thick
long-dashed line) as well as representative cases for $R_L < R_0$,
$R_0 < R_L < R_f$ and $R_L > R_f$ (dotted lines).  For $R_L < R_0$
the EATS does not intersect the emission region,
and therefore the first photons start reaching the observer only at $T
= T_{\rm ej} + T_0$ from $R = R_0$ along the line of
sight. At $R_0 < R_L < R_f$
the flux typically rises (for $\Delta R \lesssim R_0$). At $T = T_{\rm
ej}+T_f$ the last photons from the line of sight (at $R
= R_f$) reach the observer, while for $T > T_{\rm ej}+T_f$ 
the front part of the EATS, which would otherwise contribute the
most to the observed flux, sticks outside of the emission radius
resulting in a sharp decay in the observed flux, which is then
dominated by emission from large angles relative to the line of sight
(HLE).  }}
\label{fig_EATS}
\end{center}
\end{figure}

The observer is assumed to be located at a distance from the source
that is much larger than the source size, so that the angle extended
by the source as seen by the observer is very small and the observer
effectively at ``infinity''.  In order to calculate the flux density
$F_{\nu}$ that reaches the observer at an observed time $T$ we
integrate the luminosity $L'_{\nu'}$ over the Equal Arrival Time Surface 
(EATS; see Figure \ref{fig_EATS}), i.e. the locus of points from which 
photons that are emitted by the shell at a radius $R$, angle $\theta$ 
relative to the line of sight, and a lab frame time $t$, reach the 
observer simultaneously at an observed time $T$ (for full derivation see 
Appendix \ref{ap_dyn_flux}).

\subsection{Expected parameters values for internal shocks}

The internal shocks model is the most popular model for the prompt GRB
emission. Moreover, our model is very suitable for internal shocks.
Therefore, we consider it in the following. Here we calculate
the scalings of the various quantities with radius, that are expected
for the internal shocks model. First, when different shells
(i.e. parts of the outflow with different Lorentz factors) collide,
they are expected to be in the coasting phase, corresponding to $m =
0$. Moreover, for the simplest case of uniform shells, the strength of
the shocks going into the two shells, as characterized by the relative
upstream to downstream Lorentz factor, $\Gamma_{ud}$, is expected to
be roughly constant with radius while the shock are crossing the
shells. The electrons are expected to be fast cooling, i.e. cool
significantly on a timescale much shorter than the shell crossing time
of the shock, and therefore most of the emission is expected to arise
from a thin cooling layer behind the shock. Therefore our thin shell
approximation is expected to be reasonably valid. Admittedly, we use
one emitting thin shell, corresponding to a single shock front, while
the shock going into the other shell is not explicitly modeled. One
could always model such a second shock by adding another thin emitting
shell that turns on and off at the same radii ($R_0$ and $R_f$,
respectively) but has a slightly smaller or larger Lorentz factor. This
will not introduce a big difference in the overall result, so for the
sake of simplicity we do not include this here.

Now we turn to find the expected scaling of $L'_{\nu'_p}$ and $\nu'_p$
with radius, under the assumption that the observed soft gamma-ray
range is dominated by synchrotron emission. For fast cooling, the peak
frequency $\nu'_p$ of the $\nu F_{\nu}$ spectrum is $\nu'_m \approx (e
B'\gamma_m^2)/(2\pi m_e c^2)$ where $dN_e/d\gamma_e \propto
\gamma_e^{-p}$ for $\gamma_e > \gamma_m$ where $\gamma_m =
(p-2)/(p-1)(\epsilon_e/\xi_e)(m_p/m_e)(\Gamma_{ud}-1)$, while
$\epsilon_e$ is the fraction of the internal energy behind the shock
in the power law distribution of the relativistic electron, and
$\xi_e$ is the fraction of all electrons taking part in this power
energy distribution (and an electron-proton plasma is assumed for the
composition of the outflow). As mentioned above, $\Gamma_{ud}$ is
expected to be roughly constant during the shell crossing (for roughly
uniform colliding shells), and therefore $\gamma_m$ would also be
approximately constant, so that $\nu'_p \propto B'$. The magnetic field
is expected to be predominantly normal to the radial direction, so
that $B' \approx B/\Gamma \propto B$ for $m = 0$. Moreover, $B \propto
R^{-1}$ is expected both for a magnetic field convected from the
central source, and for a field generated at the shock that hold some
constant fraction ($\epsilon_B$) of the internal energy behind the
shock. Therefore, one expects the peak frequency to evolve as 
$\nu'_p \propto R^{-1}$. We have also assumed $L'_{\nu'_p} \propto
(R/R_0)^a$. For synchrotron emission $L'_{\nu',{\rm max}}
\propto N_e B' \propto R^0$ as the number of emitting electron is proportional
to
the radius, $N_e \propto R$. Since the cooling break frequency scales
as $\nu'_c \propto R$, we have $L'_{\nu'_p} \approx L'_{\nu',{\rm
max}}(\nu'_m/\nu'_c)^{-1/2} \propto R^1$, implying $a = 1$.

More generally (without specifying the emission mechanism) for roughly
uniform shells with constant $\Gamma_{ud}$ both the rate at which
particles cross the shock and the average energy per particle are
constant with radius, implying a constant rate of internal energy
generation ($dE'_{\rm int}/dt' \propto R^0$), and therefore for fast
cooling this also applies for the total comoving luminosity, $L' \sim
\nu'_p L'_{\nu'_p} \propto R^0$, and therefore $d + a = 0$. This is
indeed satisfied for synchrotron emission for which $d = -1$ and $a =
1$, and holds more generally for other emission mechanisms in the fast
cooling regime.

For now on the values $m=0$ and $d=-1$ derived in this part will be used 
throughout the paper. However, since the expressions do not become much 
simpler by specifying the value of $a$, we leave $a$ in the simpler 
expressions, and use the value of $a = 1$ for figures only.
In particular, all the figures showing lightcurves in this paper 
use these parameter values, as well as the mean BATSE values 
for the Band function spectral slopes: $b_1 = -0.25$ and 
$b_2 = -1.25$ (Preece et al., 2000).

\subsection{Relevant Times and Timescales}

A photon emitted from the source (at
the origin) when the shell is ejected from it (i.e. at a lab frame
time $t_{\rm ej}$ when the shell radius is $R = 0$) arrives at the 
observer at an observer time $T_{\rm ej}$ which can be 
thought of as the observed ejection time of the shell. We define 
$T_0$ the initial radial time by $T = T_{\rm ej} + T_0$
 being the time at which the first photons
emitted reach the observer (that is, photons emitted at a radius $R_0$
along the line of sight). We also define $T_f$ the final angular time
 by $T = T_{\rm ej} + T_f$ 
being the time at which the last photons that are emitted along the line
 of sight (from $R_f$ and $\theta = 0$) reach the observer.

For a constant Lorentz factor with radius ($m = 0$), as expected for
internal shocks, the expressions for $T_0$ and $T_f$ are simple:
\begin{equation}\label{eq_TfT0IS}
T_0 = \frac{(1+z)R_0}{2c \Gamma_0^2}\ , \qquad \qquad  T_f
= T_0 \left(1+\frac{\Delta R}{R_0}\right) \ .
\end{equation}

We also define two normalized times (and their corresponding values at $T_f$) 
that will be used in the following:
\begin{equation}\label{eq_Ttildebar}
\tilde{T} \equiv 1+\bar{T} \equiv \frac{T-T_{\rm ej}}{T_0}\ , 
\qquad \qquad  
\tilde{T}_f \equiv 1+\bar{T}_f \equiv \frac{T_f}{T_0} = 1+\frac{\Delta R}{R_0}\ ,
\end{equation}
where $\tilde{T}=1$ (or $\bar{T} = 0$) corresponds to the onset of the
spike -- the very first photon that reaches the observer (emitted at
$R_0$ on the line of sight). The main motivation for defining
these two times is that they correspond to the two most natural
choices for the zero to, $\tilde{T} = 0$ corresponding to the ejection
time of the shell, and $\bar{T} = 0$ corresponding to the onset of the
spike in the lightcurve. The choice of the zero time is important for
the definition of the temporal index in \S\;\ref{subsec_instslopes},
where we explore these two choices in detail. Moreover, it is more
convenient to use $\bar{T}$ in some expressions and $\tilde{T}$ in
others.

\section{Results for Internal shocks with a Band function spectrum} \label{sec_res_flux}

\subsection{Emission from a single radius} \label{subsec_band_R0}

Before to turn to the more generic case of emission from a range of 
radii, we first consider the limiting case of emission from a single 
radius $R_0$. The peak frequency is then $\nu'_p = \nu'_0$, and the 
luminosity is
\begin{equation}\label{eq_L_DR0_maintext}
L'_{\nu'} = L'_0 S\left(\frac{\nu'}{\nu'_p}\right) R_0 \delta(R-R_0)\ ,
\end{equation}
which after some algebra (see appendix \ref{ap_dyn_flux} for details,
 and in particular section \ref{ap_band_gen_singleR}) we obtain the 
flux:
\begin{eqnarray}\label{F_nu_Band_R0_main}
F_{\nu}(\tilde{T} \geq 1) = \frac{(1+z)}{4 \pi d_L^2} L_0
\tilde{T}^{-2} S\left(\frac{\nu}{\nu_0}\tilde{T}\right)\ ,
\end{eqnarray}
where $d_L$ and $z$ are the luminosity distance and cosmological 
redshift of the source, $L_0 \equiv 2\Gamma_0 L'_0$ and 
$\nu_0 \equiv 2\Gamma_0\nu'_0/(1+z)$. 
Denoting $F_s \equiv L_0(1+z)/(4\pi d_L^2)$ and using the explicit
expression for the Band function (eq.~[\ref{eq_bandfunction_lumin}]),
the observed flux reads
\begin{eqnarray}
\frac{F_{\nu}(\tilde{T} \geq 1)}{F_s} & = &
\left\{ \begin{array}{ll}
\tilde{T}^{b_1-2} (\nu/\nu_0)^{b_1}
e^{(1+b_1)[1-\tilde{T}\nu/\nu_0]}
&  \tilde{T} \leqslant x_b \nu_0/\nu\ ,\quad
\\ \\
\tilde{T}^{b_2-2} (\nu/\nu_0)^{b_2} x_b^{b_1-b_2} e^{1+b_2}
&  \tilde{T} \geqslant x_b \nu_0/\nu\ .
\end{array} \right.
\end{eqnarray}

\subsection{Emission from a finite range of radii} \label{subsec_band_R0_Rf}

Integrating the luminosity (eq. (\ref{eq_bandfunction_lumin})) over
the Equal Arrival Time Surface (for details of the calculation see 
appendix \ref{ap_dyn_flux}, and in particular its section 
\ref{ap_band_gen_rangeR}) leads to the following expression 
for the flux:
\begin{eqnarray}\label{eq_bandDR_finalcondensed}
F_{\nu}(\tilde{T} \geq 1) = F_0 \tilde{T}^{-2}
\left[\min \left(\tilde{T},\tilde{T}_f\right)^{2+a}-1\right]
S\left(\frac{\nu}{\nu_0}\tilde{T}\right)\ ,
\end{eqnarray}
where $F_0 \equiv (1+z)L_0/[(2+a)4\pi d_L^2]$.
This can be explicited as:
\begin{equation}
\frac{F_{\nu}(\tilde{T} \geq 1)}{F_0} =
\left\{ \begin{array}{ll}

(\nu/\nu_0)^{b_1} \tilde{T}^{b_1-2}
\left(\tilde{T}^{2+a}-1\right) e^{(1+b_1)(1-\tilde{T}\,\nu/\nu_0)}
 & \tilde{T} < \min\left[\tilde{T}_f,x_b\nu_0/\nu\right]\ ,\\
\\
(\nu/\nu_0)^{b_1} \tilde{T}^{b_1-2}
\left[\tilde{T}_f^{2+a}-1\right]
e^{(1+b_1)(1-\tilde{T}\,\nu/\nu_0)}
 & \tilde{T}_f < \tilde{T} < x_b\nu_0/\nu\ ,\\
\\
(\nu/\nu_0)^{b_2} \tilde{T}^{b_2-2}
\left(\tilde{T}^{2+a}-1\right)
x_b^{b_1-b_2} e^{1+b_2}
 &  x_b\nu_0/\nu < \tilde{T} <  \tilde{T}_f\ ,\\
\\
(\nu/\nu_0)^{b_2} \tilde{T}^{b_2-2}
\left[\tilde{T}_f^{2+a}-1\right]
x_b^{b_1-b_2} e^{1+b_2}
\label{F_expanded_latetime}
 & \tilde{T} > \max\left[\tilde{T}_f,x_b\nu_0/\nu\right]\ .
\end{array} \right.
\end{equation}
Note that the observed function has exactly the same shape as the
local spectral emissivity -- a pure Band function. This occurs only
for $m = 0$ and $d = -1$.

In terms of number of photons $N$ per unit photon energy $E$, area $A$
and observed normalized time $T$ (which is simply equal to $F_\nu/hE$), this can
be expressed as
\begin{eqnarray}
\frac{d N}{dE dA dT}(E,\tilde{T}\geq 1) =
\tilde{T}^{-1}
\left[\min \left(\tilde{T},\tilde{T}_f\right)^{a+2}-1\right]
B\left(\frac{E}{E_0}\tilde{T}\right)\ ,
\end{eqnarray}
where
\begin{eqnarray}\label{eq_bandfunction_energy}
B(z) = B_{\rm norm} \left\{ \begin{array}{ll}
z^{b_1-1} e^{-z}  &  z \leqslant b_1-b_2\\
z^{b_2-1} (b_1-b_2)^{b_1-b_2} e^{-(b_1-b_2)}  &  z \geqslant b_1-b_2
\end{array} \right.
\end{eqnarray}
is the familiar Band function with a normalization constant 
$B_{\rm norm}$, where $z = (E/E_0)\tilde{T} = (1+b_1)x$, 
while $E = h\nu$ and $E_0 = h\nu_0$ are the corresponding 
photon energies (the more common notation is $\alpha_{\rm Band} 
= b_1-1$ and $\beta_{\rm Band} = b_2-1$).

\section{Properties of the single pulse emission} \label{sec_emission_single}

Now that we have derived the observed flux for a single emission
episode (or single pulse in the light curve), we study its temporal
and spectral behaviour for any radial width $\Delta R \geqslant 0$ of
the emitting region.  We remind the reader that we consider only
internal shocks, and use the corresponding model parameter values
($a=1$, $m=0$ and $d=-1$) for fast cooling synchrotron emission, with
a Band function emission (and observed) spectrum (except in some cases
where the discussion can stay general without much complication).
Some of the results may not hold for more general parameter values of
$m$ or $d$, and we point this out when relevant.  For all figures
showing lightcurves (throughout the whole paper), the panels or
figures with a linear scale show $F_{\nu}/F_{\rm max}$ where $F_{\rm
max} \equiv F_{\nu}(\tilde{T}_f)$, while panels or figures with a
logarithmic scale show $F_{\nu}/F_0$ where we remind the reader that
$F_0(a = 1) = (1+z)L_0/(12 \pi d_L^2)$. All figures showing temporal
evolution of parameters or lightcurves with a logarithmic time axis in
this section use $\bar{T}$, as this shows the early behaviour
much more clearly than for $\tilde{T}$.

From eq.~(\ref{eq_bandDR_finalcondensed}), for reasonable values 
of the parameters $\tilde{T}_f$, $b_1$, $b_2$, $\nu/\nu_0$, and
$a$, the pulse peaks at $T = T_{\rm ej}+T_f$ ($\tilde{T} =
\tilde{T}_f$). While this is generally the case,
for some combinations of parameters (often involving relatively
large values of $\tilde{T}_f$) the pulse has a round peak and
starts decaying before $\tilde{T}_f$.

For $\tilde{T}<1$, the Equal Arrival Time Surface (EATS) does
not intersect the emission region and no photons reach
the observer (its outermost radius $R_L$ is smaller than $R_0$):
 $F_\nu (\tilde{T}<1) = 0$. When $1 \leqslant \tilde{T}
 \leqslant \tilde{T}_f$ ($R_0 \leqslant R_L \leqslant R_f$), 
the EATS intersects the emission
region but does not yet encounter its outer edge (in particular the
observed flux is independent of the radial extension $\Delta R$ of the 
emission region); the fraction of the EATS within
the emission region increases with time, as does the maximal angle
$\theta_{\rm max}$ relative to the line of sight from which photons
reach the observer, $(\theta_{\rm max}\Gamma_0)^2 = (\tilde{T}-1)$.
When $\tilde{T} > \tilde{T}_f$ ($R_L > R_f$), the
front part of the EATS is outside the emission region, and its parts
inside the emission region are at increasing angles from the line of
sight. In particular, photons reach the observers from $\theta_{\rm
min} \leqslant \theta \leqslant
\theta_{\rm max}$ where $(\theta_{\rm min}\Gamma_0)^2 =
(\tilde{T}-\tilde{T}_f)\tilde{T}_f^{-1}$.  Note that at
$\tilde{T} \gg \tilde{T}_f$, well into the tail of the pulse, $\theta_{\rm
max}/\theta_{\rm min} \approx \tilde{T}_f^{1/2}$, so that the
emission comes from a rather narrow range of angles $\theta$, whose
typical value increases linearly with $\tilde{T}$. Moreover, for
$\tilde{T} > \tilde{T}_f$, the flux ratio for two identical sets of
emission parameters that differ only in their $\tilde{T}_f$ (denoted by subscripts 
1 and 2), is constant in time and equal to
\begin{equation}\label{eq_fluxratio}
\frac{F_\nu(\tilde{T}>\tilde{T}_{f,2} > 1)}{F_\nu(\tilde{T}>\tilde{T}_{f,1}>1)} =
\frac{\tilde{T}_{f,2}^{2+a}-1}
{\tilde{T}_{f,1}^{2+a}-1}\ ,
\qquad
\frac{F_\nu(\tilde{T}>\tilde{T}_{f,2}>1)}{F_\nu(\tilde{T}>\tilde{T}_{f,1} = 1)} =
\frac{\tilde{T}_{f,2}^{2+a}-1}{2+a}\ .
\end{equation}
The first ratio approaches $\Delta R_2/\Delta R_1$ for $\Delta R_{1,2}
\ll R_0$, since this corresponds to the thin shell limit, while the
overall emitted energy is proportional to $\Delta R$, since
$L'_{\nu'}(R) \approx L'_{\nu'}(R_0)$ is almost independent of $R$
within the very thin emission region. For the second ratio, the
denominator is the flux for a delta function emission with radius, for
which the total emitted energy is held fixed, and therefore the ratio
approaches $\Delta R_2/R_0 \ll 1$ in the limit of a thin emission
region. The fact that the flux ratio is constant in time at $\tilde{T} >
\tilde{T}_f$ holds only for $m = 0 $ and $d = -1$, and means that the
flux at these late times (typically after the peak of the spike, which
is usually at $\tilde{T}_f$) has the same time dependence regardless the
width of the emitting region ($\Delta R$). This can simplify the
calculation of the flux for a family of pulses that differ only in
$\Delta R$: one can calculate the flux for $\Delta R = 0$ 
($\tilde{T}_f=1$) and apply it to $\tilde{T} \geqslant \tilde{T}_f$,
 multiplied by a factor $[\tilde{T}_f^{2+a}-1]/(2+a)$ for any value 
$\Delta R > 0$ ($\tilde{T}_f>1$). Moreover, it is
also sufficient to calculate the flux for $\Delta R \to \infty$ and
apply it to $\tilde{T} \leqslant \tilde{T}_f$ (this holds much more
generally; Granot, Cohen-Tanugi \& do Couto e Silva 2008).

Figure~\ref{fig_LC_onepulse_selon_freq} shows light curves for a
single pulse in both linear and logarithmic scales, for different
values of the normalized frequency $\nu/\nu_0$. The peak
time is at $\tilde{T}_f=2$ (equivalent to $\bar{T}_f = 1$).
 The light curves sample the two
parts of the Band function both before and after the peak time. The
differences between the light curves for different frequencies reflect
the spectral evolution, and in particular the evolution of the
spectral break frequency $\nu_p$. At higher observed frequencies $\nu$
the change in the spectral and temporal indexes associated with the
passages of $\nu_p$ occurs earlier. The shape of a pulse 
(left panel of figure~\ref{fig_LC_onepulse_selon_freq}) can vary 
from being very spiky ({\it dotted line}) to a rounder peak 
({\it dot-dashed line}), depending on the frequency. It may 
thus provide some latitude 
in the fitting of actual observed pulses.

Figure \ref{fig_LC_spectra_DR} shows the dependence of the same pulse
on $\bar{T}_f$ for three values of the normalized
frequency $\nu/\nu_0$ ($0.01$, $0.1$ and $1$).
 It is evident from the logarithmic scale figures that at
$\bar{T} \leqslant \bar{T}_f$ the flux is independent of $\Delta R$ (and therefore of
$\bar{T}_f$), and that at $\bar{T} \geqslant \bar{T}_f$ all the light
curves have the same time dependence (i.e. their flux ratio is
constant in time). At any given time the spectrum is independent of
$\Delta R$ (this is valid only for $m = 0$ and $d = -1$). The bottom 
right panel of this figure shows linear scale to help visualise 
a case where the peak of the pulse is before $\tilde{T}_f$.

Figure \ref{fig_var_a} shows the dependence of the same pulse on the 
parameter $a$  for three values of the normalized frequency 
$\nu/\nu_0$ ($0.01$, $0.1$ and $1$). We can see that, compared 
to the case for $a=1$, when $a$ increases the peak
is at $\bar{T_f}$ and becomes sharper. When $a$ decreases the pulse becomes
larger, the slope for $\bar{T} \geqslant \bar{T_f}$ becoming closer to zero
up to a point where is is zero. For values of $a$ even smaller, the peak of the
pulse occurs before $\bar{T_f}$ and becomes rounder; in this case at $\bar{T_f}$ only
a sharp break is observed.

\subsection{Local temporal and spectral indexes} \label{subsec_instslopes}

It is natural to define the local values of the spectral and temporal
indexes as the logarithmic derivatives of the flux density with
respect to frequency and time, respectively. For the spectral index,
there is no ambiguity and $\beta \equiv -d \log F_{\nu}/d \log
\nu$. For the temporal index, however, we must choose a reference
time, and the choice is not obvious. for this reason we consider two
alternative definitions: $\alpha_{ej} \equiv -d\log F_\nu/d\log\tilde{T}$,
 that uses the ejection time as the reference time, and $\alpha_{on} \equiv -d \log
F_{\nu}/d\log \bar{T}$ that uses the onset of the spike as the reference time.
The figures in this subsection use the observed frequency $\nu$
instead of its normalized value $\nu/\nu_0$, in order to
provide a more realistic example that could be at least qualitatively
compared with data, and include the BAT and XRT energy ranges. For
these figures we consider $E_0 = 2\Gamma_0E'_0/(1+z) = 300\;$keV,
which could for example correspond to $E'_0 = 1\;$keV, $\Gamma_0 =
300$ and $z = 1$.

Figure \ref{fig_tempslopes} shows the evolution of the temporal
indexes $\alpha_{ej}$ and $\alpha_{on}$ during a pulse (See 
appendix \ref{ap_temp_spec_slopes} for the detailed evolution 
of the temporal and spectral slopes). The
temporal index $\alpha_{ej}$ starts at very negative values and gradually
increases, until at $\bar{T}_f$ it makes an abrupt jump to its value during
the decaying part of the pulse, which is exactly $2+\beta$ (see
eq.~[\ref{eq_relation_alpha_beta}]). The temporal index $\alpha_{on}$
starts at early times, $\bar{T} \ll 1$, either at $-1$ for $\bar{T}_f
> 0$ and $\bar{T} < \bar{T}_f$, or from $0$ for
$\bar{T}_f \to 0$. Moreover, for $0 < \bar{T}_f \ll 1$,
$\alpha_{on} \approx -1$ for $\bar{T} < \bar{T}_f$ and $\alpha_{on}
\approx 0$ for $\bar{T}_f < \bar{T} \ll 1$ (see eqs.~[\ref{eq_DR0_alphas}]
and [\ref{eq_alpha_4cas_barT}]). Note that when $\alpha_{on}$ jumps
from its negative value to a positive value at $\bar{T} = \bar{T}_f$
(i.e. at the peak of the spike), it reaches the same function of
$\bar{T}$, independent of the time of the jump, $\bar{T}_f$, and
therefore the same function also holds for $\bar{T}_f = \Delta R/R_0 =
0$ (see eqs.~[\ref{eq_DR0_alphas}] and [\ref{eq_alpha_4cas_barT}]).
At late times, $\bar{T} \gg 1$ and $\bar{T} > \bar{T}_f$, the HLE
relation is approached, $\alpha_{on} \approx 2-b_2$.

The left panel of figure~\ref{fig_specslope} shows the evolution of $2+\beta$ (where $\beta$
 is the spectral index) with the temporal
indexes $\alpha_{ej}$ and $\alpha_{on}$. The spectral index
naturally softens ($\beta$ increases with time), similar to what is
typically observed (at least qualitatively), until it reaches $-b_2$
at late times ($\bar{T} \geq x_b\nu_0/\nu-1$). The change
in $\beta$ occurs earlier at higher photon energies. At $\bar{T} \ge
\bar{T}_f$, $\alpha_{ej} = 2+ \beta$ while $\alpha_{on}$ only approaches
$2+ \beta$ at late times.

In order to get a better idea of how the observed spectral index
$\beta$ is expected to behave in {\it Swift} XRT observations, we
calculate its average values over the XRT energy range
(0.2--10$\;$keV). We define two average values, by integrating over
either the frequency $\nu$ or its logarithm $\log\nu$:
\begin{eqnarray} \label{eq_calc_swift_averaged_specslopes}
\langle\beta\rangle_\nu \equiv \frac{1}{(\nu_{\rm max}-\nu_{\rm min})}
\int_{\nu_{\rm min}}^{\nu_{\rm max}}d\nu\,\beta(\nu)\ , \qquad \qquad 
\langle\beta\rangle_{\log\nu} \equiv
\frac{1}{\log(\nu_{\rm max}/\nu_{\rm min})}\int_{\nu_{\rm min}}^{\nu_{\rm max}}
\frac{d\nu}{\nu}\,\beta(\nu) =
-\frac{\log(F_{\nu_{\rm max}}/F_{\nu_{\rm min}})}{\log(\nu_{\rm max}/\nu_{\rm
min})}\ .
\end{eqnarray}
The middle panel of figure~\ref{fig_specslope} shows the evolution of these two
averages as well as the values of $\beta$ at $\nu_{\rm min} =
0.2\;$keV, $\nu_{\rm max}/2 = 5\;$keV and $\nu_{\rm max} =
10\;$keV. As expected, $\langle\beta\rangle_\nu$ gives a larger weight
to higher frequencies compared to $\langle\beta\rangle_{\log\nu}$, and
its value it is usually very close to the spectral slope at $\nu_{\rm
max}/2$ ($5\;$keV), except when the break frequency $\nu_p$ of the
Band spectrum passes through the XRT range, and the change in $\beta$
within this range is the largest. Therefore,
$\langle\beta\rangle_{\log\nu}$ appears to better reflect the spectral
slope measured over a finite frequency range.\\

\subsection{Spectrum}

The local spectral emissivity in the comoving frame is taken to be a
Band function. We have seen previously that for the
parameter values relevant for internal shocks ($m = 0$, $d = -1$), the
observed spectrum is also a pure Band function. This is evident in
the right panel of figure~\ref{fig_specslope}, which shows the temporal
 evolution of the
observed spectrum in our model. It results from the fact that for
these parameter values the observed peak frequency $\nu_p$ is constant
along the EATS. We have $\nu_p/\nu_0=E_p(T)/E_0 =
1/\tilde{T} = 1/(1+\bar{T})$ (see eq.~(\ref{F_nu_Band_R0})) which is 
independent of $\bar{T}_f$. This behaviour is evident in the right panel 
of  figure~\ref{fig_specslope}, where $E_p/E_0$ is $1$ at the onset of the
spike ($\bar{T} = 0$), $E_p/E_0 = 1/2$ at the peak of the spike
($\bar{T} = \bar{T}_f = 1$), and $E_p/E_0$ decreases roughly linearly with
$\bar{T}$ at later times, during the tail of the pulse.

\section{Combining pulses to obtain the prompt emission} \label{sec_prompt}

There is good observational evidence that the steep decay phase is the
tail of the prompt emission (O'Brien et al. 2006). Within our model,
the prompt emission is the sum over a finite number of pulses, and
therefore the steep decay phase is the sum of their tails. In this
section we provide examples of combining several pulses to model the
prompt emission, and study the effect of varying the different pulse
parameters. To this end, we start with a simple prompt emission model
consisting of six pulses that are identical except for their ejection
time $T_{\rm ej}$ (see Fig.~\ref{fig_var_param}a). Each pulse 
corresponds to a single emission episode of a particular shell that 
was ejected at $T_{{\rm ej},i}$ (for $i$'th pulse), has an initial 
radial time $T_{0,i}$, and a final angular time of $T_{f,i}$.
Then, we study the effect of changing the other model parameters one
by one among the pulses. All lightcurves in this section are drawn 
against $T$, as the ejection time is different for each pulse (and 
then the definition of a $\tilde{T}$ $\bar{T}$ would differ for each 
pulse). In Fig.~\ref{fig_var_param}b the peak flux $F_{\rm peak}$ is
varied.  Next, we vary $T_0$ and/or $T_f$. In Fig.~\ref{fig_var_param}c, 
$T_0$ is varied while $T_f/T_0$ remains constant. In
Fig.~\ref{fig_var_param}d, $T_f$ and $\Delta R/R_0$ vary while $T_0$ and 
$R_0$ remain constant. In Fig.~\ref{fig_var_param}e, $T_0$ and 
$\Delta R/R_0$ vary while $T_f$ and $R_f$ remain constant. 
Each of these panels show the light curve in
logarithmic scales, $T = 0$ is set to the onset time of
the first pulse, which means that $T_{\rm ej,1} = -T_{0,1}$, thus
showing the modeled prompt from a time close to what would be the
trigger time for an observed burst. The red solid line represents the
total prompt emission (the sum of all the pulses), while the black non solid
lines are the individual underlying pulses. All the examples shown
here of the prompt emission are for $\nu/\nu_0 = 0.1$.

In the case of six equal pulses (Fig.~\ref{fig_var_param}a), later pulses
appear to decay much more steeply just after their peak in a
logarithmic scale with the zero time near the beginning of the first
pulse. At very late times the relative contribution from the different
pulses becomes almost the same. As the only parameter that varies
between pulses is the ejection time, $T_{\rm ej}$, this change in
slope must depend only on it. Noting that the temporal slope is 
$\alpha \equiv -d \log F_{\nu}/d\log T = \alpha_{ej}/(1-T_{\rm ej}/T)$,
 and that the value of $\alpha_{ej}$
just after the peak is independent of $T_{\rm ej}$ (it depends only on
$\tilde{T}_f$; see eq.~[\ref{eq_alpha_4cas_t}]), we can see that the 
value of $\alpha$ just after the peak scales as $\alpha_{\rm peak} = 
\alpha_{ej,{\rm peak}}(1+T_{\rm ej}/T_f)$, since $T = T_{\rm ej}+T_f$ is the 
time of the peak of the pulse.  Since the pulses are
equal they have the same $T_f$ and $ \alpha_{ej,{\rm peak}}$, it is
clear that $\alpha_{\rm peak}$ increases with $T_{\rm ej}$. At
late times when $T \gg T_{\rm ej}$, $\alpha$ approaches
$\alpha_{ej} = 2+\beta$.

When varying $F_{\rm peak}$ while fixing the other parameters (see
Fig.~\ref{fig_var_param}b), the relative flux from each pulse at very
late times is proportional to its $F_{\rm peak}$, so that the largest
contribution is from the pulse with the largest $F_{\rm peak}$.

At late times the observed flux density of a single spike scales as
$F_\nu \propto \tilde{T}^{\,b_2-2}$ (see, e.g.,
eq.~[\ref{F_expanded_latetime}]). If at the peak time of the spike,
which for simplicity is assumed here to be at $T = T_{\rm ej}+T_f$ (as
is usually the case), the observed photon energy is at the high-energy
part of the Band function, $E \geq E_* \equiv E_0(T_0/T_f)x_b$ or $\nu
\geq \nu_* \equiv \nu_0(T_0/T_f)x_b$, then
(using eq.~[\ref{F_expanded_latetime}]) the flux from the peak onwards
is simply given by
\begin{equation}\label{F_nu(F_peak1)}
F_{\nu \geq \nu_*}(T \geq T_{\rm ej}+T_f) =
F_{\nu,{\rm peak}}\left(\frac{T-T_{\rm ej}}{T_f}\right)^{b_2-2}\ ,
\end{equation}
while for $E/E_* = \nu/\nu_* < 1$ the expression is slightly more complicated,
\begin{equation}\label{F_nu(F_peak2)}
\frac{F_{\nu < \nu_*}}{F_{\nu,{\rm peak}}} =
\left\{ \begin{array}{ll}
\left(\frac{T-T_{\rm ej}}{T_f}\right)^{b_1-2}
e^{-(1+b_1)(\nu/\nu_0)(T-T_{\rm ej}-T_f)/T_0}
 & T_f \leq T-T_{\rm ej} \leq T_0 x_b \nu_0/\nu\ ,\\
\\
\left(\frac{T_0 x_b\nu_0}{T_f\nu}\right)^{b_1-b_2}
e^{b_2-b_1+(1+b_1)(T_f/T_0)(\nu/\nu_0)}
\left(\frac{T-T_{\rm ej}}{T_f}\right)^{b_2-2}
 &  T-T_{\rm ej} \geq T_0 x_b \nu_0/\nu\ ,
\end{array} \right.
\end{equation}
but the qualitative behaviour is still rather similar.  Therefore, the
flux ratio of two pulses with ejection times $T_{\rm ej,1} \leq T_{\rm
ej,2}$ and a comparable $E_p(\bar{T}_f) = (T_0/T_f)E_0$ (as is usually
the case for different pulses in the prompt emission of the same GRB),
at late times ($T > \max(T_{\rm ej,1}+T_{f,1},T_{\rm ej,2}+T_{f,2})$
and $T - T_{\rm ej,2} \gg T_{\rm ej,2}-T_{\rm ej,1}$) is approximately
\begin{equation}\label{eq_ratio_flux_Fpeak_tr}
\frac{F_{\nu,1}(T)}{F_{\nu,2}(T)} \sim \frac{F_{\rm peak,1}}{F_{\rm peak,2}}
\left(\frac{T_{f,1}}{T_{f,2}} \right)^{2+\beta}\ ,\quad{\rm for}\ \
\min[\bar{T}_{f,1},\bar{T}_{f,2}]>1\ \ {\rm and}\ \
T - T_{\rm ej,2} \gg T_{\rm ej,2}-T_{\rm ej,1}\ ,
\end{equation}
where $\beta = -b_2$ for $\nu \geq \nu_*$ while $\beta$ is generally
intermediate between $-b_2$ and $-b_1$ for $\nu < \nu_*$.

Fig.~\ref{fig_var_param}c demonstrates this nicely for a series of six
pulses with the same $F_{\rm peak}$ but decreasing $T_f$, so that the
later pulses with a smaller $T_f$ decay faster and become sub-dominant
at late times. At the latest times the first spike, which has the
largest $T_f$, dominates the observed flux in the tail emission.  A
similar behaviour is also seen in Fig.~\ref{fig_var_param}d.  In
Fig.~\ref{fig_var_param}e both $F_{\rm peak}$ and $T_f$ are the same
between the different pulses, and therefore their tail fluxes at late
times are similar. In Fig.~\ref{fig_var_param}c, $T_0$ and $T_f$ are varied
while $T_f/T_0$ is constant, and it can be seen that this corresponds
to a rescaling of the pulse width (its typical duration) without
effecting its shape. In Fig.~\ref{fig_var_param}d, $T_f$ and $T_f/T_0$ are
varied while $T_0$ is constant, and this nicely demonstrates how the
shape of the pulse depends on $T_f/T_0$. Typically,
the rise time of a pulse is $T_f-T_0$ while its decay time is $T_f$,
so that the ratio of the rise and decay time is $1-T_0/T_f$. 
In Fig.~\ref{fig_var_param}e, $T_0$ and
$T_f/T_0$ are varied while $T_f$ is constant. In this case the rise
time varies considerably between the different pulses while the decay
timescale and the late time tail of the pulses are practically the
same. This arises since the tail is dominated by emission from $R \sim
R_f$, that in this case is very similar for all the pulses.  Moreover,
for the particular choice of parameters in Fig.~\ref{fig_var_param}e,
where $E_p(\bar{T}_f) = (T_0/T_f)E_0$ and $E_* = x_bE_p(\bar{T}_f)$ remain 
constant for all the pulses, their late time
tails have the same flux normalization. This can be understood from
eq.~(\ref{F_nu(F_peak2)}), where the flux for $\tilde{T} >
\max(\tilde{T}_f,x_b\nu_0/\nu)$ can be written as
$F_\nu/F_{\nu,{\rm peak}} =
(E/E_*)^{b_2-b_1}\exp[(b_1-b_2)(E-E_*)/E_*]\tilde{T}^{\,b_2-2}$.

Fig.~\ref{fig_var_param}f shows a more realistic example of
the prompt emission, in which a larger number of model parameters is
varied between the different pulses. This example contains only 
three pulses in order to be clearer. It can be seen that the flux 
during the decaying phase is initially dominated by
the last pulse just after its peak ($T > 27\;$s), but the second peak
becomes dominant (even if only by a small factor) as early as $T \sim
37\;$s, and finally at $T \sim 140\;$s the first pulse becomes the
dominant one. This demonstrates that different pulses can dominate the
observed flux during the course of the steep decay phase. Which pulses
would contribute more to the steep decay phase can be estimated
according to their typical width (or duration), peak flux, and peak
time.  The peak time is most important at the beginning of the steep
decay phase, where the last spike always dominates just after its peak
if its peak is above the flux from the other spikes. Later on the
relative contribution of the different spikes can be estimated
according to eq.~(\ref{eq_ratio_flux_Fpeak_tr}).
Since the late time flux scales as $F_{\rm peak}T_f^{2+\beta}$ and
usually $0 \lesssim \beta \lesssim 2$, the power of $T_f$ (which
corresponds to the typical width of the spike) is higher than that of
$F_{\rm peak}$, so that wider spikes tend to dominate over narrower
spikes, even if their peak flux is somewhat lower.

One should be very careful when fitting actual data with such a model.
Fig.~\ref{fig_compfit} shows what can happen if because of noisy data
or coarse time bins, a prompt emission (red solid line) which is
actually composed by several pulses (three, six or twelve in the cases
shown; black non-solid lines) is fitted by a single broad pulse (green
solid line). In this case the tail of the prompt emission can be
significantly overestimated at late times, by a factor that tends to
increase with the true number of underlying pulses. This can be
understood by the simple example of comparing a single spike with $N$
identical spikes with the same peak flux but a duration smaller by a
factor of $N$, for which the sum of their late time tail flux would be
smaller than that of the single pulse by a factor of $\sim
N^{1+\beta}$. However, in more realistic examples, the late time flux
would often be dominated by the widest underlying pulse, so that its
width would be more important than the total number of narrower
underlying spikes. It is important to keep this effect in
mind when confronting such a model with actual data.

\section{Discussion and conclusions} \label{sec_concl}

We have presented and explored a model for the prompt GRB emission and
its high latitude emission (HLE) tail. This model is physically
motivated and realistic: it consists of a finite number of emission
episodes, each of which corresponds to a single spike in the prompt
light curve, and is modeled by a relativistically expanding thin
spherical uniform shell emitting isotropically in its own rest frame
within a finite range of radii. Our model thus describes the prompt
emission and the steep decay pahse as a whole from its very start to
its late tail. Yet this model is easy to use (fully analytic in its
simplest form described here), making it particularly suitable for
detailed combined temporal and spectral global fits to the prompt GRB
emission and the following steep decay phase (SDP). Such fits can
provide a stricter test of the HLE model for the SDP compared to most
previous models, since we use a single self-consistent model to fit
both the prompt emission and the SDP, while most previous models fit
only the SDP and are largely decoupled from the details of the prompt
emission. Moreover, our model is also physically motivated, and more
realistic than previous models. We have derived analytic expressions
for the flux in the realistic case of a Band function spectrum
(eqs.~[\ref{eq_bandDR_finalcondensed}] and
[\ref{F_expanded_latetime}]), which consists of two power laws that
smoothly join at some typical photon energy.

The temporal evolution of the instantaneous values of the temporal
($\alpha$) and spectral ($\beta$) indexes for a single emission
episode was studied, corresponding to a single observed pulse in the
light curve. The definition of $\alpha$ is not unique as it depends on
the choice of reference time. We explored two options for the
reference time, either the ejection time ($\alpha_{ej}$) or the onset
time of the spike ($\alpha_{on}$), and found that for the former the
HLE relation ($\alpha_{ej} = 2 + \beta$) is satisfied from immediately
after the peak of the spike ($\bar{T} > \bar{T}_f$), while for the
former it is only approached at late times ($\alpha_{on} \approx 2 +
\beta$ at for $\bar{T} > \bar{T}_f$ and $\bar{T} \gg 1$).

We have intentionally chosen a simple model to describe the pulses, in
order to reduce the number of free parameters.
For a single emission episode (or pulse), in the most generic case 
there are ten free
parameters: the power $m = -2d\log\Gamma/d\log R$, $d = d\log
\nu'_p/d\log R$, $a = d\log L'_{\nu'_p}/d\log R$, the normalization
factor $F_0$ (or $L_0$), three additional parameters for the
Band function (the two spectral slopes, $b_1$ and $b_2$, as well as
the peak energy at the onset of the pulse $E_0$), the two timescales
$T_0$ and $T_f$, and the ejection time $T_{\rm ej}$. We have the
general constraint $\Delta R > 0$, which implies $\tilde{T}_f= 1 + \Delta
R/R_0 > 1$. Focusing on the internal shocks model fixes some 
of these parameters: as the outflow is typically in the coasting
phase, $m=0$, while for synchrotron emission from fast cooling
electrons $d = -1$ and $a = 1$.  Since we expect $\Delta R/R_0 =
\bar{T}_f \sim 1$ we can fix $\bar{T}_f
\sim 1$ (although a wider range, such as $0.2 \Delta R/R_0 \lesssim 5$,
may still be considered as plausible). Fixing $m$, $d$, $a$, and
$T_f/T_0$ in this manner would leave only six free parameters. For a
prompt emission with several pulses, one may be able in some cases to
neglect the spectral evolution and use the same values of $b_1$,
$b_2$, and $E_0$ for all the different pulses (or at least two of
them, e.g. $b_1$ and $b_2$), which leads to a total number of free
parameter of $3(N_{\rm pulses}+1)$ (or $4N_{\rm pulses}+2$ if $E_0$
cannot be fixed for all the pulses) for a burst with $N_{\rm pulses}$
pulses.

\begin{figure}
\begin{center}
\includegraphics[width=0.465\textwidth,
height=0.465\textwidth]{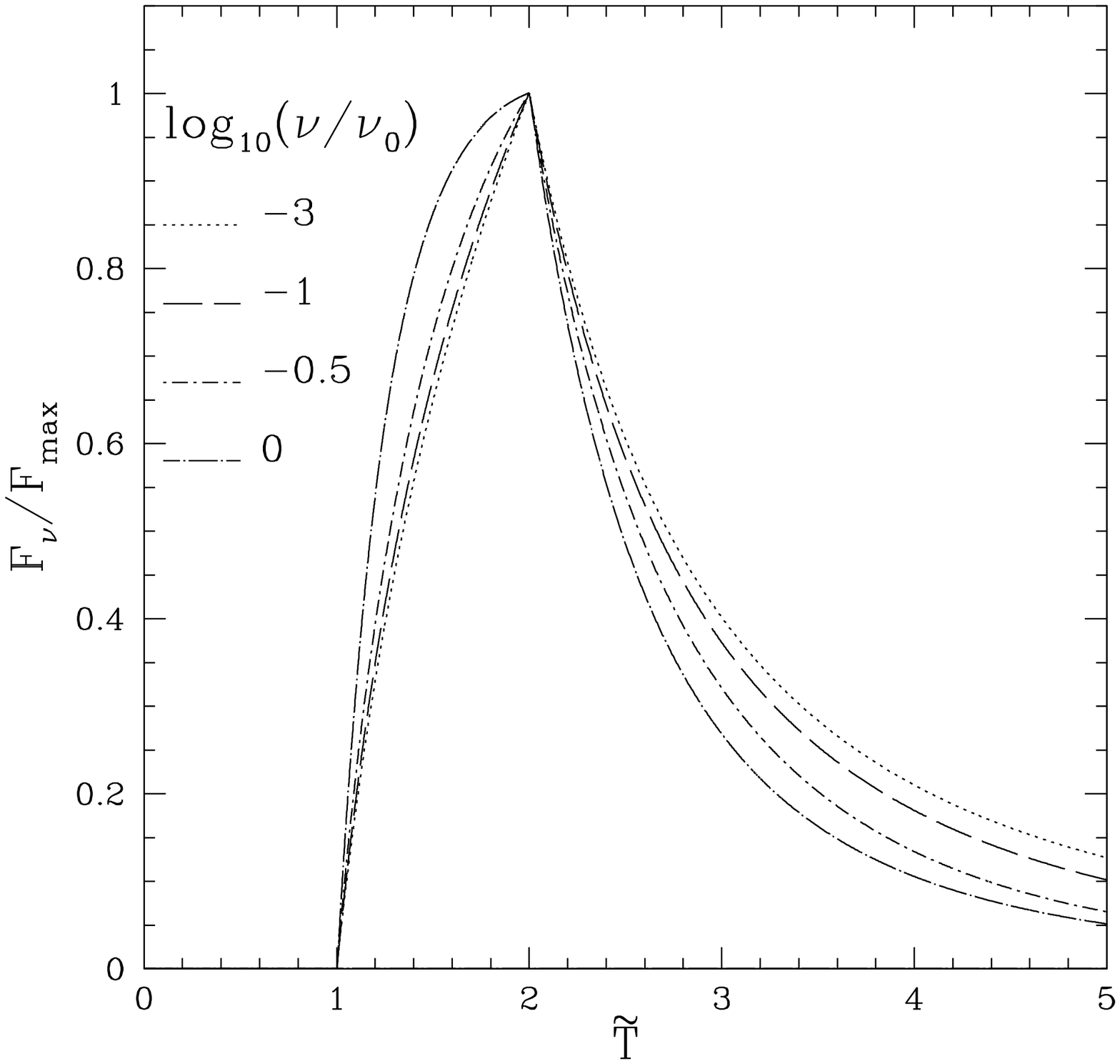}
\hspace{0.8cm}
\includegraphics[width=0.465\textwidth,
height=0.465\textwidth]{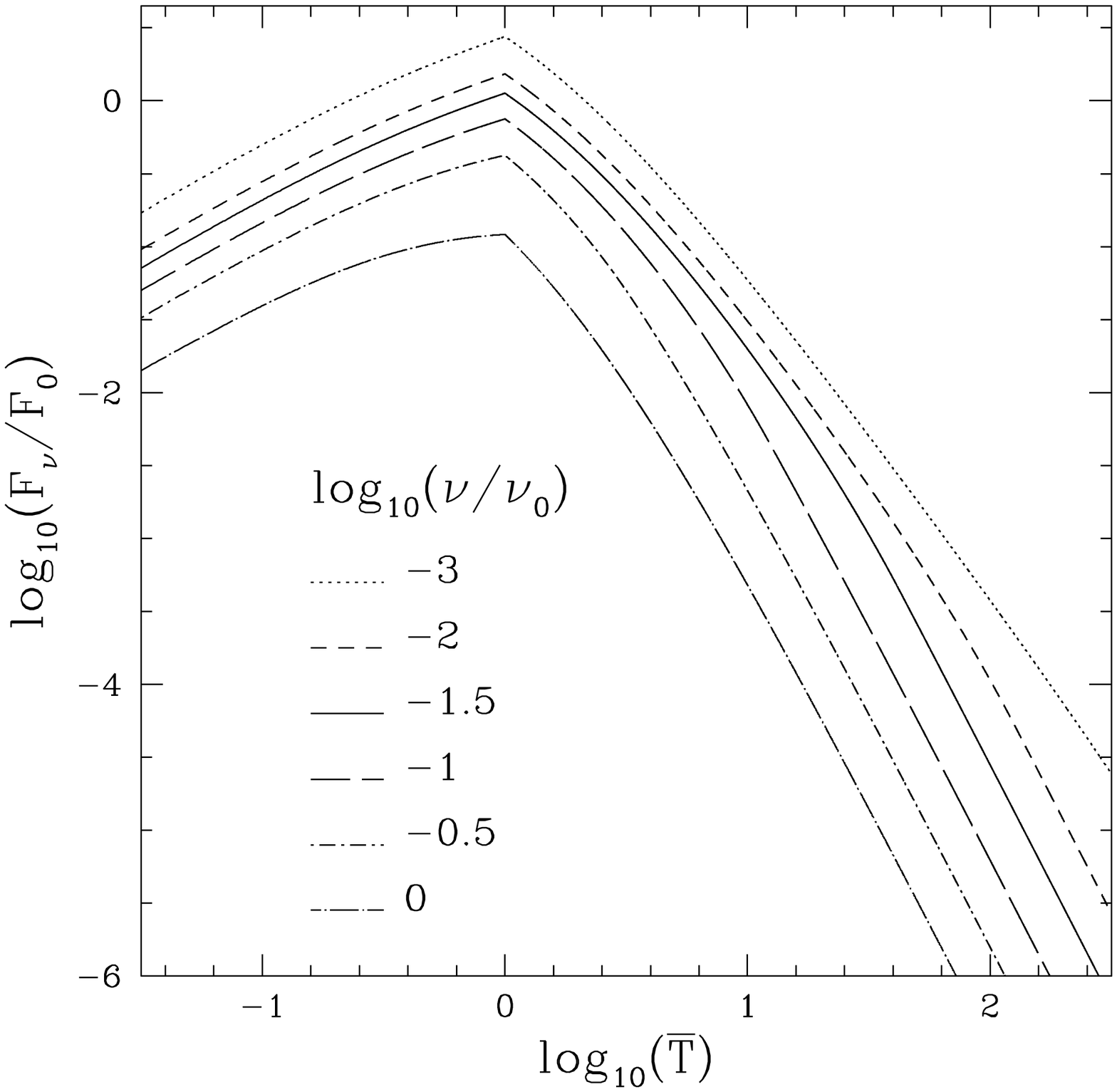}
\end{center}
\caption[LC at different frequencies]{
\emph{Lightcurves of a single pulse at different normalized frequencies,
$\nu/\nu_0$. The low and high energy slopes of the spectrum are
$b_1=-0.25$ and $b_2=-1.25$, while $a=1$. $\Delta R/R_0=1$, so that 
$\tilde{T}_f=2$ and $\bar{T}_f=1$.} {\bf Left}: \emph{Normalized flux 
density shown as a function of $\tilde{T}$ in linear scale.} 
{\bf Right}: \emph{Flux density shown as a function of $\bar{T}$ in 
logarithmic scale.}}
\label{fig_LC_onepulse_selon_freq}
\end{figure}

\begin{figure}
\begin{center}
\includegraphics[width=0.465\textwidth,
height=0.4\textwidth]{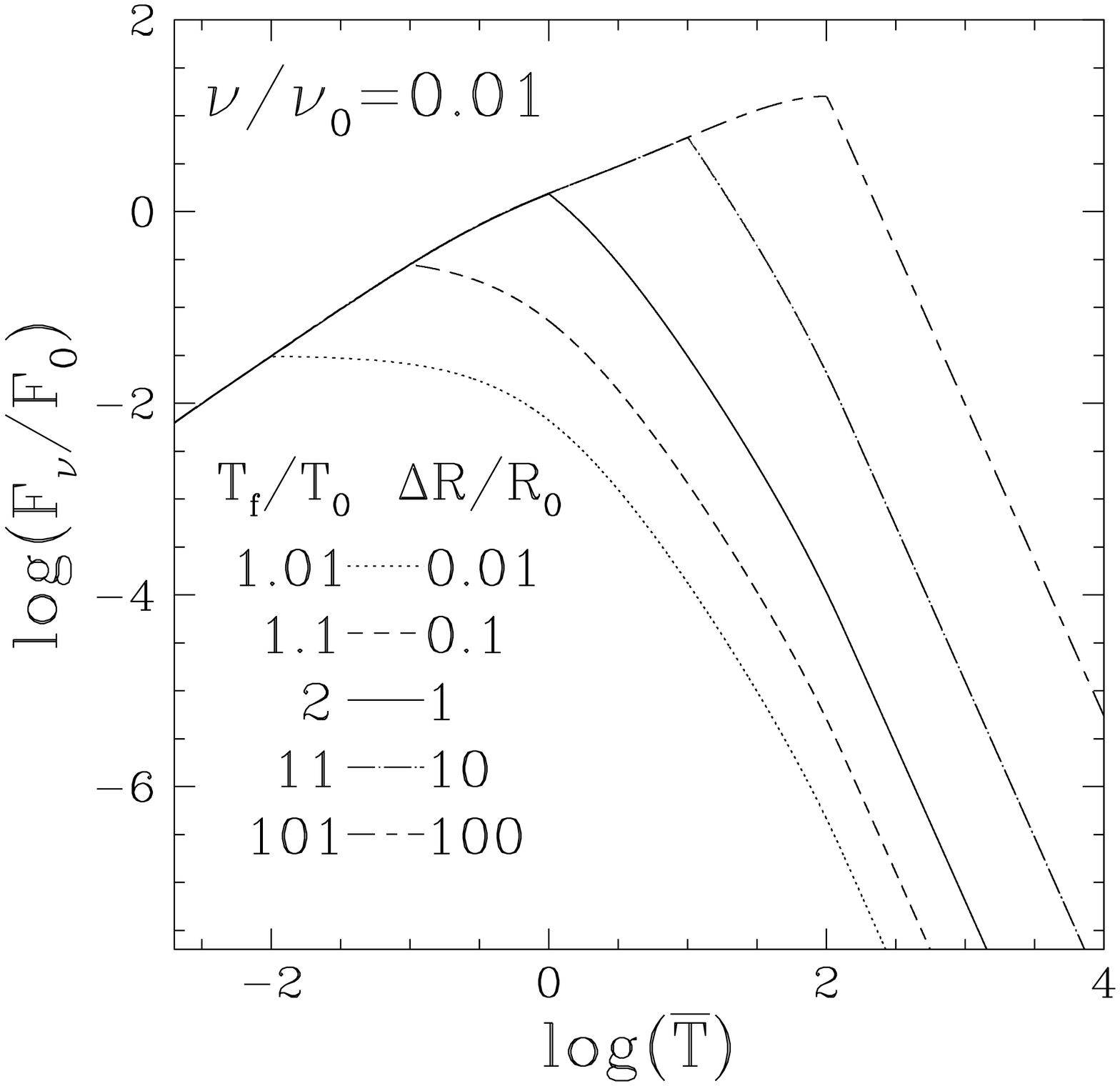}\includegraphics[width=0.465\textwidth,
height=0.4\textwidth]{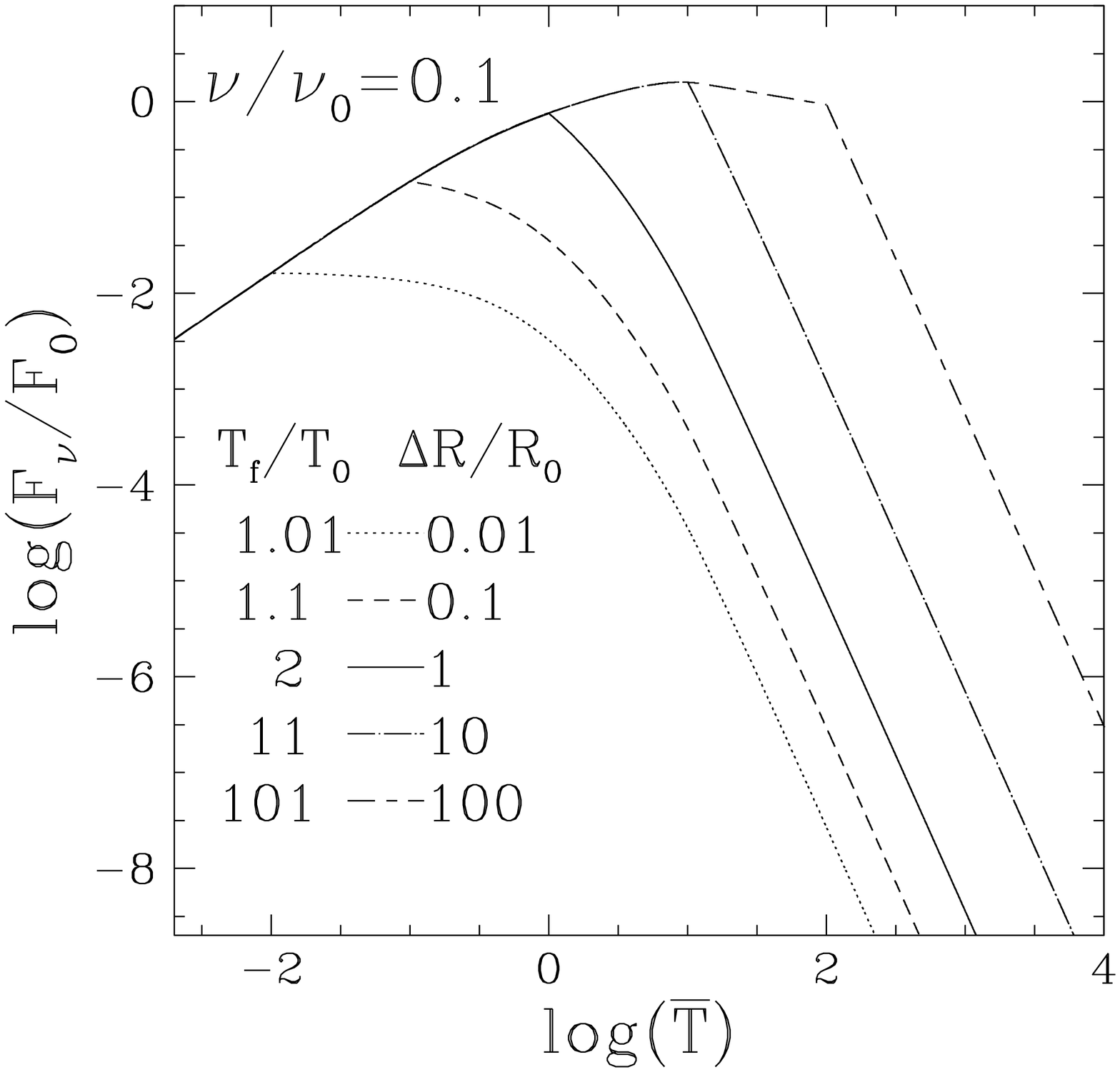}
\includegraphics[width=0.465\textwidth,
height=0.4\textwidth]{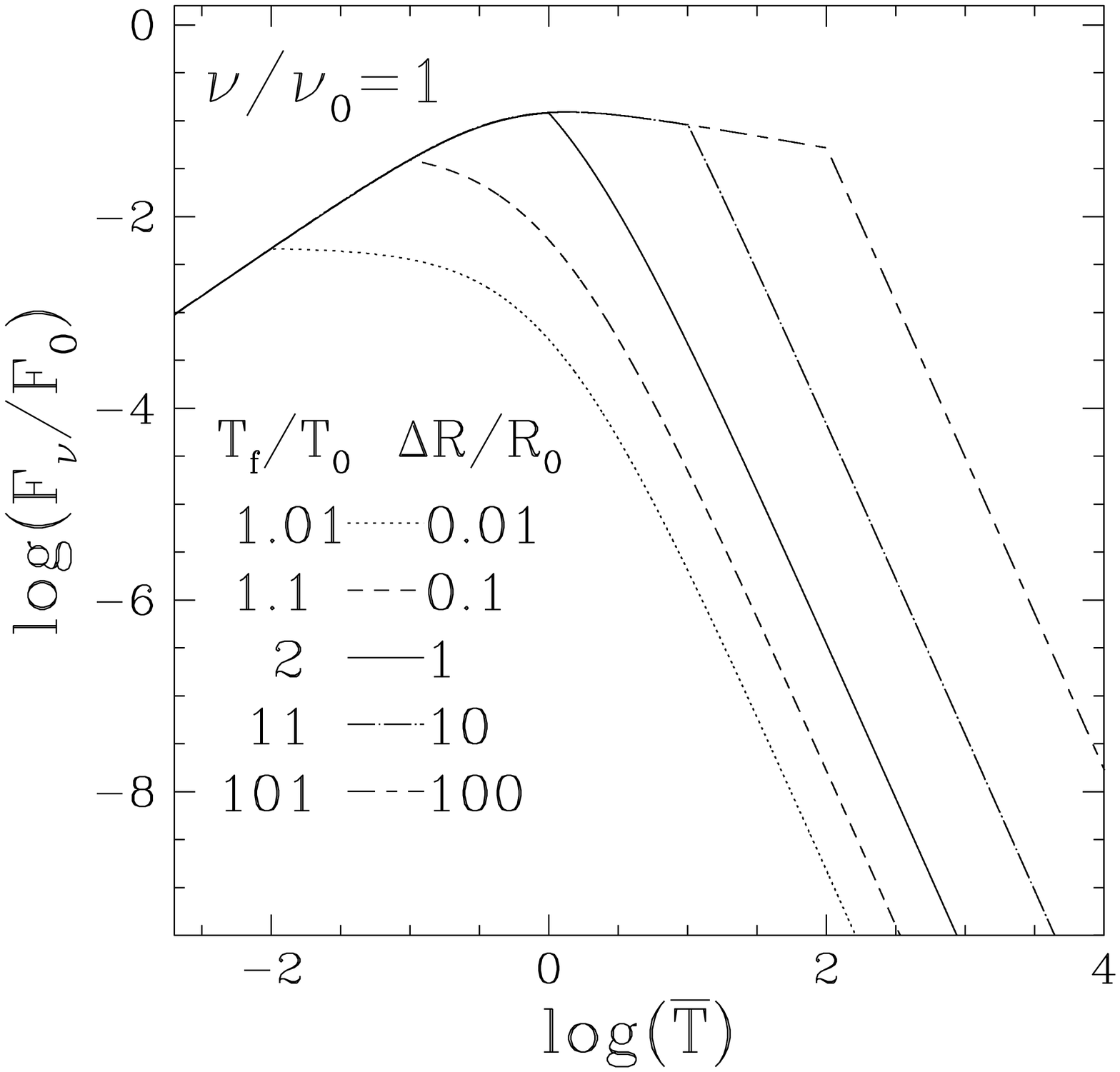}\includegraphics[width=0.465\textwidth,
height=0.4\textwidth]{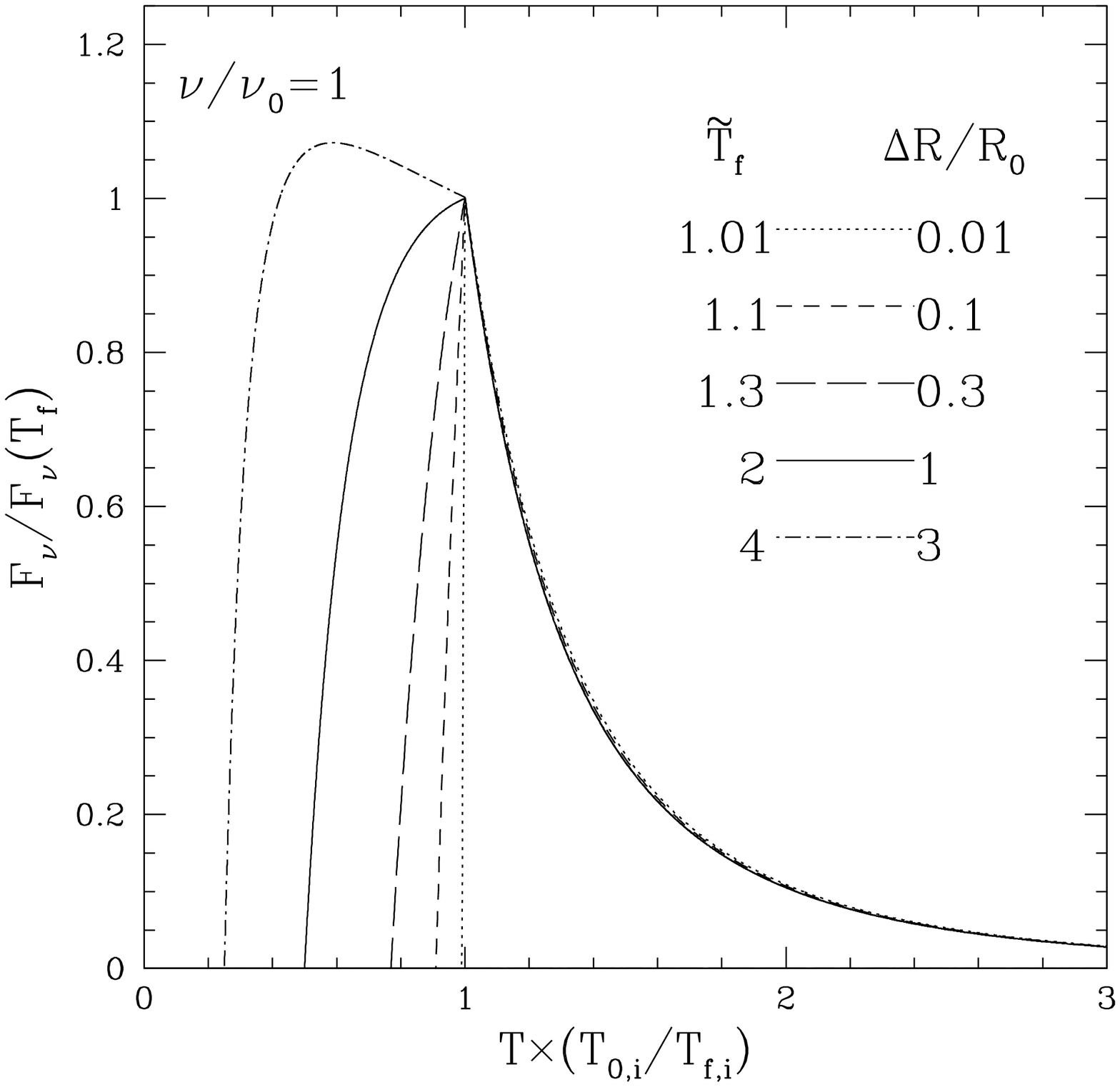}\\
\end{center}
\caption[evolution of LC with DR]{
\emph{The same pulse as in figure~\ref{fig_LC_onepulse_selon_freq}
is shown for different values of $\bar{T}_f$ for
$(\nu/\nu_0) = 0.01$, $0.1$, and $1$ in the first, second and third panel
respectively
(in logarithmic scale). The fourth panel shows the case $(\nu/\nu_0) = 1$ in
linear
scale in order to show the shape of a pulse having its peak before $T=T_f$. The
normalized flux density is shown as a function of $T\times
T_{0,i}/T_{f_i}$ where the subscript $i$ denotes the $i$'th pulse, so
that all the $T_{f,i}$ would appear to coincide, and the decay times
of the different pulses would appear to be the same.}}
\label{fig_LC_spectra_DR}
\end{figure}

\begin{figure}
\begin{center}
\includegraphics[width=0.33\textwidth,
height=0.4\textwidth]{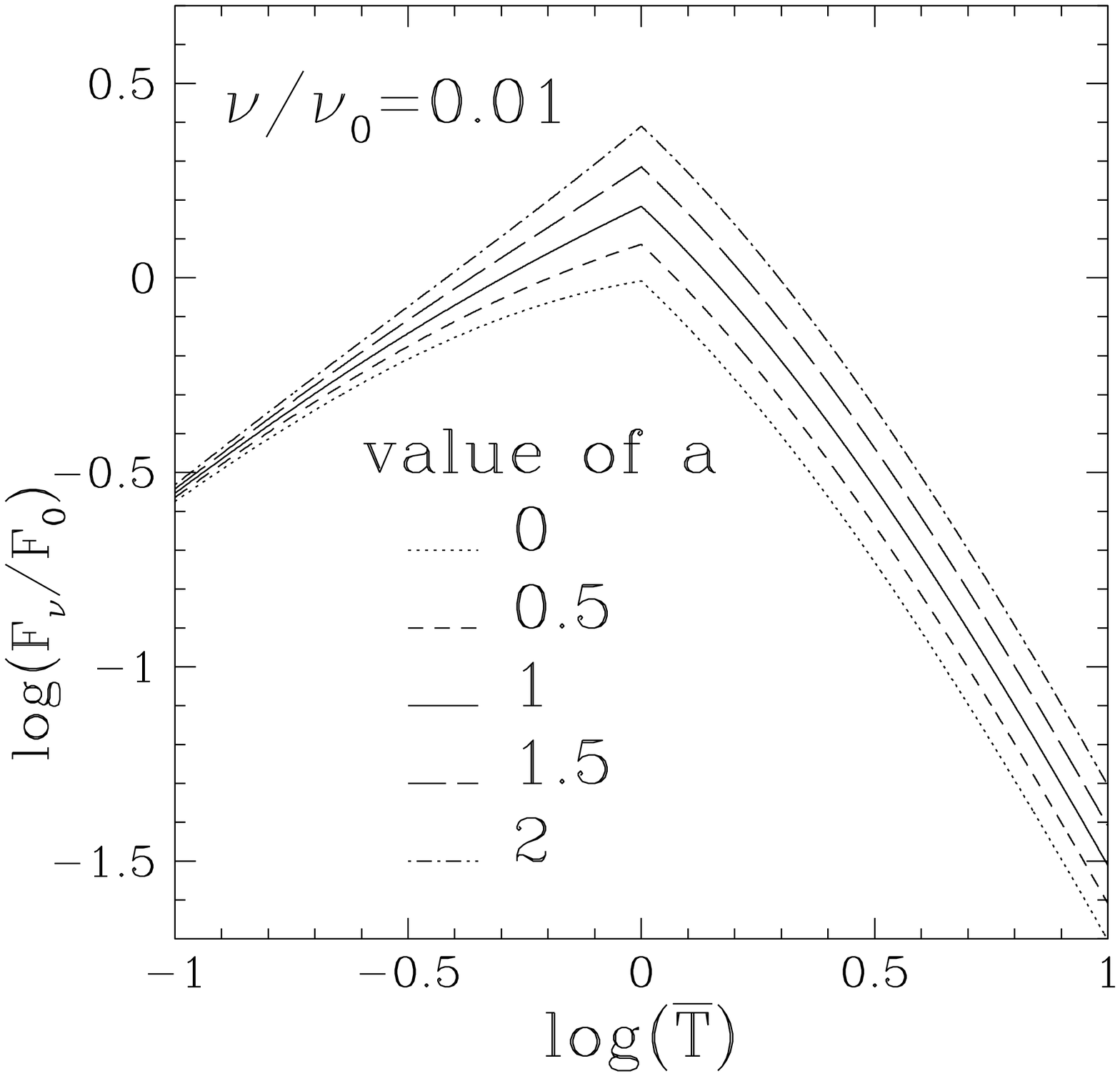}\includegraphics[width=0.33\textwidth,
height=0.4\textwidth]{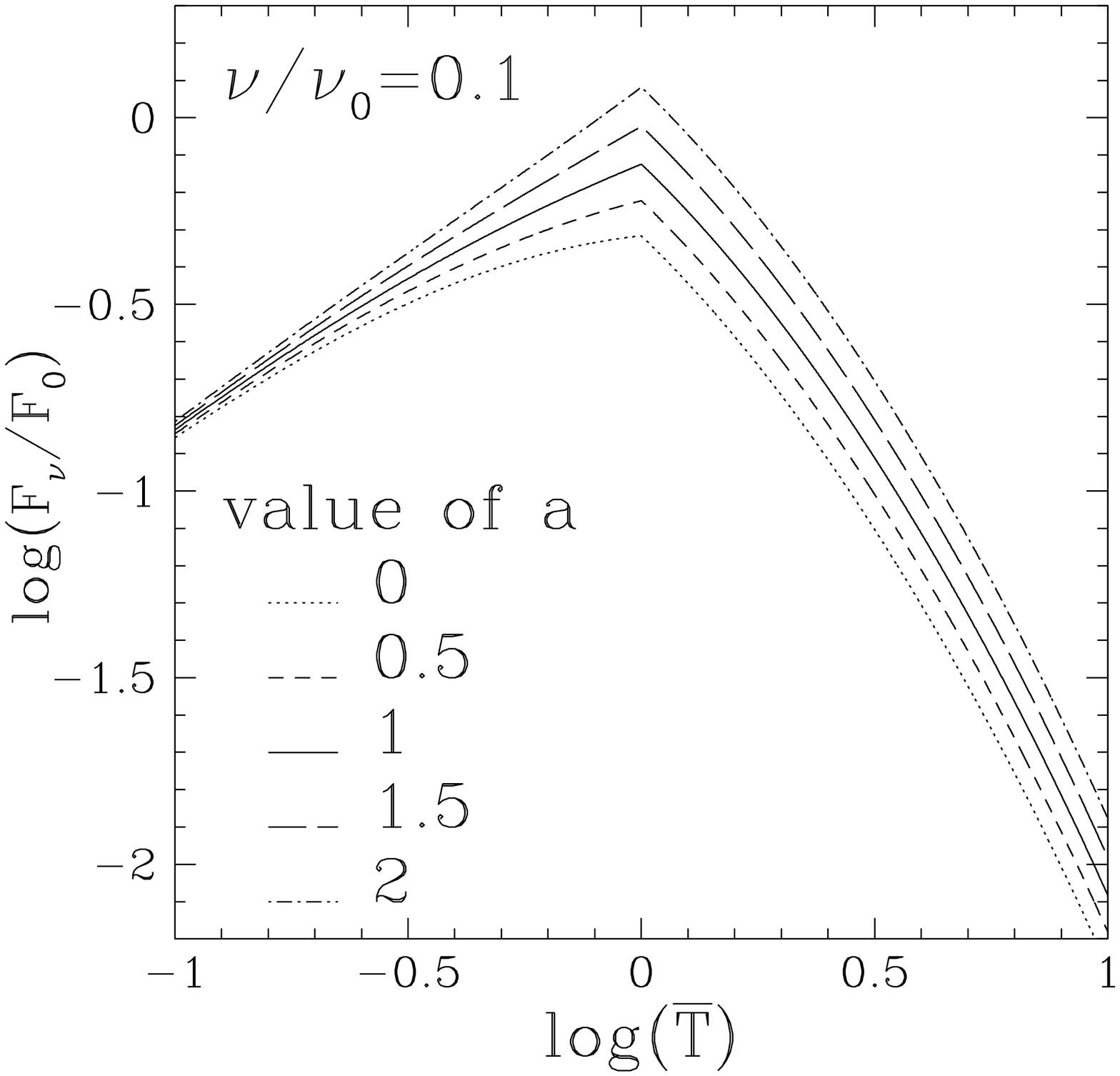}\includegraphics[width=0.33\textwidth,
height=0.4\textwidth]{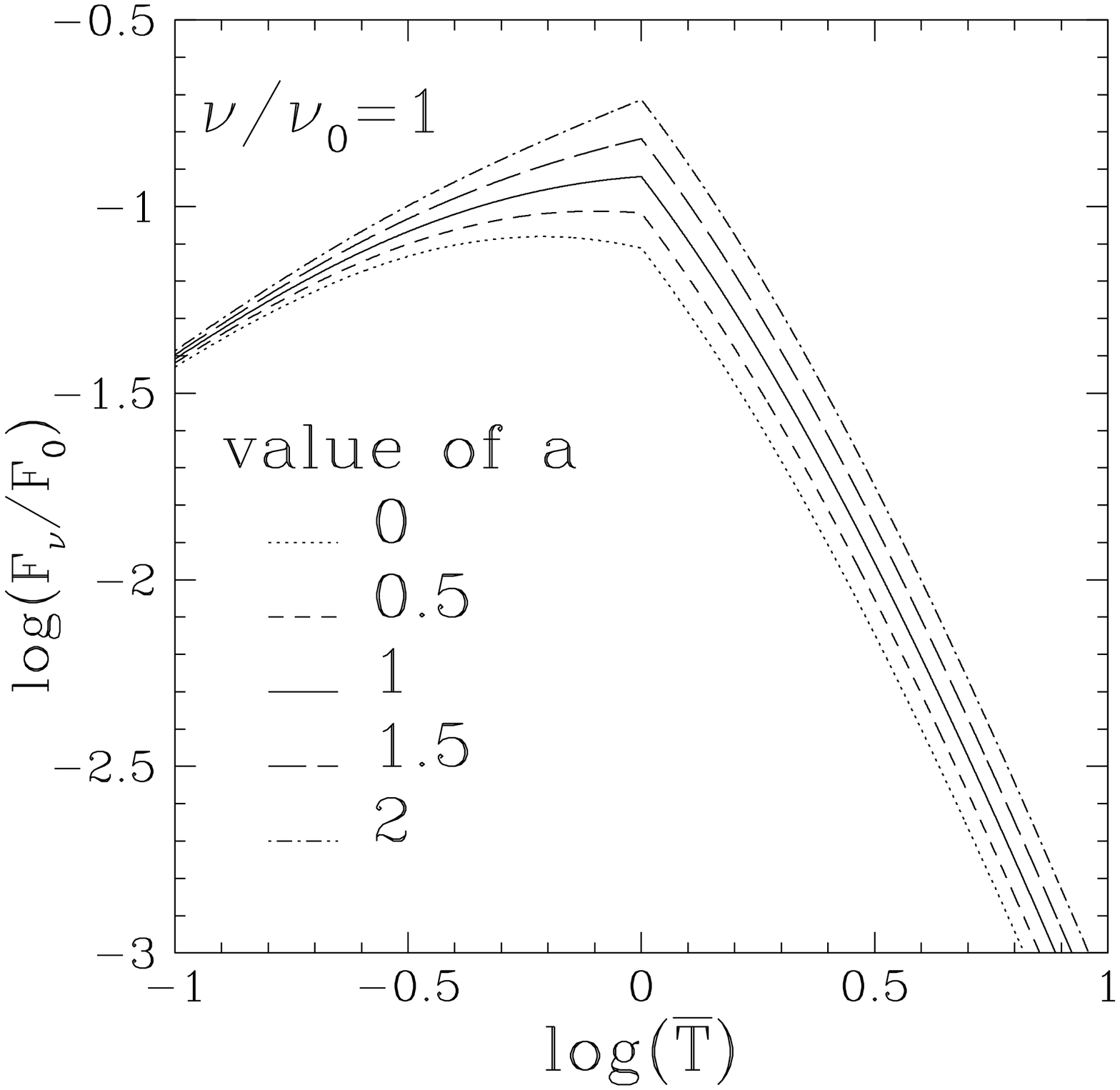}
\end{center}
\caption{\emph{Effect of the variation of $a$ on the shape of a pulse for
$(\nu/\nu_0) = 0.01$, $0.1$, and $1$ in the first, second and third panel
respectively
(in logarithmic scale). We can see that increasing $a$ makes the pulse sharper.
The constant parameters are $T_0=1$ s and $T_f=2$ s and the low and
high energy spectral slopes are $b_1=-0.25$ and $b_2=-1.25$.}}\label{fig_var_a}
\end{figure}

\begin{figure}
\begin{center}
\includegraphics[width=0.465\textwidth,
height=0.465\textwidth]{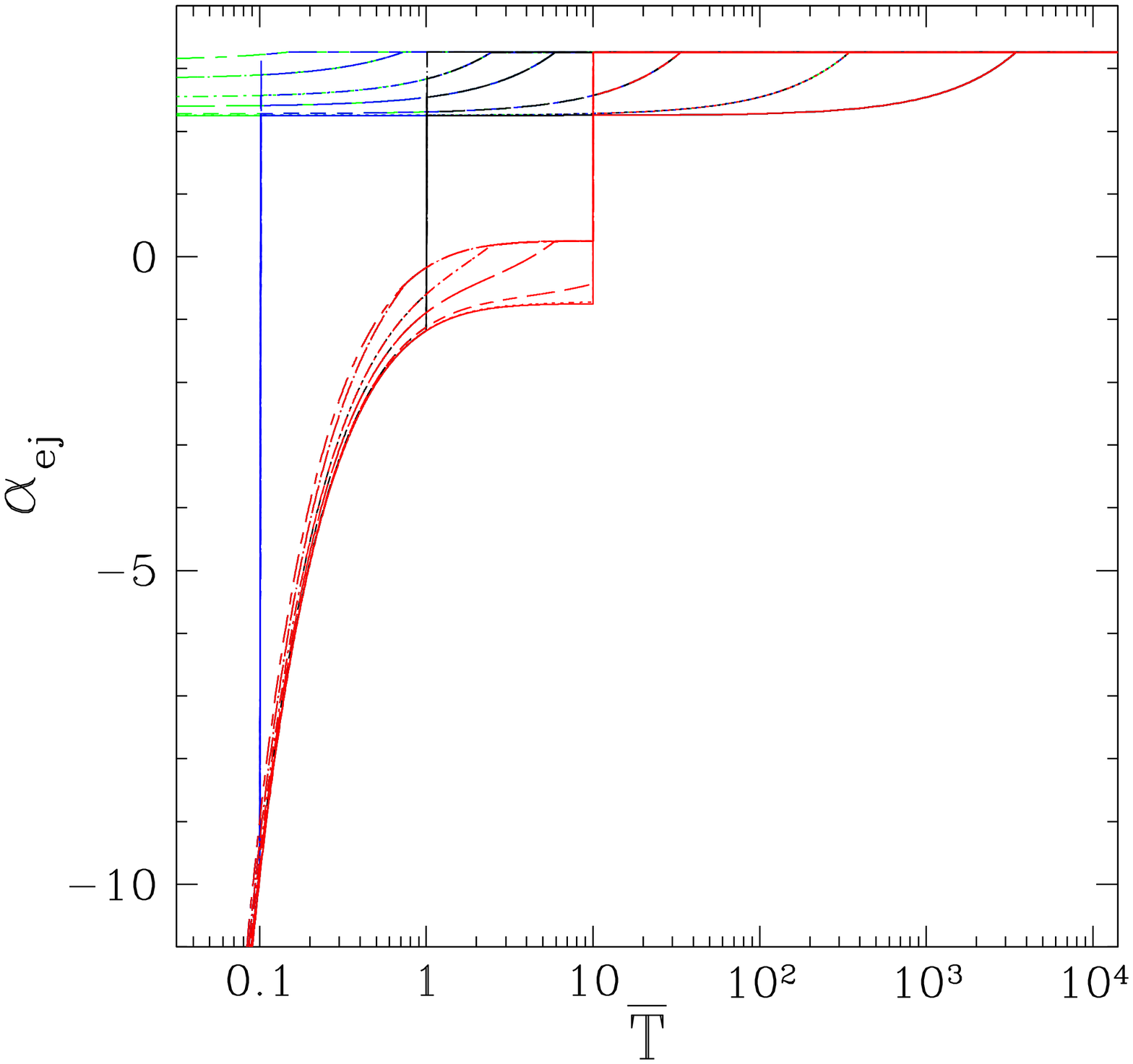}
\includegraphics[width=0.465\textwidth,
height=0.465\textwidth]{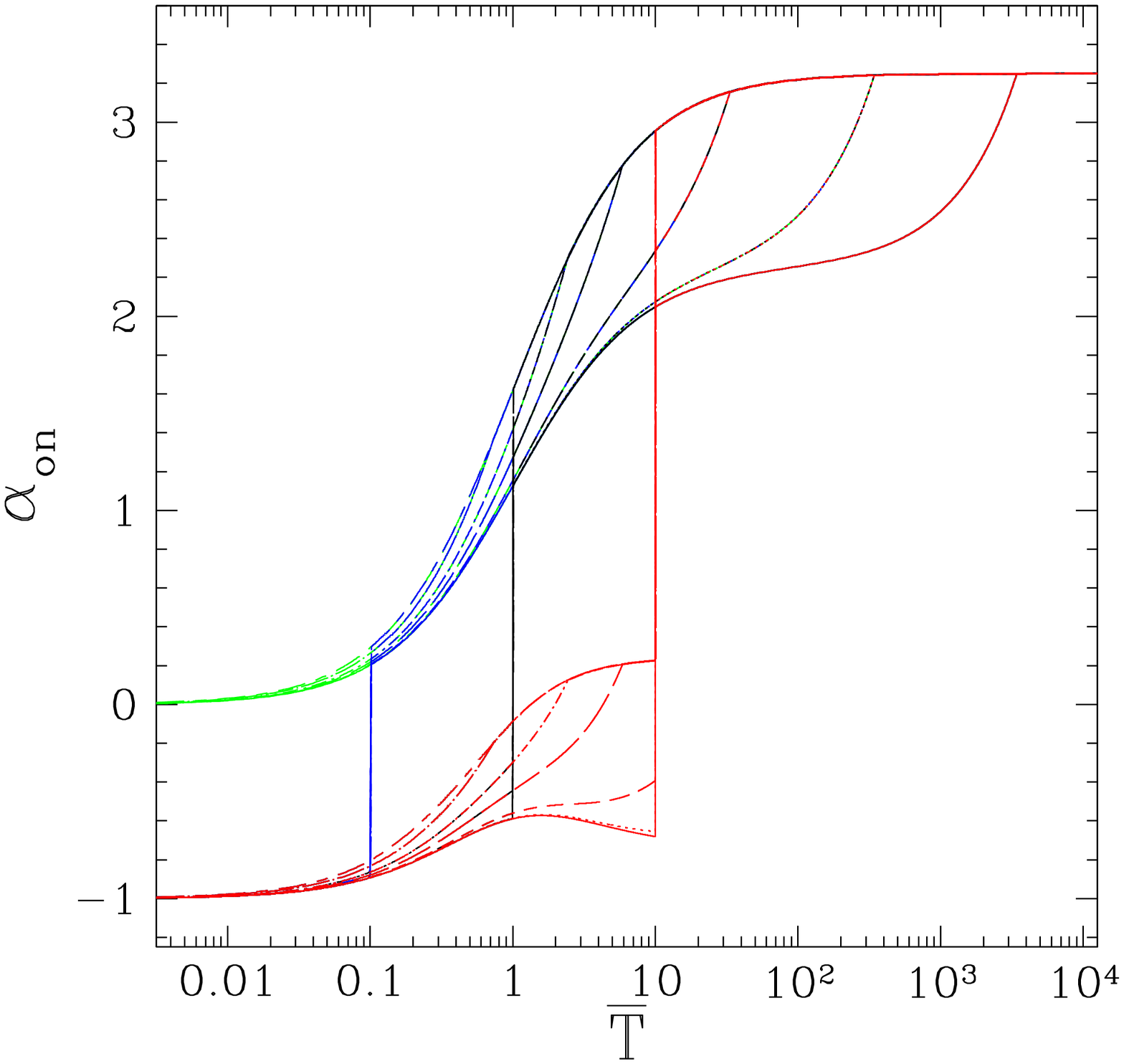}
\end{center}
\caption[temporal slopes]{
\emph{Evolution of the temporal indexes $\alpha_{ej} = -d\log
F_\nu/d\log\tilde{T}$
({\rm left panel}) and $\alpha_{on} = -d\log F_\nu/d\log\bar{T}$
({\rm right panel}) with normalized observed time $\bar{T}$, at
different observed photon energies (for $E_0 = 300\;{\rm
keV}$). Different line styles are used for the different
energies. The color coding shows the temporal indexes for several values of
$\bar{T}_f = \Delta R/R_0$: $0$ (green), $0.1$ (blue), $1$ (black) and
$10$ (red). The low and high energy spectral slopes are
$b_1=-0.25$ and $b_2=-1.25$, while $a=1$.}} \label{fig_tempslopes}
\end{figure}

\begin{figure}
\begin{center}
\includegraphics[width=0.33\textwidth,
height=0.4\textwidth]{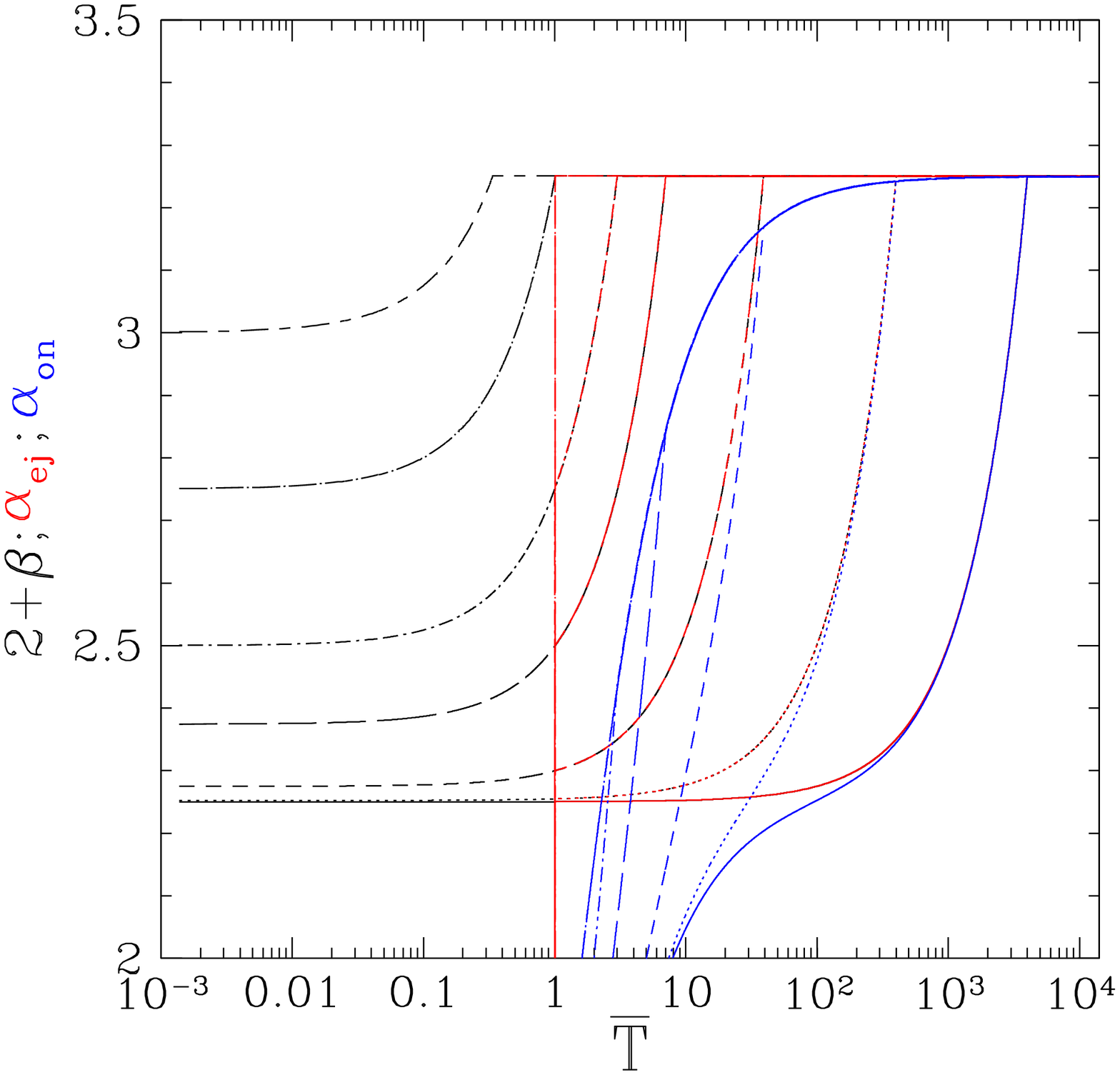}\includegraphics[width=0.33\textwidth,
height=0.4\textwidth]{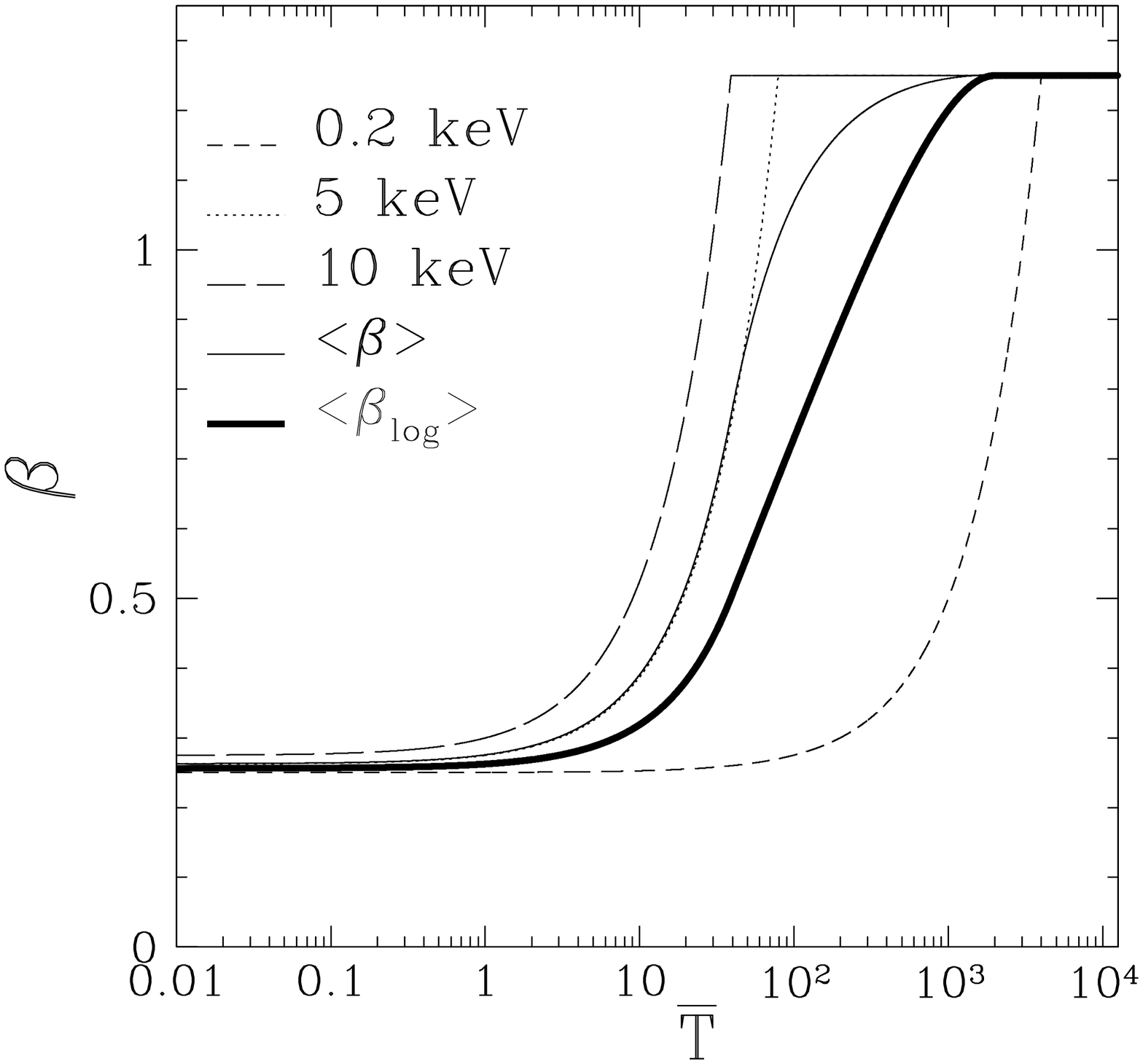}\includegraphics[width=0.33\textwidth,
height=0.4\textwidth]{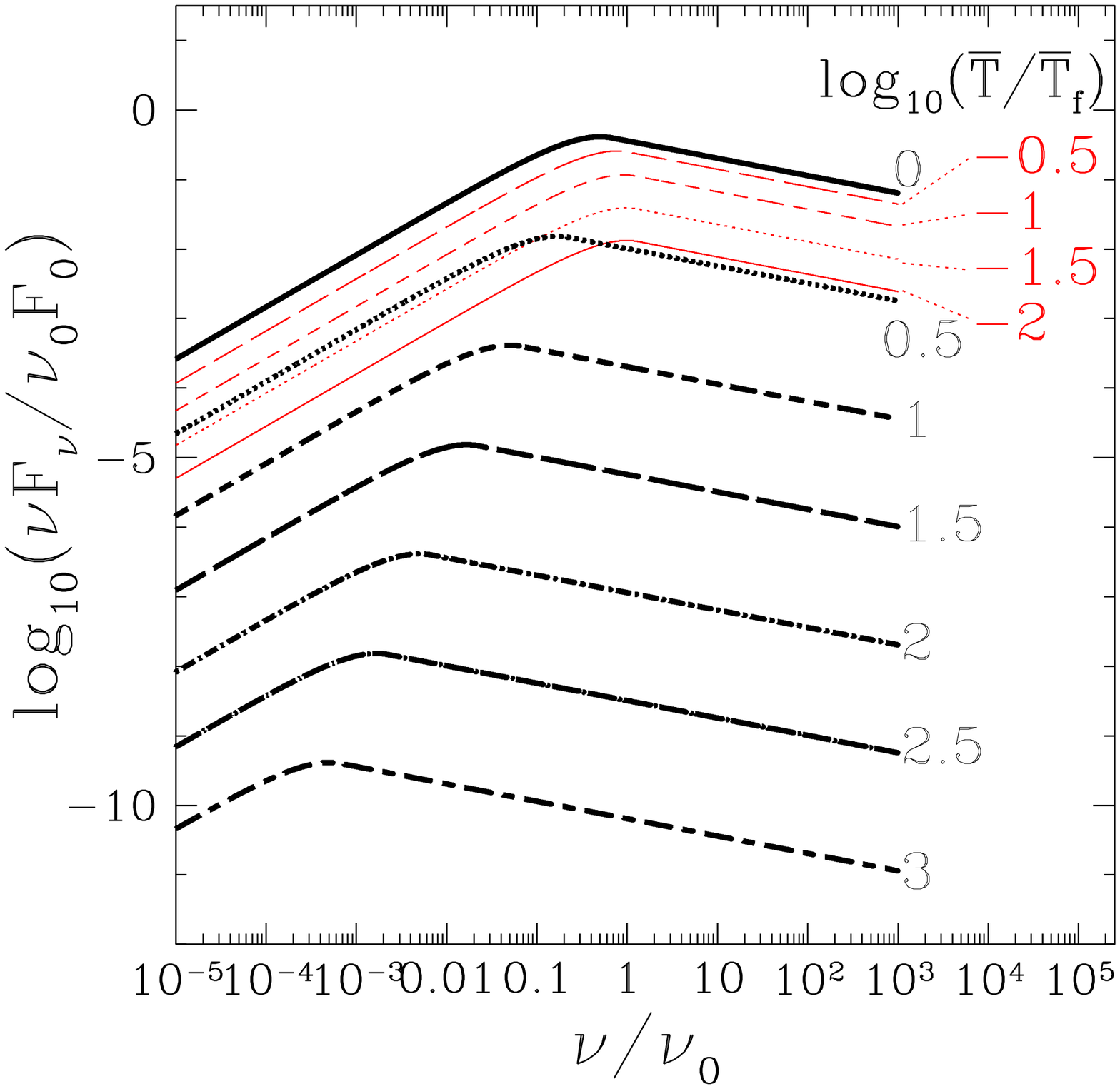}
\end{center}
\caption[spectral slopes]{
{\bf Left panel}: \emph{Comparison of the evolution of the spectral ($\beta$ -- thin lines)
and temporal ($\alpha_{ej}$ -- thick lines -- and $\alpha_{on}$ -- very thick lines) 
indexes at different photon 
energies ($E_0=300\;$keV). The low and high energy spectral slopes are $b_1 = -0.25$
and $b_2=-1.25$, while $a=1$.} \label{fig_specslope}
{\bf Middle panel}: \emph{Evolution of spectral index $\beta$ over the {\it
Swift}
XRT energy range ($\nu_{\rm min} < \nu < \nu_{\rm max}$ with $\nu_{\rm
min} = 0.2\;{\rm keV}$ and $\nu_{\rm max} = 10]\;{\rm keV}$. Shown are
the local values of $\beta$ at $\nu_{\rm min}$ ({\rm short dashed
line}), $\nu_{\rm max}/2$ ({\rm dotted line}) and $\nu_{\rm max}$
({\rm long dashed line}), as well as the average values of $\beta$
over the XRT range, taken either over $\nu$ ({\rm thin solid line}) or
over $\log\nu$ ({\rm thick solid line}). The low and high energy
spectral slopes are $b_1=-0.25$ and $b_2=-1.25$, while $a=1$.}
\label{fig_specslope_swift}
{\bf Right panel}: \emph{Evolution of the observed spectrum with time. The
spectrum,
$\nu F_\nu/(\nu_0 F_0)$, is shown as a function of the normalized
frequency, $\nu/\nu_0$, for different values of the normalized time,
$\log_{10}(\bar{T}/\bar{T_f})$, where we have used $\bar{T}_f = 1$. 
The red thin lines correspond to the rising stage of the
pulse $(\bar{T} < \bar{T}_f)$, while the black thick lines are for its
peak $(\bar{T} = \bar{T}_f)$ and decaying stage $(\bar{T} >
\bar{T}_f)$. The low and high energy spectral slopes of the spectrum
are $b_1=-0.25$ and $b_2=-1.25$, while $a=1$.} \label{fig_spectrum}
}
\end{figure}

\begin{figure}
\begin{center}
\includegraphics[width=0.31\textwidth,
height=0.4\textwidth]{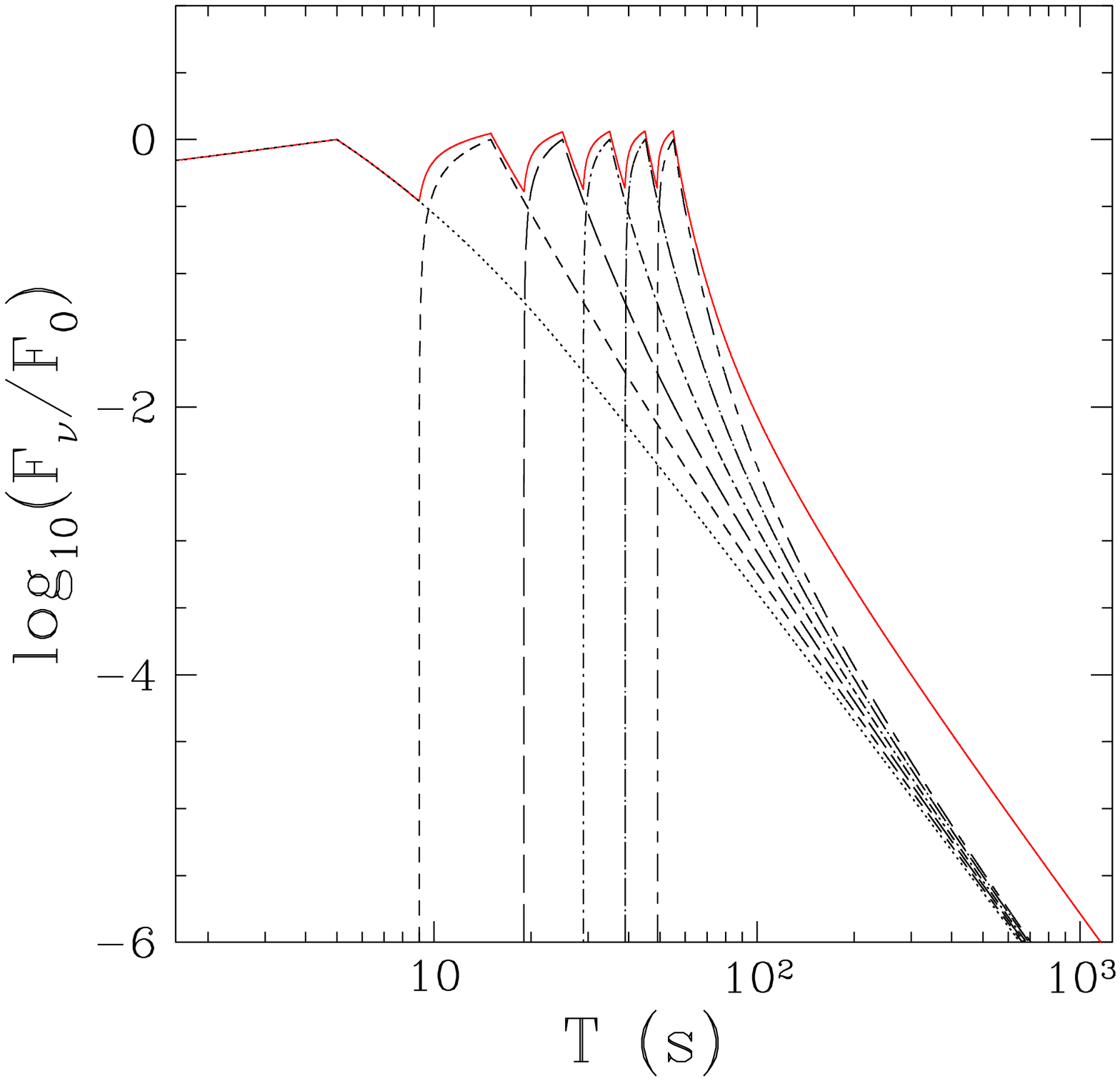}\includegraphics[width=0.31\textwidth,
height=0.4\textwidth]{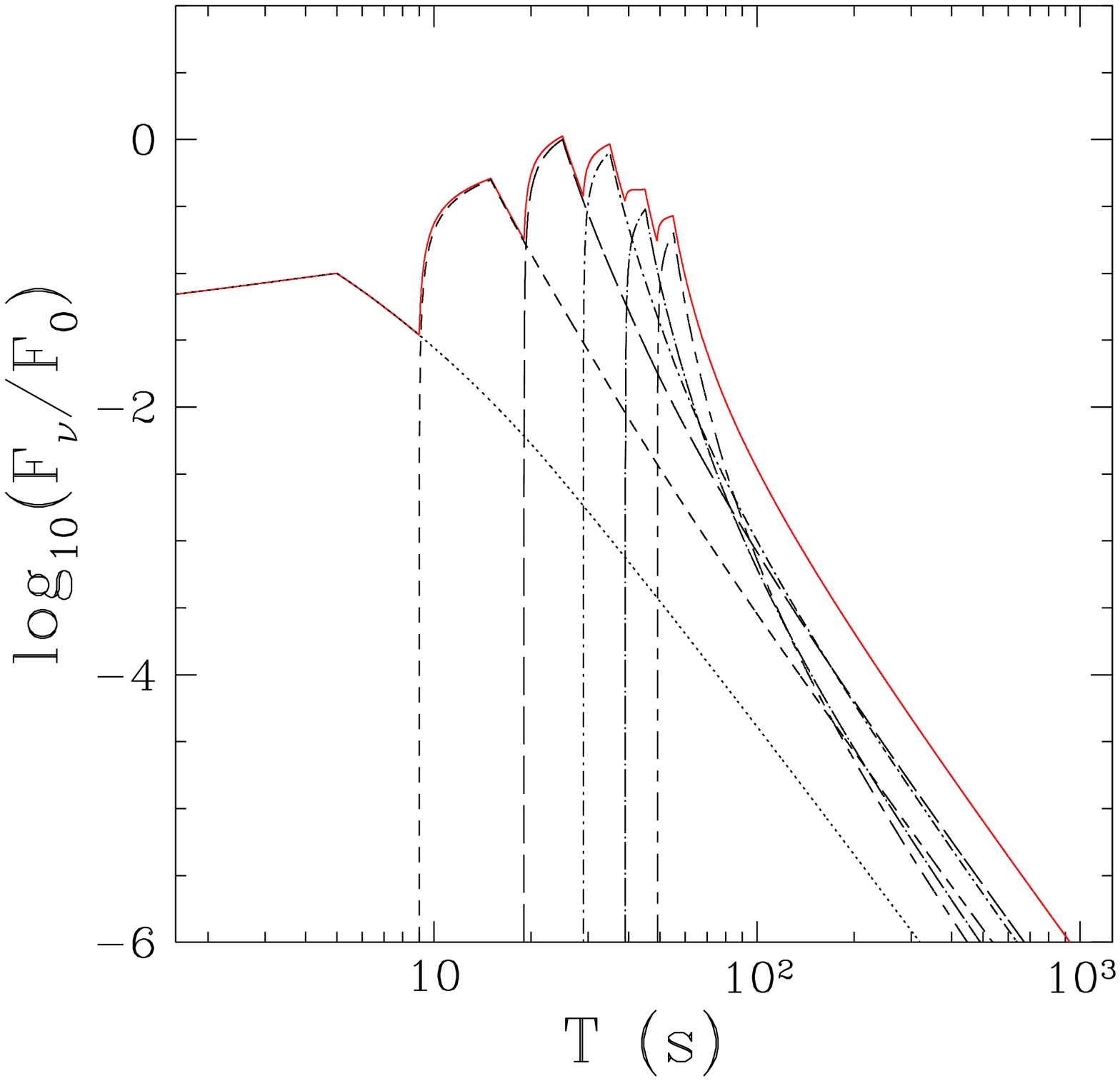}\includegraphics[width=0.31\textwidth,
height=0.4\textwidth]{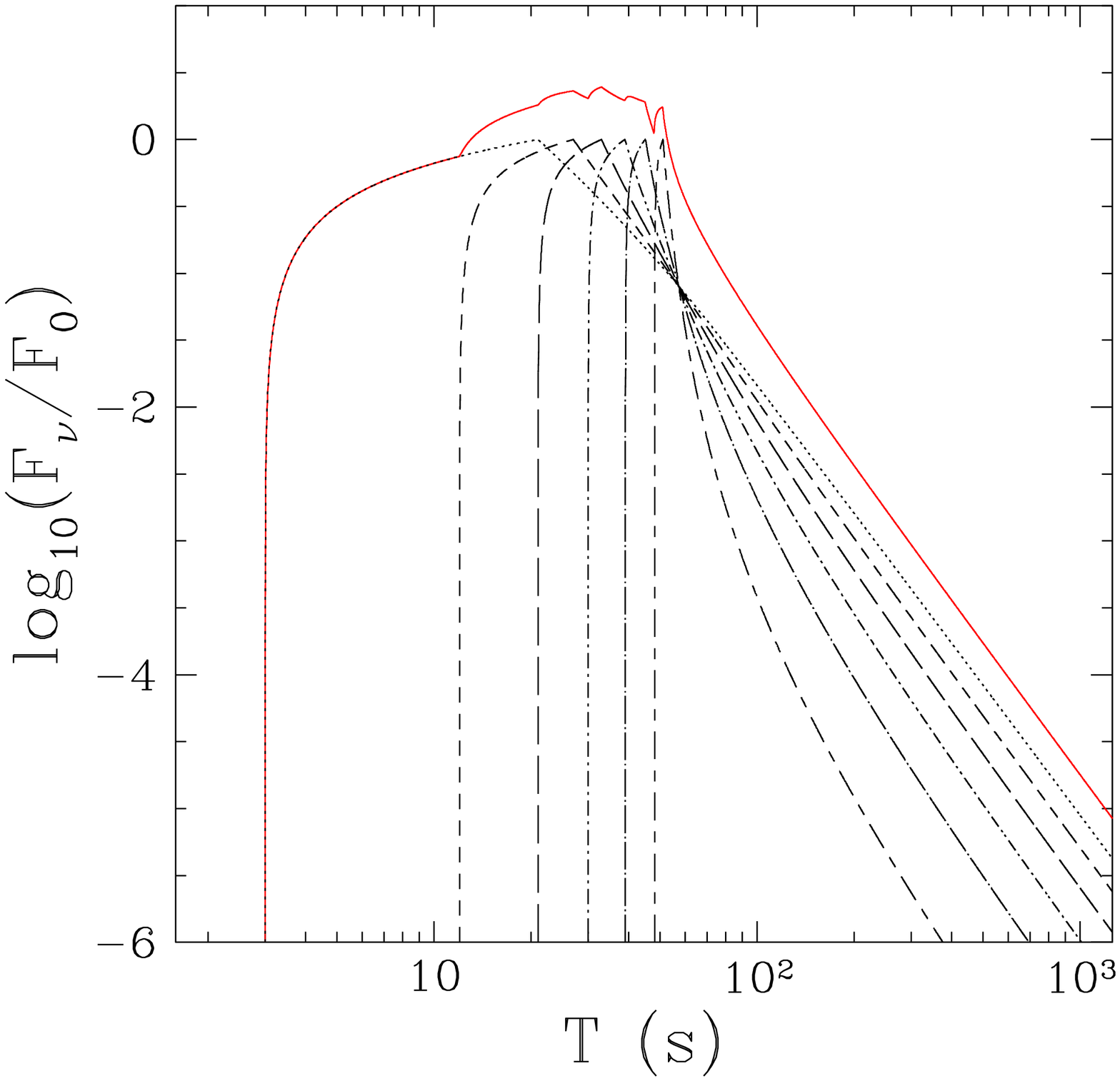}\\
\includegraphics[width=0.31\textwidth,
height=0.4\textwidth]{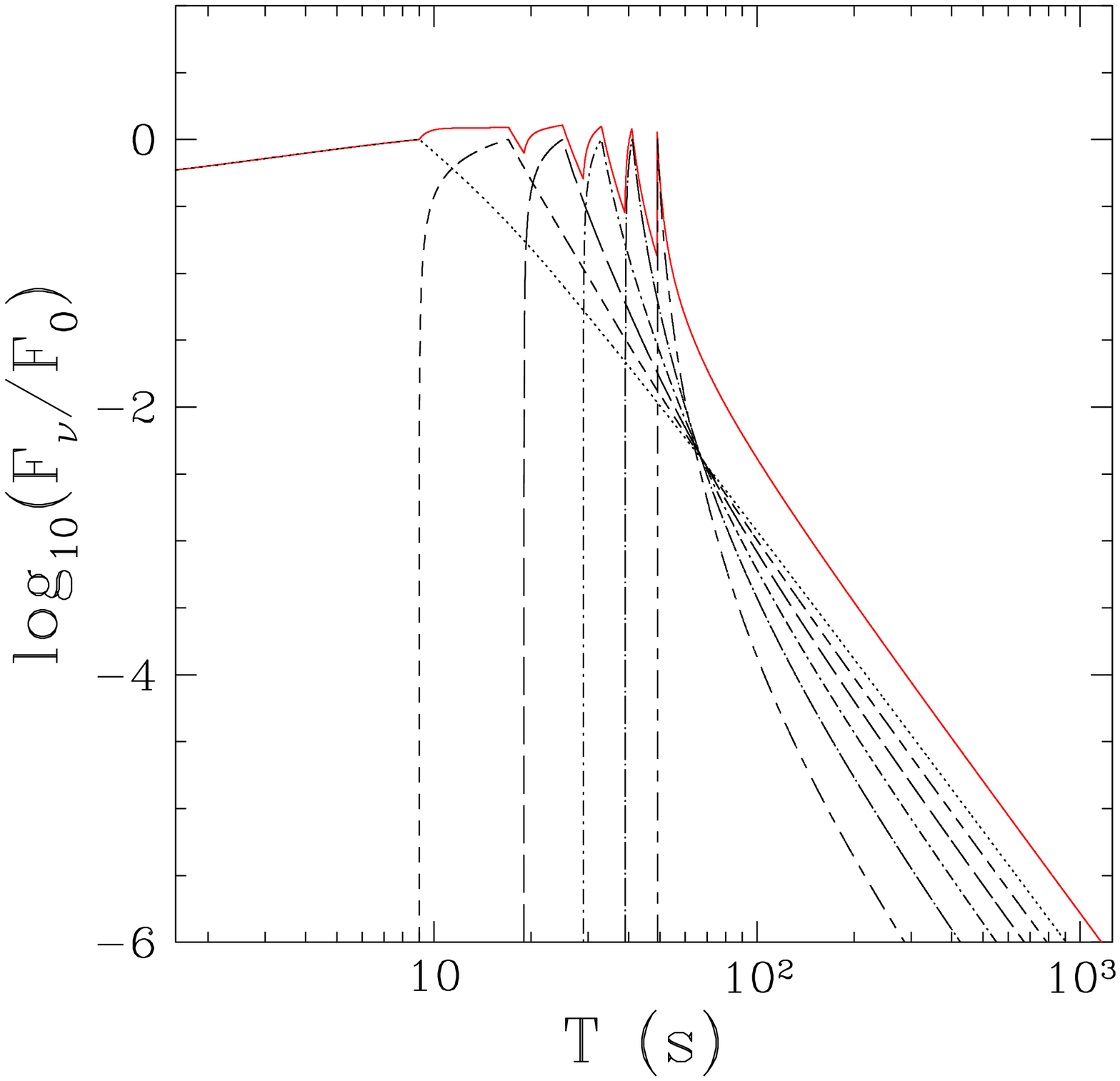}\includegraphics[width=0.31\textwidth,
height=0.4\textwidth]{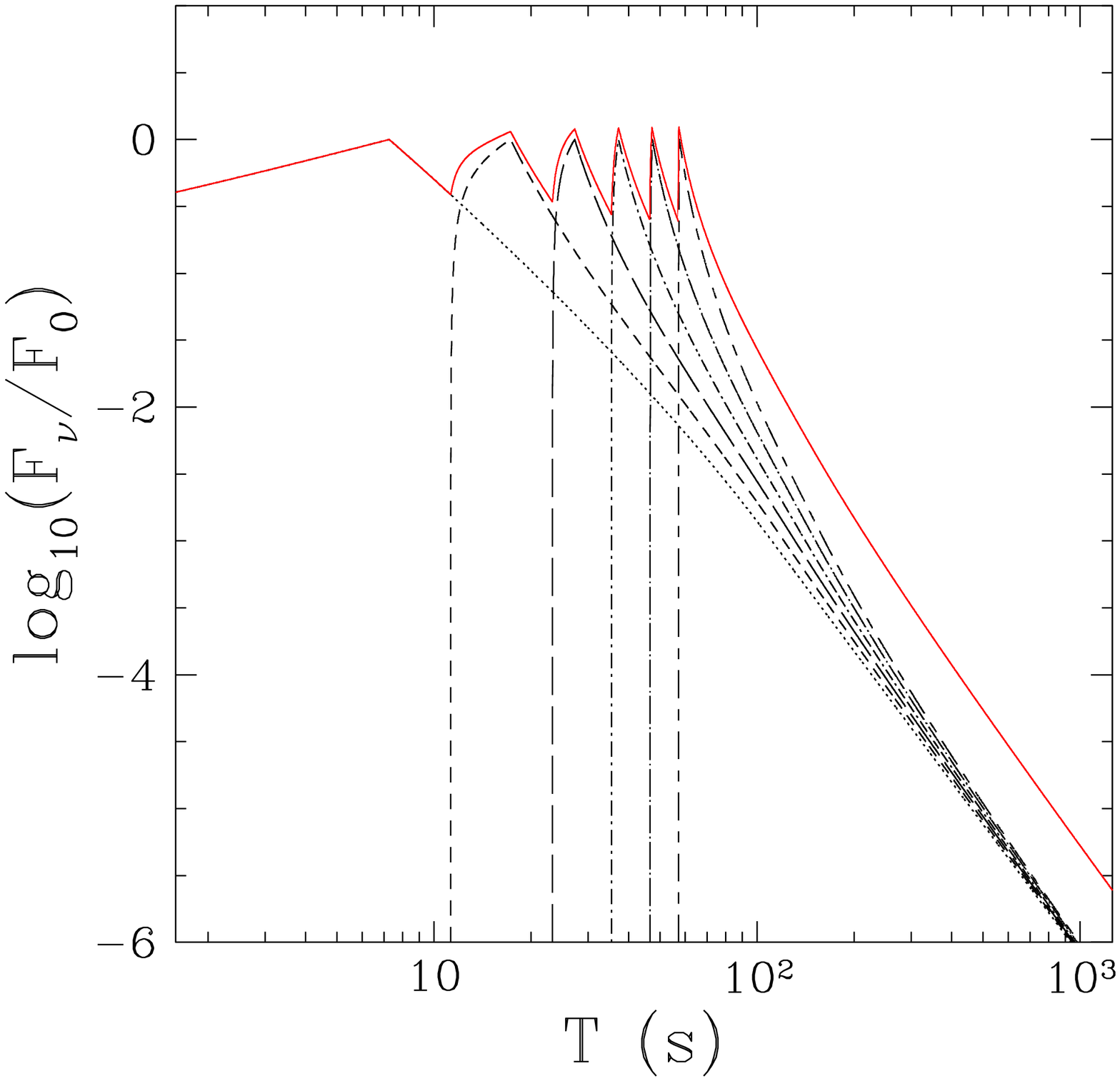}\includegraphics[width=0.31\textwidth,
height=0.4\textwidth]{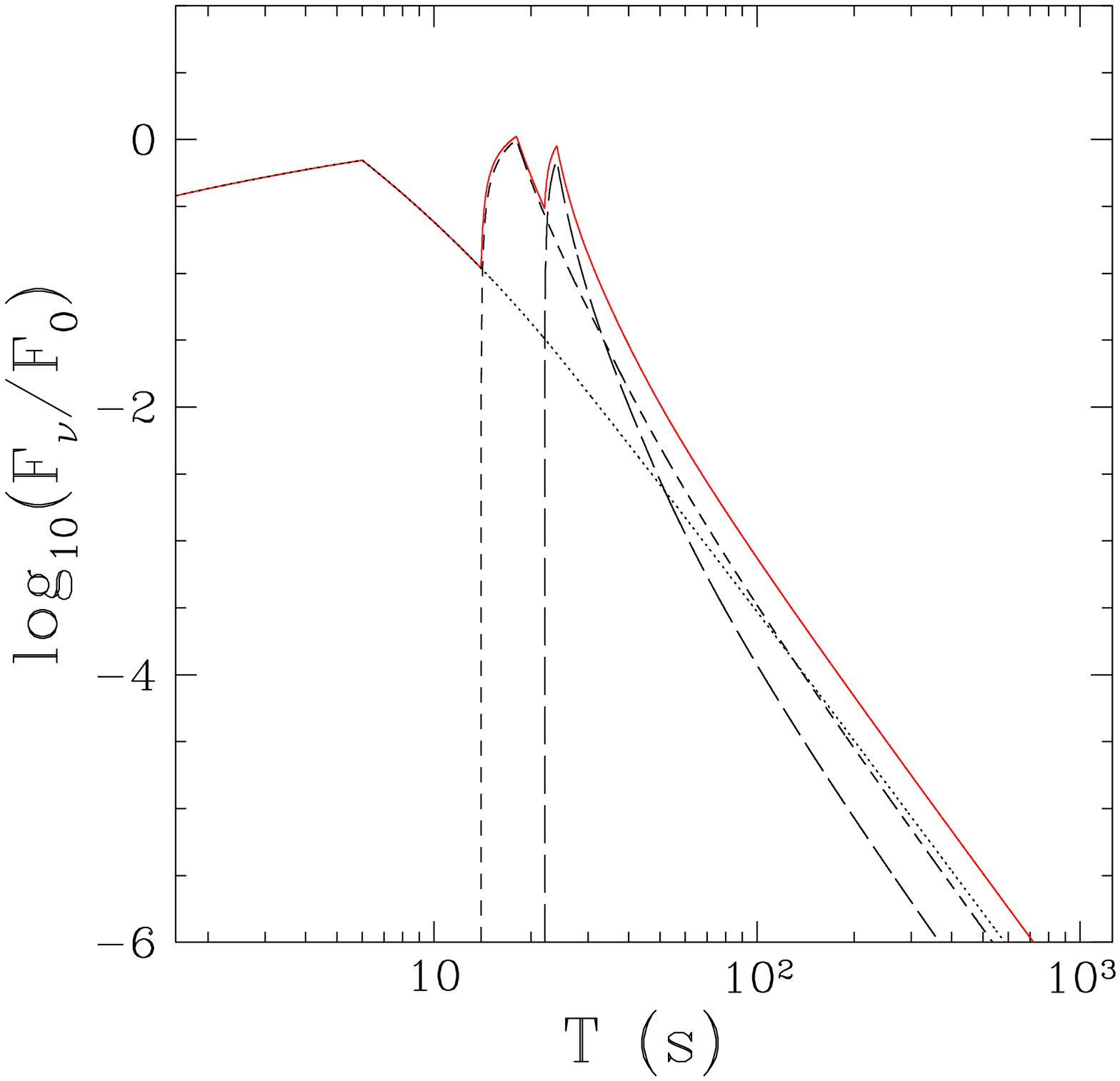}
\end{center}
\caption[prompt variations of one parameter and more realistic exemple]{
\emph{For all panels, The black lines show the individual pulses, while the red
line shows the total prompt emission. The normalized observed frequency is
$\nu/\nu_0 = 0.1$}
{\bf (a)}: \emph{Prompt emission with six pulses, all having the
same following
parameters: $m=0$, $d=-1$, $a=1$, $b_1=-0.25$, $b_2=-1.25$, $T_0 =
2\;$s, $\bar{T}_f = \Delta R/R_0 = 3$ and $F_{\rm peak}/F_0 = 1$. The ejection
times
$T_{\rm ej}$ are (from the first to the last pulse): $-2\;$s, $8\;$s, $18\;$s,
$28\;$s,
$38\;$s and $48\;$s.}
{\bf (b)}: \emph{Same as top left panel, except for varying
$F_{peak}/F_0$, which is from the first to the last pulse: $0.1$,
$0.5$, $1$, $0.8$, $0.3$, and $0.2$.}
{\bf (c)}: \emph{Same as top left panel except for varying
$T_0$ while $\Delta R/R_0 = 3$ remains constant, whose
values are (from first to last pulse): $6\;$s, $5\;$s, $4\;$s, $3\;$s,
$2\;$s, $1\;$s, which correspond to $T_f =$ $24\;$s, $20\;$s, $16\;$s, $12\;$s,
$8\;$s, $4\;$s. To keep $t_{ej,1} = -T_{0,1}$ the ejection times in this case
are: $-6\;$s, $4\;$s, $14\;$s, $24\;$s, $34\;$s and $44\;$s.}
{\bf (d)}: \emph{Same as top left panel except for varying
$\Delta R/R_0$ while keeping $R_0$ and therefore $T_0$ constant.
The values of
$\Delta R/R_0$ are (from first to last pulse): $5$, $4$, $3$, $2$,
$1$, $0$. Since $T_0 = 2\;$s, this corresponds to $T_f = (1+\Delta
R/R_0)T_0 =$ $12\;$s, $10\;$s, $8\;$s, $6\;$s, $4\;$s, and $2\;$s,
respectively.}
{\bf (e)}: \emph{Same as top left panel except for varying
$\Delta R/R_0$
while keeping $R_f$ constant and therefore $T_f$ and $T_0/R_0$ also 
remain constant, while both $R_0$ and
$T_0$ vary. From first to last pulse, $\Delta
R/R_0 =$ $10$, $3$, $1$, $0.3$, $0.1$, and $0.03$, and since $T_f =
8\;$s this corresponds to $T_0 =$ $0.727\;$s, $2\;$s, $4\;$s,
$6.15\;$s, $7.27\;$s, and $7.77\;$s. The final peak frequency
$\nu_p(\tilde{T}_f) = (T_0/T_f)\nu_0$ at $T_f$ is
also kept constant, so that from the first to the last pulse
$\nu/\nu_0 =$ $ 0.0091$, $0.025$, $0.05$, $0.0769$, $0.0909$,
$0.0971$.}
{\bf (f)}: \emph{example of a more realistic prompt emission
consisting
of three pulses with $T_{\rm ej} = -1\;$s, $13\;$s, $21\;$s, $T_0 = 2\;$s for
all three
pulses, $\Delta R/R_0 = 3$, $2$, $1$, and $F_{\rm peak}/F_0 = 0.7$, $1$,
$0.7$.}
}
\label{fig_var_param}
\end{figure}

\begin{figure}
\begin{center}
\includegraphics[width=0.465\textwidth,
height=0.35\textwidth]{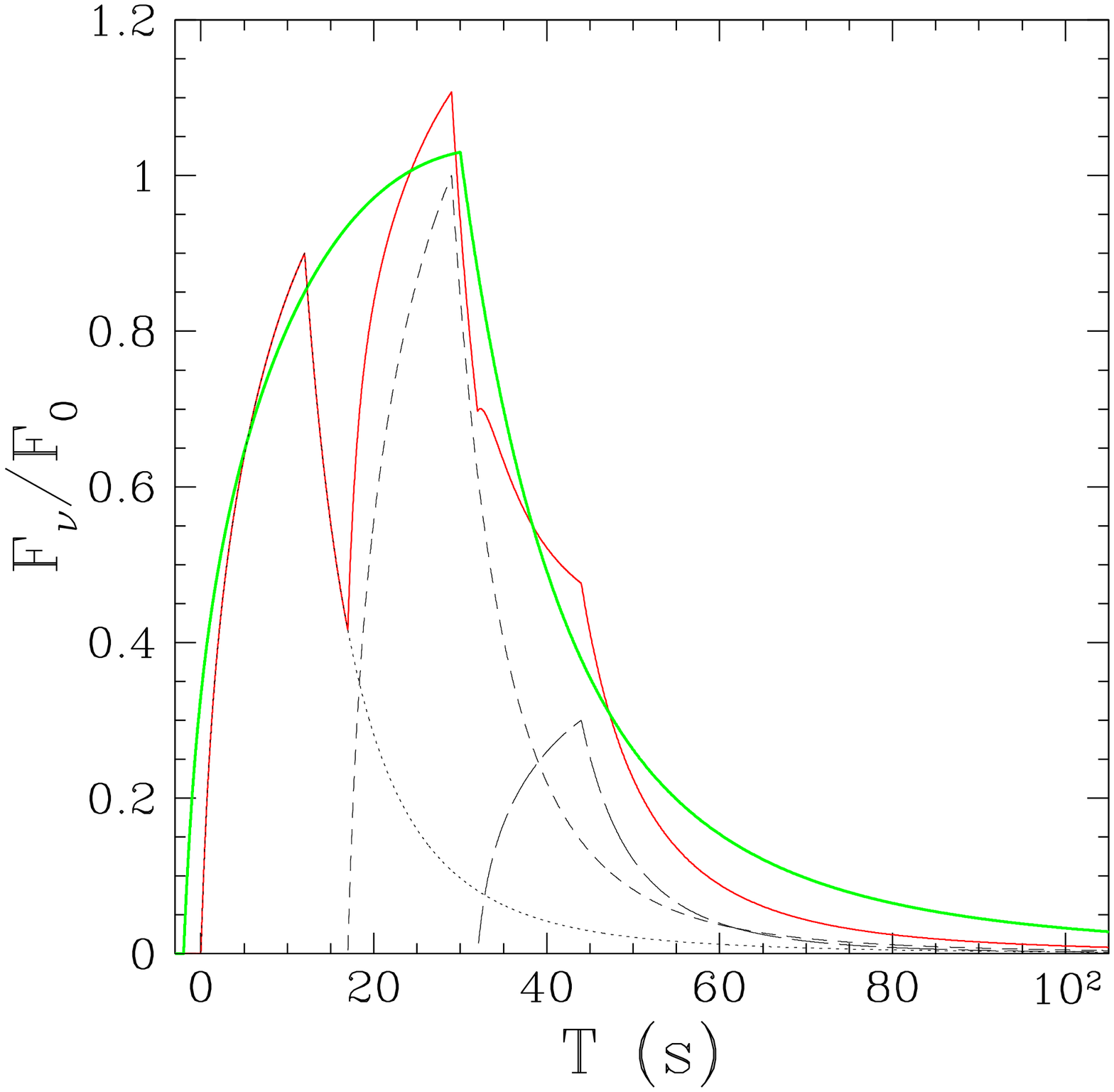}
\hspace{0.8cm}
\includegraphics[width=0.465\textwidth,
height=0.35\textwidth]{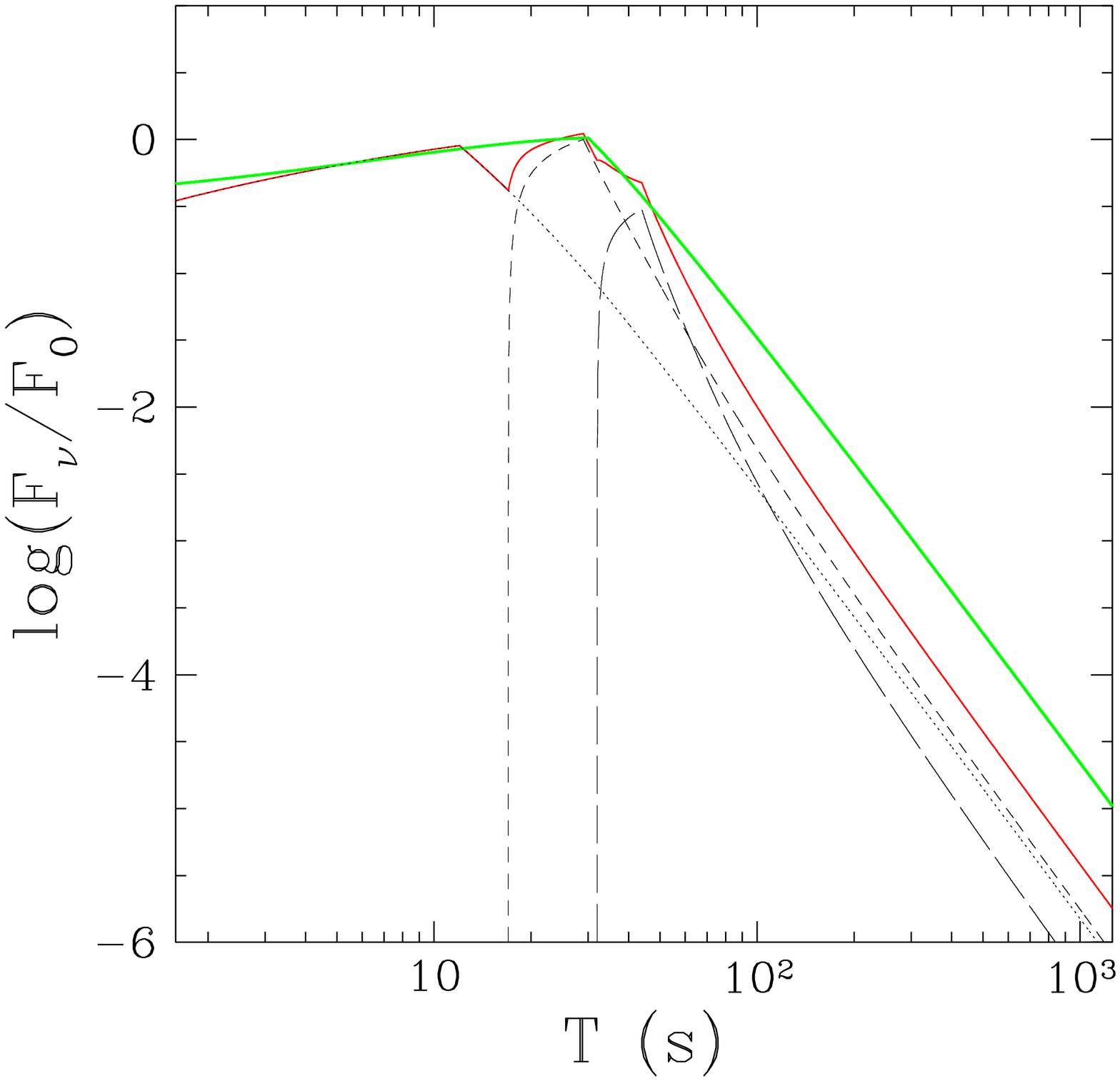}\\
\includegraphics[width=0.465\textwidth,
height=0.35\textwidth]{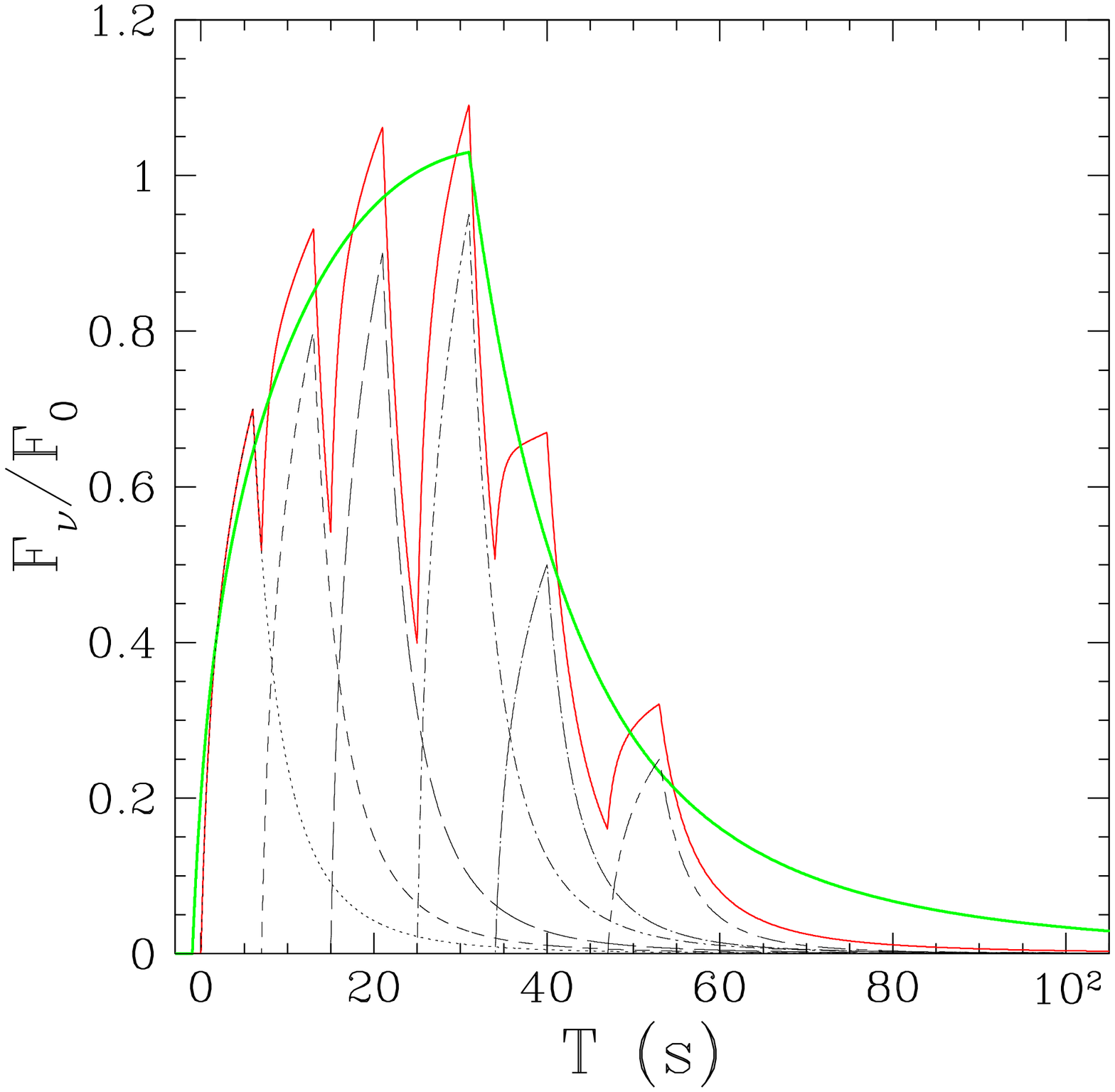}
\hspace{0.8cm}
\includegraphics[width=0.465\textwidth,
height=0.35\textwidth]{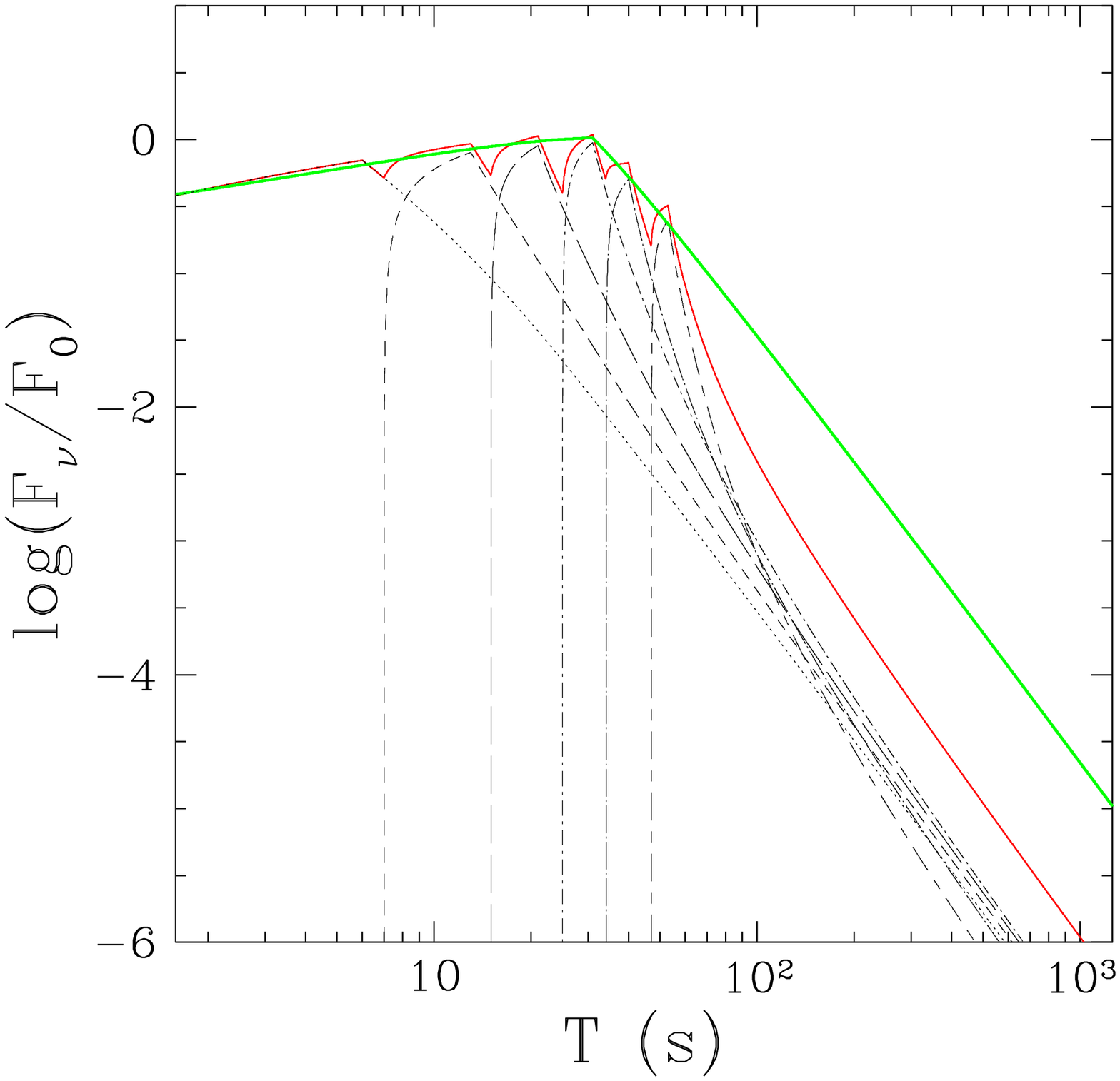}\\
\includegraphics[width=0.465\textwidth,
height=0.35\textwidth]{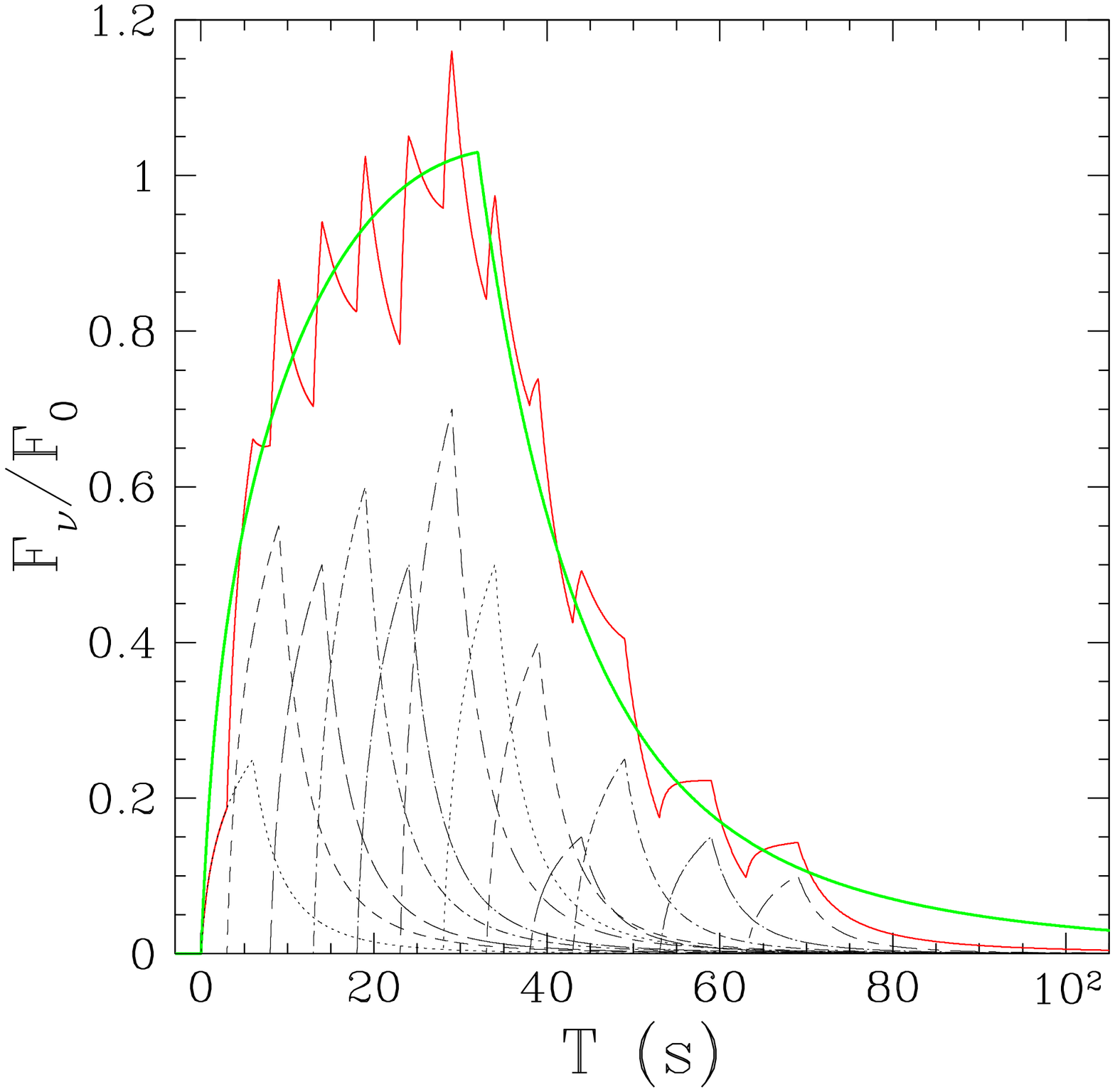}
\hspace{0.8cm}
\includegraphics[width=0.465\textwidth,
height=0.35\textwidth]{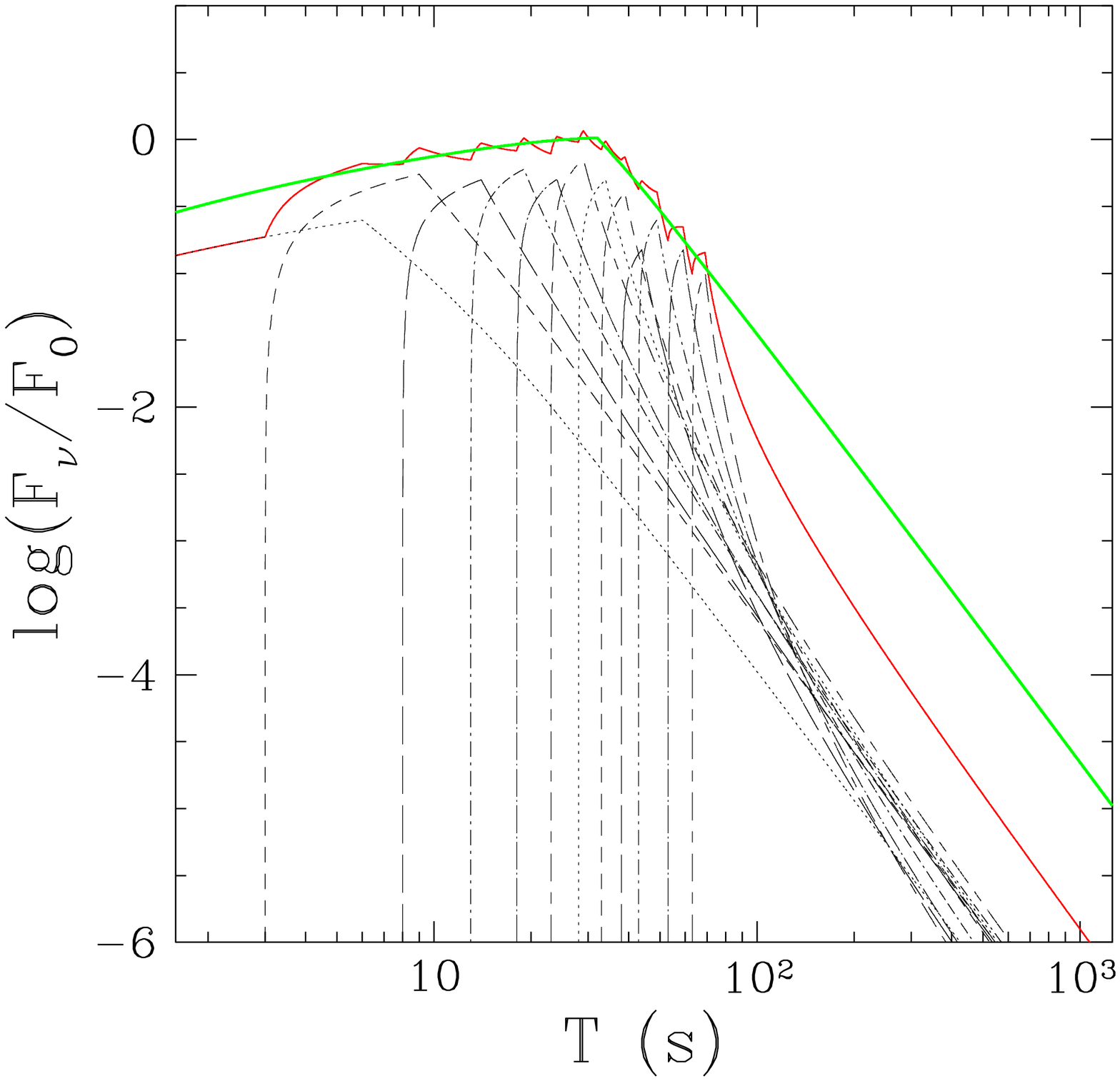}
\end{center}
\caption[prompt multi pulses]{
\emph{Comparison between the emission from several partially temporally
overlapping pulses (individual pulses are in non-solid black lines,
while the total prompt emission is in a solid red line), and a
tentative fit to these pulses using a single broad pulse (thick solid
green line). The same single broad pulse is used as a tentative fit
for three different prompt emissions, with $3$ ({\rm top panels}), $6$
({\rm middle panels}) and $12$ ({\rm bottom panels}) pulses. The parameters of
the
pulses are shown tables \ref{table_compfit_3pulses},
\ref{table_compfit_6pulses} and \ref{table_compfit_12pulses}. The
normalized observed frequency is $\nu/\nu_0 = 0.1$} {\bf Left panels:}
\emph{linear scale.}
{\bf Right panels:} \emph{Logarithmic scale.}} \label{fig_compfit}
\end{figure}

The shape of the pulses in our model can vary considerably, from very
spiky peaks to rounder ones, from a very sharp rise to shallower rise,
and so on (see Figs.~\ref{fig_LC_onepulse_selon_freq} --
\ref{fig_var_a}). This can help reproduce some of the observed
diversity in the shape of spike in the prompt light curve. This
appears to be a promising feature of our model. However, we have
an abrupt change in the temporal index at $\bar{T_f}$, that 
usually corresponds to a sharp peak of the
spike. This is caused by our model assumption that the emission
abruptly shuts off at the outer emission radius $R_f$. Therefore, we
also consider an alternative and more realistic assumption, which
leads to a rounder peak for the spikes, where the emission more
gradually turns off at $R > R_f$. This is done by introducing and
exponential turn-off with radius of the comoving spectral luminosity,
$L'_{\nu'}(R)$, and is examined in Appendix \ref{ap_expcutoff}.
The more gradual the
turn-off of the emission with radius the rounder the peak of the pulse
in the light curve. This can help fit the observed variety of pulse
shapes even better (at the cost of adding an additional free
parameter).

In the particular case of synchrotron emission from internal shocks,
we find that the observed spectrum has the same shape as the emitted
one, which in our case is modeled as a Band function. The observed
peak photon energy of the Band function decreases with time, 
$E_p(\tilde{T}) = E_0/\tilde{T}$, naturally leading to a softening of the
spectrum with time, similar to what is observed by {\it Swift}.
Thus, our model can at least qualitatively reproduce the main temporal
and spectral features observed by {\it Swift}. The spectral index
$\beta$ evolves from its value below $E_p$ ($\beta = -b_1$) to its
value above $E_p$ ($\beta = -b_2$), where the transition that
corresponds to the passage of $E_p$ through the observed energy band
occurs at earlier times for higher observed photon energies (or
frequencies).

When modeling the prompt emission by combining several pulses, the SDP
is initially dominated by the last pulse (just after its peak, if it
is above the flux fro the other pulses), but can later be dominated by
the tail of other pulses. The relative contribution of a pulse to the
late time flux scales as $\sim F_{peak}T_f^{2+\beta}$, and therefore
wider pulses (with a larger $T_f$), and to a lesser extent pulses with
a larger peak flux ($F_{\rm peak}$), tend to dominate the late time
flux, deep into the SDP. Moreover, often the contribution to the total
flux from the tails of several pulses can be comparable, so it cannot
be adequately modeled using a single pulse model. Therefore, we
caution here that modeling the steep decay phase using the HLE of a
single pulse, $F_\nu \propto (T-T_{\rm ref})^{-(2+\beta)}$, 
may lead to wrong conclusions, and all the more so if the reference
time $T_{\rm ref}$ is arbitrarily set to the GRB trigger time. 
Even if $T_{\rm ref}$ is set to the onset time of the last
spike, this may still be a bad approximation in many cases since (i)
we find that $\alpha_{ej} = 2 + \beta$ (with $T_{\rm ref} = T_{\rm ej}$)
rather than $\alpha_{on}$ (with $T_{\rm ref} = T_{\rm ej}+T_0$, corresponding
to the onset of the spike) while $\alpha_{on}$ approaches
$\alpha_{ej} = 2+\beta$ only at late times well after the peak of
the last pulse, and (ii) at such late times the flux often becomes
dominated by the tails of earlier pulses.

Our model can produce different shapes for the tail of the prompt
emission, from close to a power law (which can have a different
temporal index than its asymptotic late time value) to a curved shape
with decreasing temporal index $\alpha$. This is qualitatively
consistent with observations, where these type of behaviour are
observed. We have demonstrated that just after the peak of the last
pulse, the decay index of the prompt emission tail can reach very
large values, far greater than the typical average value observed
during the SDP by {\it Swift}, of $3 \lesssim \alpha
\lesssim 5$ (Nousek et al. 2006). Larger values for the temporal
index, however, are sometimes observed close to the end of the prompt
emission (for example in GRB050422, GRB050803 or GRB050916; see figure
2 from O'Brien et al. 2006), in accord with our model.

Because of the large number of free parameters, the fitting of actual
data should be handled with care, and there may be various
degeneracies involved. The results of such fits to data should also be
taken cautiously because of the difficulty in properly resolving
distinct pulses in the prompt emission. For different reasons (such as
noisy data, coarse time bins, pulse overlap, etc.), a group of
distinct pulses may be fitted by a single broader pulse, resulting in
an over-prediction of the flux during the SDP, as well different
spectral and temporal evolution, which might lead to a
misinterpretation of the SDP. Nevertheless, when handled with care, a
fit of our model to a good combined data set of the prompt GRB
emission and its SDP tail can serve as a powerful test of the HLE
model for the SDP, and thus help distinguish between different models
for the prompt GRB emission.\\

\vspace{0.5cm}
\noindent
J.~G. gratefully acknowledges a Royal Society Wolfson Research Merit
Award.

\newpage

\newpage

\begin{table}
\begin{center}
\begin{tabular}{|l|l|l|l|l|l|l|l|}
\hline
pulse number           & 1     & 2   & 3     & broad pulse\\
\hline
$T_{\rm ej}$ [s]       & -2    & 15  & 35    & -4\\
\hline
$T_0$ [s]              & 2     & 4   & 5     & 4\\
\hline
$T_f$                  & 16    & 16  & 25    & 36\\
\hline
$F_{\rm peak}/F_0$     & 0.85  & 1   & 0.12  & 1.03\\
\hline
\end{tabular}
\end{center}
\caption{\emph{Parameters of the pulses for figure \ref{fig_compfit} (top
panels)}}
\label{table_compfit_3pulses}
\end{table}

\begin{table}
\begin{center}
\begin{tabular}{|l|l|l|l|l|l|l|l|}
\hline
pulse number     & 1    & 2   & 3   & 4   & 5   & 6   & broad pulse\\
\hline
$T_{\rm ej}$ [s] & -2   & 1   & 16  & 26  & 36  & 46  & -4\\
\hline
$T_0$ [s]        & 2    & 2   & 2   & 1.5 & 2   & 2   & 4\\
\hline
$T_f$            & 6    & 10  & 6   & 6   & 8   & 8   & 36\\
\hline
$F_{\rm peak}/F_0$       & 0.25 & 0.8 & 0.9 & 1   & 0.4 & 0.2 & 1.03\\
\hline
\end{tabular}
\end{center}
\caption{\emph{Parameters of the pulses for figure \ref{fig_compfit} (middle
panels)}} \label{table_compfit_6pulses}
\end{table}

\begin{table}
\begin{center}
\begin{tabular}{|l|l|l|l|l|l|l|l|l|l|l|l|l|l|}
\hline
pulse number     & 1    & 2   & 3    & 4    & 5    & 6     & 7    & 8    & 9
& 10   & 11   & 12   & broad pulse\\
\hline
$T_{\rm ej}$ [s] & -2   & -1  & 5    & 11   & 19   & 20    & 26   & 31   & 36
& 44   & 51   & 66   & -4\\
\hline
$T_0$ [s]        & 2    & 2   & 2    & 2    & 1    & 2     & 2    & 2    & 3
& 2    & 2    & 3    & 4\\
\hline
$T_f$            & 4    & 6   & 6    & 6    & 2    & 5     & 6    & 6    & 7.5
& 8    & 6    & 6    & 36\\
\hline
$F_{\rm peak}/F_0$      & 0.25 & 0.5 & 0.75 & 0.85 & 0.75 & 0.85  & 0.95 & 0.55
& 0.35 & 0.25 & 0.11 & 0.11 & 1.03\\
\hline
\end{tabular}
\end{center}
\caption{
\emph{Parameters of the pulses for figure \ref{fig_compfit} (bottom panels)}}
\label{table_compfit_12pulses}
\end{table}


\clearpage

\clearpage
\newpage

\appendix
\renewcommand{\theequation}{A\arabic{equation}}
\setcounter{equation}{0}

\begin{center}
{\bf APPENDIX}
\section{Detailed calculation of the flux} \label{ap_dyn_flux}
\end{center}

In order to calculate the flux density $F_{\nu}$ that reaches the
observer at an observed time $T$, we closely follow Granot,
Cohen-Tanugi and DoCouto e Silva 2008: we integrate 
over the Equal Arrival Time Surface (EATS), i.e. the locus 
of points from which photons that
are emitted by the shell at a radius $R$, angle $\theta$ relative to
the line of sight, and a lab frame time $t$, reach the observer
simultaneously at an observed time $T$. The lab frame time and
the shell radius are related by
\begin{eqnarray} \label{eq_relat_t_R}
t - t_{\rm ej} = \int_0^R \frac{dr}{\beta c}
\approx \frac{R}{c} + \frac{R}{2(m+1)c\Gamma^2}\ .
\end{eqnarray}
From simple geometrical considerations, the EATS is given by
\begin{equation}\label{eq_T_EATformula}
\frac{T-T_{\rm ej}}{(1+z)} = t - t_{\rm ej} - \frac{R\cos \theta}{c}
\approx \frac{R}{c}\left[1-\cos\theta + \frac{1}{2(m+1)\Gamma^2}\right]\ ,
\end{equation}
Since $\Gamma \gg 1$ we can consider only small emission angles
$\theta \ll 1$, for which $\cos\theta \approx 1 - \theta^2/2$, so that
the EATS reads
\begin{equation}\label{EATS2}
\frac{T-T_{\rm ej}}{(1+z)} \approx \frac{R_L}{2(m+1)c\Gamma_L^2}
\left[y^{m+1} + y(m+1)(\Gamma_L\theta)^2\right]\ ,
\end{equation}
where we have introduced the normalized radius $y \equiv R/R_L$, as
well as $R_L = R_L(T)$ that is the largest radius on the EATS at time
$T$, and $\Gamma_L \equiv \Gamma(R_L)$. Since $R_L$ is always obtained
along the line of sight (at $\theta = 0$),
\begin{equation}\label{eq_RL_ROTTO}
R_L = 2(m+1)c\Gamma_L^2 \frac{T-T_{\rm ej}}{(1+z)}
= R_0 \left( \frac{T-T_{\rm ej}}{T_0} \right)^{\frac{1}{m+1}}\ , \quad
T_0 = \frac{(1+z)R_0}{2(m+1)c\Gamma_0^2}\ .
\end{equation}
Substituting eq.~(\ref{eq_RL_ROTTO}) into
eq.~(\ref{EATS2}) implies
\begin{equation}\label{dmu_dy}
1-\mu \approx \frac{\theta^2}{2} = \frac{y^{-1}-y^m}{(m+1)\Gamma_L^2}\ ,\quad
\frac{d\mu}{dy} = \frac{y^{-2} + my^{m-1}}{2(m+1)\Gamma_L^2}\ ,
\end{equation}
where $\mu \equiv \cos\theta$.
The Doppler factor between the comoving frame and the lab frame is given by
\begin{equation}\label{eq_delta_m}
\delta \equiv (1+z)\frac{\nu}{\nu'} = \frac{1}{\Gamma(1-\beta\mu)}
\approx \frac{2\Gamma}{1+(\Gamma\theta)^2} =
\frac{2(m+1)\Gamma_L y^{-m/2}}{m + y^{-m-1}}\ .
\end{equation}

Remembering the reader that $T = T_{\rm ej} + T_f$ is the time at 
which the last photons that are emitted along the line of sight 
(from $R_f$ and $\theta = 0$) reach the observer (which can be 
defined here by $R_L(T_{\rm ej} + T_f) \equiv R_f$), from 
equation (\ref{eq_RL_ROTTO}) its general value is
\begin{equation} \label{eq_Tf}
T_f = T_0\left(\frac{R_f}{R_0}\right)^{m+1}
= T_0 \left(1+\frac{\Delta R}{R_0}\right)^{m+1}\ .
\end{equation}
In the limit $\Delta R \to 0$, $T_f \to T_0$.\\

The observed flux is then obtained by integration over the EATS
(Sari 1998; Granot 2005),
\begin{equation}\label{eq_flux_int_EATS}
F_{\nu}(T) = \frac{(1+z)}{4 \pi d_L^2} \int dL_{\nu}
= \frac{1+z}{4 \pi d_L^2} \int \delta^3 dL'_{\nu'}
= \frac{1+z}{8 \pi d_L^2} \int_{y_{\rm min}}^{y_{\rm max}}
dy\,\frac{d\mu}{dy}\,\delta^3(y)\,L'_{\nu'}(y)\ ,
\end{equation}
where $dL'_{\nu'} = L'_{\nu'}(R)d\nu d\phi/4\pi \to L'_{\nu'}(r)d\mu/2
= \frac{1}{2}L'_{\nu'}(R)(d\mu/dy)dy$ due to symmetry around the line
of sight (no dependence of the emission on the azimuthal angle
$\phi$), $L'_{\nu'}(R)$ is the total comoving spectral luminosity of
the shell (the emitted energy per unit time and frequency), $\nu' =
\nu(1+z)/\delta$, and $d_L(z)$ is the luminosity distance of the
source. The limits of integration over $y$ are
\begin{eqnarray}\label{eq_ymin_ymax}
y_{\rm min} &=& \min\left(1, \frac{R_0}{R_L(T)} \right)
= \left\{ \begin{array}{ll} 1 & T \leq T_{\rm ej}+T_0\ ,\\
\left(\frac{T-T_{\rm ej}}{T_0}\right)^{-1/(m+1)}  &  T \geq T_{\rm ej}+T_0\ ,\\
\end{array} \right., \nonumber\\
\\
y_{\rm max} &=& \min\left(1, \frac{R_0 + \Delta R}{R_L(T)} \right)
= \left\{ \begin{array}{ll} 1 & T \leq T_{\rm ej}+T_f\ ,\\
\left(\frac{T-T_{\rm ej}}{T_f}\right)^{-1/(m+1)}  &  T \geq T_{\rm ej}+T_f\ .\\
\end{array} \right.\ . \nonumber
\end{eqnarray}
For $T \leq T_{\rm ej} + T_0$ we have $R_L(T) \leq R_0$ and therefore
$y_{\rm min} = y_{\rm max} = 1$ and $F_\nu(T) = 0$. This is since the
EATS does not intersect the emission region for $R_L < R_0$, and only
touches it at one point, $(R,\,\theta) = (R_0,\,0)$, for $R_L = R_0$
($T = T_{\rm ej}+T_0$). The observed flux becomes non-zero for $R_L >
R_0$, corresponding to $T > T_{\rm ej}+T_0$.
Substituting eqs. (\ref{dmu_dy}) and (\ref{eq_delta_m}) into
eq.~(\ref{eq_flux_int_EATS}) finally gives
\begin{equation}\label{eq_flux_generale}
F_{\nu}(T) = \frac{(1+z)}{2 \pi d_L^2}\,\Gamma_0\left(\frac{T-T_{\rm
ej}}{T_0}\right)^{-m/[2(m+1)]}
\int_{y_{\rm min}}^{y_{\rm max}}dy\,y^{-1-m/2}\left( \frac{m+1}{m + y^{-m-1}}
\right)^2 L'_{\nu'}(y) \ .
\end{equation}

\subsection{Power-law spectrum} \label{subsec_powerlaw}

While a single power law emission spectrum is not very realistic,
 it already shows many important properties that
also appear for a Band function emission spectrum (considered
in the main text). This is the reason why this case is described
here. The luminosity is then
\begin{equation}
L'_{\nu'} = L'_0\left(\frac{R}{R_0}\right)^a\left(\frac{\nu'}{\nu'_0}\right)^b
= L'_0\tilde{T}^{(2a+mb)/[2(m+1)]}
\left(\frac{\nu}{\nu_0}\right)^b y^{a+mb/2}
\left(\frac{m+1}{m + y^{-m-1}}\right)^{-b}\ ,
\end{equation}
where the comoving spectral luminosity also scales as a power law with
radius when the emission is over a finite range of radii, $\nu'_0$ is
a fixed frequency in the comoving frame.

\subsubsection{Emission from an infinitely thin shell at radius $R_0$}
\label{PL_thin_shell}

We first study the case where the whole emission comes from a single
radius $R_0$,
\begin{eqnarray}\nonumber
L'_{\nu'} &=& L'_0 \left(\frac{\nu'}{\nu'_0}\right)^b R_0 \delta(R-R_0)
= L'_0 \left(\frac{\nu}{\nu_0}\right)^b \left(\frac{y}{y_{\rm
min}}\right)^{mb/2}
\left(\frac{m+1}{m + y^{-m-1}}\right)^{-b}y_{\rm min}\delta(y-y_{\rm min})
\\
&=& L'_0 \left(\frac{\nu}{\nu_0}\right)^b
\left(\frac{m+1}{m + y_{\rm min}^{-m-1}}\right)^{-b}y_{\rm min}\delta(y-y_{\rm
min})\ ,
\end{eqnarray}
where this is valid only for $\tilde{T} \geq 1$ that corresponds
to $R_L \geq R_0$, for which $y_{\rm min} = R_0/R_L = [(T-T_{\rm
ej})/T_0]^{-1/(m+1)}$. Eq.~(\ref{eq_flux_generale}) then implies
\begin{equation}
F_{\nu}(T \geq T_{\rm ej}+T_0) = \frac{(1+z)}{4 \pi d_L^2} L_0
\left(\frac{\nu}{\nu_0}\right)^b
\left[\frac{T-T_{\rm ej}+mT_0}{(m+1)T_0}\right]^{b-2}\ .
\end{equation}
There are two times of particular relevance here: the radial time
$T_r(R_0) = T_0 = (1+z)R_0/[2c(m+1)\Gamma_0]$, which is the time past
$T_{\rm ej}$ when the first photons start reaching the observer, and
the angular time $T_{\theta}(R_0) = (1+z)R_0/(2c\Gamma_0) =
(m+1)T_r(R_0)$ that sets the time-scale for the width of the
pulse. One can rewrite the expression for the observed flux density as
\begin{equation}
F_{\nu}(T>T_{\rm ej}+T_0) = \frac{(1+z)}{4 \pi d_L^2}L_0
\left(\frac{\nu}{\nu_0}\right)^b \left[\frac{T -
T_s}{T_{\theta}(R_0)}\right]^{b-2}\ ,
\end{equation}
where $T_s = T_{\rm ej} + T_0 - T_\theta(R_0) = T_{\rm ej} - mT_0$ is
the reference time for the power-law flux decay of the pulse, and is
exactly $T_\theta(R_0)$ before the onset of the pulse.  Since the
emission itself occurs at one particular radius ($R_0$) it depends
only on the Lorentz factor at that radius radius, and is independent
of $m$. In particular, $T_\theta(R_0) = (m+1)T_0$ and the pulse peak
flux are independent of $m$. The value of $m$ affects only the onset
time of the pulse ($T = T_{\rm ej}+T_0$) and the reference time for
the power law flux decay. For internal shocks we expect a coasting
shell ($m = 0$) for which $T_s = T_{\rm ej}$ and $T_\theta(R_0) =
T_0$. It can easily be seen that the HLE relation, $\alpha = 2+\beta$
where $F_{\nu} \propto T^{-\alpha} \nu^{-\beta}$, is satisfied here as
$\beta = -b$ and $\alpha = 2-b = 2+\beta$.

\subsubsection{Emission from a region of finite width}

We now turn to the case where the emission comes from a range of radii
between $R_0$ and $R_f=R_0+\Delta R > R_0$. The comoving spectral
luminosity in this case is $L'_{\nu'} = L'_0 \left(R/R_0\right)^{a}
\left(\nu'/\nu'_0\right)^{b}$, and the flux density is given by (Granot,
Cohen-Tanugi \& do Couto e Silva 2008):
\begin{equation}
F_{\nu}(T) = \frac{(1+z)}{4 \pi d_L^2} L_0\left(\frac{\nu}{\nu_0}\right)^b
\tilde{T}^{\frac{2a-m(1-b)}{2(m+1)}}
\int_{y_{\rm min}}^{y_{\rm max}}dy\,y^{a-1-m(1-b)/2}
\left(\frac{m+1}{m+y^{-m-1}}\right)^{2-b}\ ,
\end{equation}
which, for internal shocks ($m=0$) becomes:
\begin{eqnarray}\label{thin_shell_PL}
F_{\nu}(T \geq T_{\rm ej}+T_0) = \frac{(1+z)}{4 \pi d_L^2}
\frac{L_0(\nu/\nu_0)^b}{(2+a-b)}
\tilde{T}^{b-2}
\left[\min(\tilde{T},\tilde{T}_f)^{2+a-b}-1\right]\ .
\end{eqnarray}
It is therefore obvious that for $T \geqslant T_{\rm ej}+T_f$ the HLE
relation is valid, where the reference time is the ejection time
$T_{\rm ej}$, as in this case the spectral slope is $\beta = -b$ and
the temporal slope is $\alpha = 2-b = 2+\beta$. In this sense a finite
range of emission radii with $m = 0$ is similar to emission from a
single radius, as in both cases the HLE relation $\alpha = 2 +
\beta$ is strictly valid immediately from $T \ge T_{\rm ej}+T_f$,
for some reference time, though in the latter case the reference time
for which this is valid is equal to the observed ejection time only
for $m = 0$. For emission from a finite range of radii with $m \neq 0$
the relation $\alpha = 2 + \beta$ is approached asymptotically at $T -
T_{\rm ej} \gg T_f$.

\subsection{Band function spectrum: general case and late time dependence} \label{ap_band_gen}

In the main text we have given the flux in the specific case of internal 
shocks, $m=0$ and $d=-1$. We derive here the flux for any values of the 
parameters $m$ and $d$.

\subsection{Emission from a single radius} \label{ap_band_gen_singleR}

When the whole emission comes from a single radius $R_0$, the peak
frequency is $\nu'_p = \nu'_0$, and the luminosity is thus
\begin{equation}\label{eq_L_DR0}
L'_{\nu'} = L'_0 S\left(\frac{\nu'}{\nu'_p}\right) R_0 \delta(R-R_0)
 = L'_0 S\left(\frac{\nu'}{\nu'_p}\right) y_{\rm min} \delta(y-y_{\rm min})\ ,
\end{equation}
Using this luminosity
(eq.~[\ref{eq_L_DR0}]) in the integral for the flux
(eq.~[\ref{eq_flux_generale}]) results in
\begin{eqnarray}\label{F_nu_Band_R0}
F_{\nu}(T \geq T_{\rm ej}+T_0) = \frac{(1+z)}{4 \pi d_L^2} L_0
\left[\frac{T-T_s}{T_\theta(R_0)}\right]^{-2} S\left(\frac{\nu}{\nu_p(T)}\right)\ ,\quad
\frac{\nu_p(T)}{\nu_0} = \frac{E_p(T)}{E_0} 
= \left[\frac{T-T_s}{T_\theta(R_0)}\right]^{-1}\ ,
\end{eqnarray}
where as in \S~\ref{subsec_powerlaw}, $T_s \equiv T_{\rm ej} - mT_0$
is the reference time for the power-law flux decay of the pulse, 
$T_\theta(R_0) = (m+1)T_0 = (1+z)R_0/2c\Gamma_0^2$ is the angular 
time at $R_0$, and $E_p(T) = h\nu_p(T)$ is the photon energy corresponding
to the peak of the Band function spectrum. Note that for $m = 0$,
 $T_s = T_{\rm ej}$. One can express the argument of $S$ as
\begin{eqnarray}
\frac{\nu}{\nu_p(T)} = \frac{E}{E_p(T)} 
= \frac{\nu}{\nu_0}\left[\frac{T-T_s}{T_\theta(R_0)}\right]\ ,\quad
\frac{\nu}{\nu_0} = \frac{E}{E_0} 
= \frac{(1+z)\nu}{2\Gamma_0\nu'_0}\ .
\end{eqnarray}
Reminding that $F_s \equiv L_0(1+z)/(4\pi d_L^2)$, we then use the explicit
expression for the Band function (eq.~[\ref{eq_bandfunction_lumin}]) to express
the observed flux as:
\begin{eqnarray}
\frac{F_{\nu}(T\geq T_{\rm ej}+T_0)}{F_s} & = & 
\left\{ \begin{array}{ll}
\left[\frac{T-T_s}{T_\theta(R_0)}\right]^{b_1-2} \left(\frac{\nu}{\nu_0}\right)^{b_1} 
e^{(1+b_1)[1-\frac{\nu}{\nu_0}(T-T_s)/T_\theta(R_0)]}
&  \frac{T-T_s}{T_\theta(R_0)} \leqslant \frac{x_b}{\frac{\nu}{\nu_0}}\ ,\quad
\\ \\
\left[\frac{T-T_s}{T_\theta(R_0)}\right]^{b_2-2} \left(\frac{\nu}{\nu_0}\right)^{b_2} 
x_b^{b_1-b_2} e^{1+b_2}
&  \frac{T-T_s}{T_\theta(R_0)} \geqslant \frac{x_b}{\frac{\nu}{\nu_0}}\ .
\end{array} \right.
\end{eqnarray}

\subsection{Emission from a range of radii} \label{ap_band_gen_rangeR}

In the case of emission with a Band function spectrum over 
a finite range of radii, $R_0 < R < R_f = R_0 + \Delta R$,
we remind that the comoving luminosity is:
\begin{equation} \label{eq_Lnu_band_DRdiff0}
L'_{\nu'} = L'_0 \left(\frac{R}{R_0}\right)^a
S\left(\frac{\nu'}{\nu'_p(R)}\right)
\end{equation}

Introducing (\ref{eq_Lnu_band_DRdiff0}) into (\ref{eq_flux_generale})
we obtain the general expression of the flux:
\begin{eqnarray}\label{F_nu_Band_R0_Rf}
\frac{F_{\nu}}{F_s} & = & y_{\rm min}^{-a+m/2}
\int_{y_{\rm min}}^{y_{\rm max}}dy\,y^{a-1-m/2}
\left( \frac{m+1}{m + y^{-m-1}} \right)^2  S\left(\frac{\nu'}{\nu'_p}\right)
\\ \nonumber
& = & \int_{1}^{\tilde{y}_{\rm max}}d\tilde{y}\,\tilde{y}^{a-1-m/2}
\left( \frac{m+1}{m +y_{\rm min}^{-m-1}\tilde{y}^{-m-1}} \right)^2
S\left(\frac{\nu'}{\nu'_p}\right)\ ,
\end{eqnarray}
where $\tilde{y} = y/y_{\rm min}$, $\tilde{y}_{\rm max} = \min[y_{\rm
min}^{-1},(T_f/T_0)^{1/(m+1)}]$, and
\begin{eqnarray}
\frac{\nu'}{\nu'_p} = \frac{\nu}{\nu_0}
\left(\frac{y}{y_{\rm min}}\right)^{m/2-d}\left(\frac{m+y^{-m-1}}{m+1}\right)
= \frac{\nu}{\nu_0}\,\tilde{y}^{m/2-d}\left(\frac{m+y_{\rm
min}^{-m-1}\tilde{y}^{-m-1}}{m+1}\right)\ ,
\end{eqnarray}
and the expression for $\nu'/\nu'_p$ assumes that $\nu'_p =
\nu'_0(R/R_0)^d = \nu'_0(y/y_{\rm min})^d$.
At late times, $T-T_{\rm ej} \gg T_f$, we have $y \ll 1$,
$\tilde{y}_{\rm max} = T_f/T_0$, and $\nu'/\nu'_p \approx
(\nu/\nu_0)\tilde{y}^{-1-d-m/2}y_{\rm min}^{-m-1}/(m+1)$ increases with
time so that $S(\nu'/\nu'_p) \propto (\nu'/\nu'_p)^{b_2}$ and $F_\nu \propto
(\nu/\nu_0)^{b_2}y_{\rm min}^{(m+1)(2-b_2)} =
(\nu/\nu_0)^{b_2}[(T-T_{\rm ej})/T_0]^{b_2-2}$, i.e. the HLE relation
$\alpha = 2+\beta$ is satisfied.\\

In the case for internal shocks, with $m=0$ and $d=-1$, $\nu'/\nu'_p$
becomes independent of $y$ and can be taken outside the integral 
($\nu'/\nu'_p = (\nu/\nu_0)/y_{\rm min} = (\nu/\nu_0) \tilde{T}$),
leading to the much simpler expression of the flux seen in the main
text (eq.~\ref{eq_bandDR_finalcondensed}).

\renewcommand{\theequation}{B\arabic{equation}}
\setcounter{equation}{0}
\begin{center}
\section{Evolution of the temporal and spectral indexes} \label{ap_temp_spec_slopes}
\end{center}

This appendix explicits the evolution of the temporal and spectral indexes with time.

\subsubsection{Single emission radius}

Where the luminosity is a delta function with radius at radius $R_0$,
we obtain
\begin{eqnarray}\label{eq_beta_DR0}
\beta &=& \left\{ \begin{array}{ll}
-b_1 + \tilde{T}(1+b_1)\nu/\nu_0 &  \tilde{T} \leqslant x_b\nu_0/\nu\\
\\
-b_2 &  \tilde{T} \geqslant x_b\nu_0/\nu
\end{array} \right.
\\ \nonumber
\\ \label{eq_DR0_alphas}
\alpha_{on} &=& \left\{ \begin{array}{ll}
(2-b_1)\bar{T}/(1+\bar{T}) + \bar{T}(1+b_1)\nu/\nu_0 &
\bar{T}\leqslant x_b\nu_0/\nu-1\\
\\
(2-b_2)\bar{T}/(1+\bar{T}) &  \bar{T} \geqslant x_b\nu_0/\nu-1
\end{array} \right.
\\ \nonumber
\\
\alpha_{ej} &=& \left\{ \begin{array}{ll}
2-b_1 +\tilde{T}(1+b_1)\nu/\nu_0    & \tilde{T} \leqslant x_b\nu_0/\nu\\
\\
2-b_2                        & \tilde{T} \geqslant x_b\nu_0/\nu
\end{array} \right.
\end{eqnarray}

We then have a very simple relation between $\alpha_{ej}$ and
$\beta$: $\alpha_{ej} = 2 + \beta$, as expected at asymptotically
late times for HLE, just that for $\alpha_{ej}$ it is satisfied all
along for the local values of the temporal and spectral indexes. At
late times $\alpha_{on}$ approaches $\alpha_{ej}$ and a similar
relation approximately holds between $\alpha_{on}$ and $\beta$
($\alpha_{on} \approx 2+\beta$).

\subsubsection{Emission from a finite range of radii: $R_0 < R < R_f$}

In this case, the spectral index $\beta$ is still given by
eq.~(\ref{eq_beta_DR0}),
while the two temporal indexes are:
\begin{eqnarray}\label{eq_alpha_4cas_barT}
\alpha_{on} =  \left\{ \begin{array}{ll}
(2-b_1)\frac{\bar{T}}{(1+\bar{T})} +\bar{T}(1+b_1)\frac{\nu}{\nu_0}
-(2+a)\frac{\bar{T}(1+\bar{T})^{1+a}}{(1+\bar{T})^{2+a}-1}\ \
 & \bar{T} < \min(\Delta R/R_0, x_b\nu_0/\nu-1)\ ,\\
\\
(2-b_1)\frac{\bar{T}}{(1+\bar{T})} +\bar{T}(1+b_1)\frac{\nu}{\nu_0}
 & \Delta R/R_0 < \bar{T} < x_b\nu_0/\nu-1\ ,\\
\\
(2-b_2)\frac{\bar{T}}{(1+\bar{T})}
-(2+a)\frac{\bar{T}(1+\bar{T})^{1+a}}{(1+\bar{T})^{2+a}-1}
 & x_b\nu_0/\nu-1 < \bar{T} < \Delta R/R_0\ ,\\
\\
(2-b_2)\frac{\bar{T}}{(1+\bar{T})}
 & \bar{T} > \max(\Delta R/R_0, x_b\nu_0/\nu-1)\ ,\\
\end{array} \right.
\end{eqnarray}
which limits at very early and very late times are
\begin{eqnarray}\label{eq_limit_baralpha}
\alpha_{on} \approx  \left\{ \begin{array}{ll}
- 1
 & \bar{T} \ll 1\ ,\\
\\
2-b_2
 & \bar{T} \gg 1\ ,\\
\end{array} \right.
\end{eqnarray}
and
\begin{eqnarray}\label{eq_alpha_4cas_t}
\alpha_{ej} =  \left\{ \begin{array}{ll}
2-b_1 - (2+a)/(1-\tilde{T}^{-a-2}) + \tilde{T}(1+b_1)\nu/\nu_0\ \
 & \tilde{T} < \min(R_f/R_0, x_b\nu_0/\nu)\ ,\\
\\
2-b_1 + \tilde{T}(1+b_1)\nu/\nu_0
 & R_f/R_0 < \tilde{T} < x_b\nu_0/\nu\ ,\\
\\
2-b_2 - (2+a)/(1-\tilde{T}^{-a-2})
 & x_b\nu_0/\nu  < \tilde{T} < R_f/R_0\ ,\\
\\
2-b_2
 & \tilde{T} > \max(R_f/R_0, x_b\nu_0/\nu)\ .
\end{array} \right.
\end{eqnarray}
According to equations (\ref{eq_beta_DR0}) and
(\ref{eq_alpha_4cas_t}) $\alpha_{ej}$ has a simple relation with
$\beta$:
\begin{eqnarray}\label{eq_relation_alpha_beta}
\alpha_{ej} = \left\{ \begin{array}{ll}
\beta+2 - (2+a)/(1-\tilde{T}^{-a-2})\ \ &
\tilde{T} < R_f/R_0 \ \ (\bar{T} < \bar{T}_f)\ ,\\
\\
\beta+2  & \tilde{T} > R_f/R_0 \ \ (\bar{T} > \bar{T}_f)\ .\\
\end{array} \right.
\end{eqnarray}
\\
Note that in the limit $\bar{T} \to 0$ ($\tilde{T} \to 1$),
at very early times, just after the onset of the spike, 
$\alpha_{ej} \to -\infty$ while
$\alpha_{on} \to -1$. Moreover, the simple HLE relation,
$\alpha_{ej} = 2+\beta$, is valid as soon as $\tilde{T} > \tilde{T}_f$,
for any value of $\tilde{T}_f$. This is a relation
between the local values of $\alpha_{ej}$ and $\beta$, that hold as
both change with time, and is strictly valid from $\tilde{T} >
\tilde{T}_f$ only for $m = 0$ and $d = -1$. For general values of $m$ or
$d$ this local HLE relation would be valid only at late times,
$\bar{T} \gg \bar{T}_f$. Note, however, that for alternative other
definitions of the temporal index, such as $\alpha_{on}$, this
relation is only approached at late time: $\alpha_{on} \approx
2+\beta$ for $\bar{T} > \bar{T}_f$ and $\bar{T} \gg 1$.

\renewcommand{\theequation}{C\arabic{equation}}
\setcounter{equation}{0}
\begin{center}
\section{Exponential turn-off of the emission with radius} \label{ap_expcutoff}
\end{center}

Throughout the paper we have assumed that the emission abruptly turns
off at $R_f$. This results in a sharp change in the temporal index at
$\bar{T}_f$, which usually corresponds to
a sharp peak for the pulses in the prompt GRB light
curve. Observations sometimes show pulses with a round peak, which may
be hard to fit with spiky theoretical spikes. Such rounder peaks for
the pulses may be obtain within the framework of our model by
introducing a more gradual turn-off of the emission at $R > R_f$.
For convenience, we parameterize this here by assuming
that the luminosity starts decreasing exponentially with radius at $R
> R_f$. For simplicity we consider here only $\Delta R > 0$, but the
results are similar for $\Delta R = 0$. Similarly, only the case for
internal shock ($m=0$, $d=-1$) is considered here. Thus, we
introduce the following comoving spectral luminosity:
\begin{eqnarray} \label{eq_expcutoff}
L'_{\nu'} = \left\{ \begin{array}{ll}
L'_0 \left(\frac{R}{R_0}\right)^a S\left(\frac{\nu'}{\nu'_p}\right) &
R_0 \leqslant R \leqslant R_f\\
\\
L'_0 \left(\frac{R}{R_0}\right)^a S\left(\frac{\nu'}{\nu'_p}\right)
e^{-\frac{q(R-R_f)}{\Delta R}} & R > R_f
\end{array} \right.
\end{eqnarray}
where $q$ is the decay constant (a larger $q$ corresponds to a sharper
turn-off of the emission).

For $1 \leqslant \tilde{T} \leqslant \tilde{T}_f$ the observed flux is 
identical to
that without introducing the gradual emission turn-off, and is
therefore given by eq.~(\ref{eq_bandDR_finalcondensed}),
\begin{eqnarray}
F_{\nu}(\tilde{T} \geq \tilde{T}_f) = F_0
\tilde{T}^{-2}
\left[\left(\min(\tilde{T},\tilde{T_f})\right)^{2+a}-1\right]
S\left(\frac{\nu}{\nu_0}\tilde{T}\right)\ .
\end{eqnarray}
The flux for $\tilde{T} > \tilde{T}_f$ is obtained by calculations very similar to
those of section \ref{subsec_band_R0_Rf}, and reads
\begin{eqnarray}\label{F_J}
F_{\nu}(\tilde{T} \geq \tilde{T}_f) &=& F_0 \tilde{T}^{-2}
S(\tilde{T}\nu/\nu_0)
\left[\tilde{T}_f^{2+a}-1 + J(\tilde{T}) \right]\ ,
\\ \nonumber
\\
J(\tilde{T}) &\equiv&
(2+a)\int_{\tilde{T}_f}^{\tilde{T}}d\tilde{y}\,\tilde{y}^{a+1}e^{-(\tilde{y}-\tilde{T}_f)/Q}\
,
\end{eqnarray}
were $Q \equiv \Delta R/(q R_0)$, and we
remind the reader that $\tilde{y} = \tilde{T} y$ for $m = 0$. The
expression for the flux is thus very similar its form for an abrupt
turn-off of the emission at $R_f$, but with the additional term $J(\tilde{T})$
that adds some flux at $\tilde{T} > \tilde{T}_f$ (representing the added
contributions from $R > R_f$). For $a = 1$ we have
\begin{equation}
J(\tilde{T}, a = 1) = 6Q^3 + 6Q^2\tilde{T}_f + 3Q\tilde{T}_f^2
-e^{-(\tilde{T}-\tilde{T}_f)/Q}\left(6Q^3 + 6Q^2\tilde{T}+3Q\tilde{T}^2\right)\
.
\end{equation}
At late times $J(\tilde{T}, a = 1)$ approaches a constant value,
\begin{eqnarray}
J_{\infty}\equiv 6Q^3 + 6Q^2\tilde{T}_f + 3Q\tilde{T}_f^2 \sim  \left\{
\begin{array}{ll}
6 \left(\frac{\Delta R}{qR_0}\right)^3                            &     q \ll
\frac{\Delta R/R_0}{1+\Delta R/R_0}\\
3 \frac{\Delta R}{qR_0} \left(1+\frac{\Delta R}{R_0}\right)^2     &     q \gg
\frac{\Delta R/R_0}{1+\Delta R/R_0}
\end{array} \right.
\end{eqnarray}
where we have replaced $Q$ and $\tilde{T}_f$ by their dependence on
$q$ and $\Delta R/R_0$. Since $J(\tilde{T})$ appears in eq.~(\ref{F_J}) in a
sum with $\tilde{T}_f^{2+a}-1$, it will dominate the observed flux at
late times if $J_\infty > \tilde{T}_f^{2+a}-1$ or equivalently if $q <
q_{\rm crit}$ where $J_\infty(q_{\rm crit}) \equiv
\tilde{T}_f^{2+a}-1$.

\begin{figure}
\begin{center}
\includegraphics[width=0.465\textwidth,
height=0.465\textwidth]{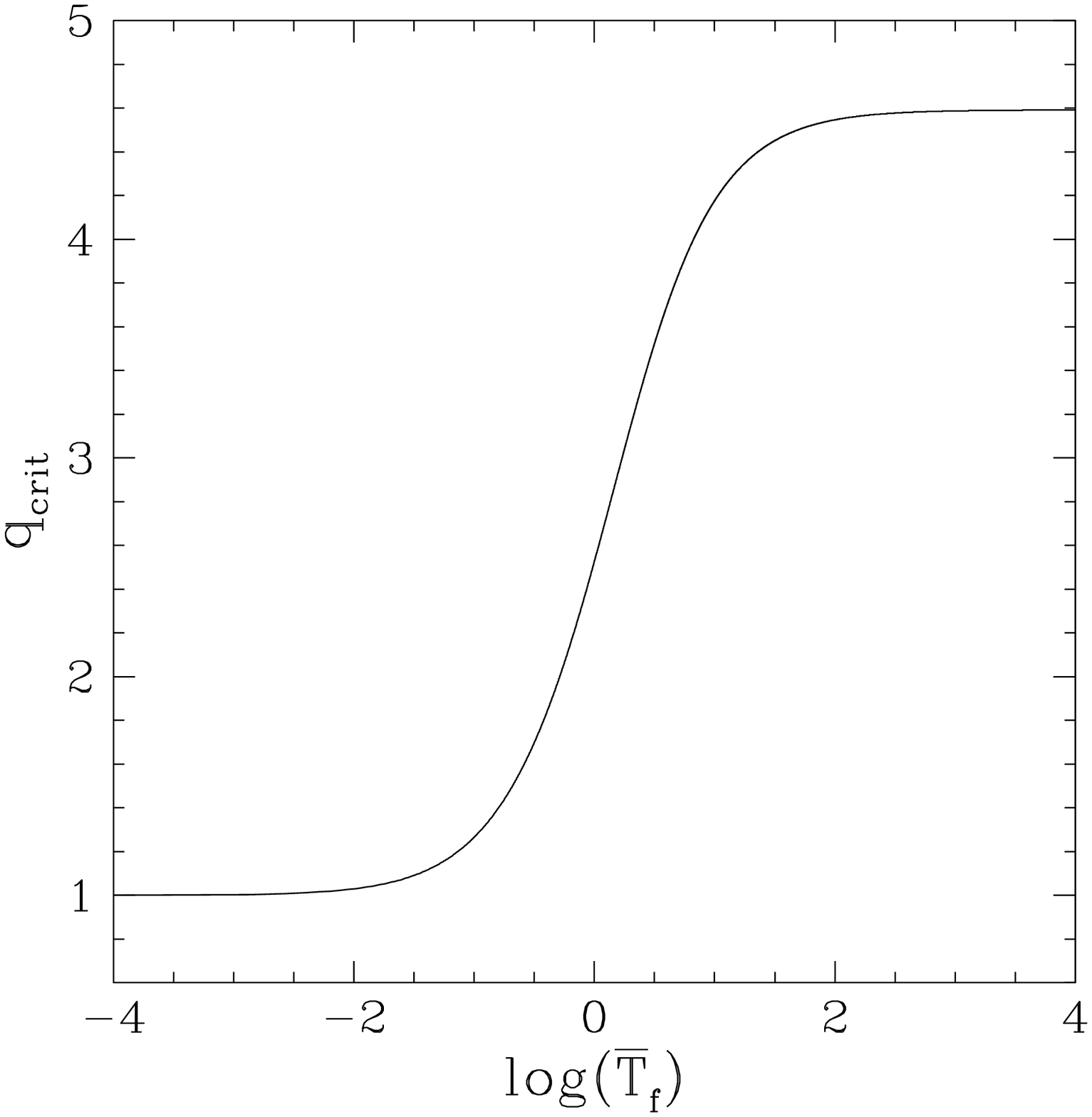}
\hspace{0.8cm}
\includegraphics[width=0.465\textwidth,
height=0.465\textwidth]{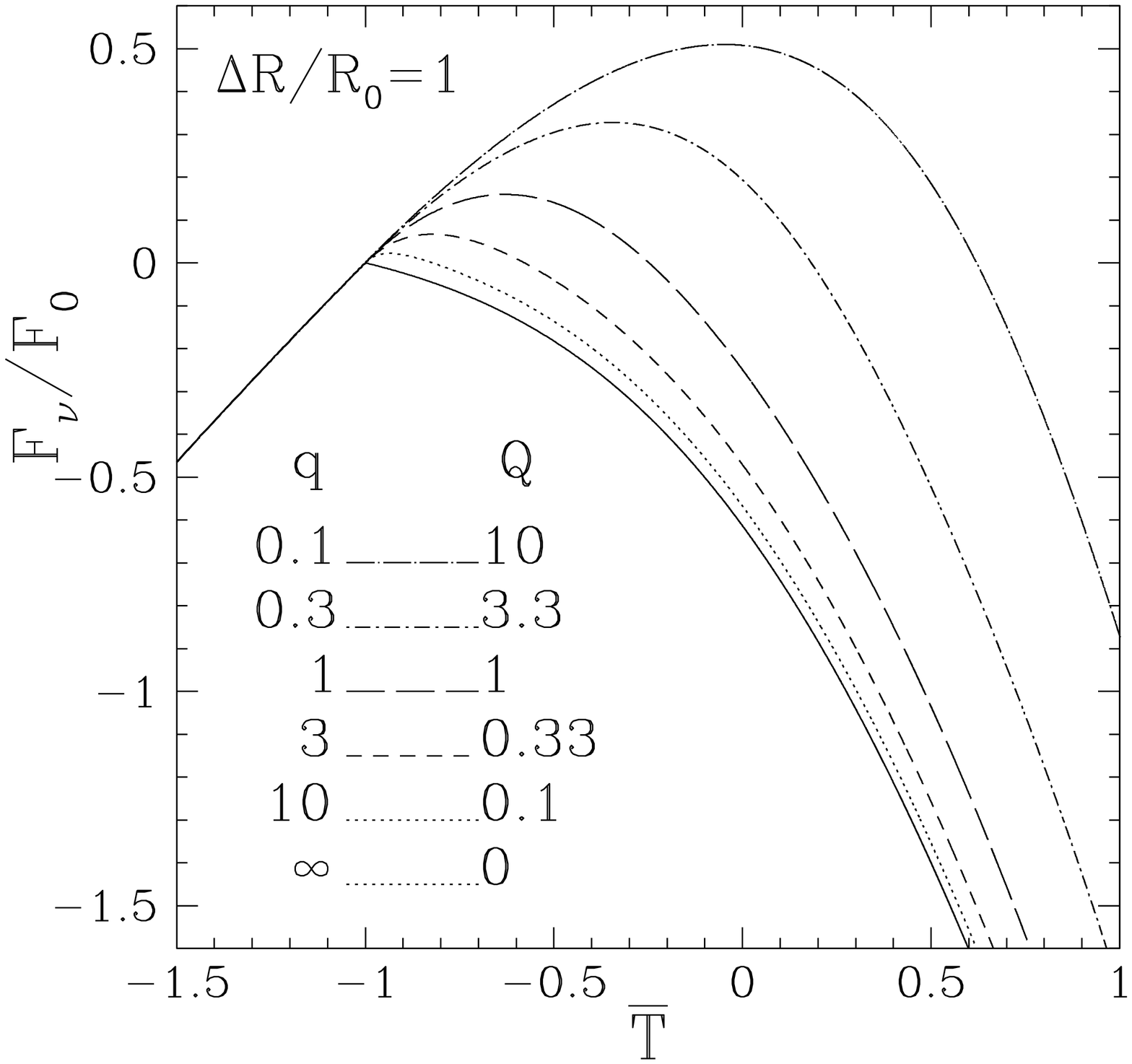}
\end{center}
\caption[Exponential cutoff]{
{\bf Left:} \emph{Dependence of the critical decay index $q_{crit}$ of the exponential
cut-off on $\bar{T}_f = \Delta R/R_0$, in semi-logarithmic scale. It is defined by 
$J_{\infty}(q_{crit}) = \tilde{T}_f^3-1$, i.e. at late time, the contribution to the flux from the
exponential cut-off is equal to the contribution from the emitting region
between $R_0$ and $R_f$.}
{\bf Right:} \emph{Comparison of the shape of pulses with and without the exponential
turn-off of the luminosity for a ratio $\Delta R/R_0 = 1$ in logarithmic scale. The solid
line shows the shape of the pulse for an abruptly stopping luminosity
(no exponential turn-off), the other lines show the pulse shape for
different values of the decay constant $q=0.1$ of the exponential turn-off.}
}
\label{fig_comp_ecoff}
\end{figure}

The left panel of figure \ref{fig_comp_ecoff} shows $q_{crit}$ as a function of $\bar{T}_f =
\Delta R/R_0$ for $a = 1$, and it can be seen that the limiting values
of $q_{\rm crit}$ are $1$ for $\bar{T}_f \ll 1$, and
$(7+22^{1/2})^{1/3}+1+3/(7+22^{1/2})^{1/3} \approx 4.59$ for
$\bar{T}_f \gg 1$, so that $q_{\rm crit}$ is always of order
unity. Therefore, for $q \ll q_{\rm crit} \sim 1$ the late time flux
is dominated by contributions from $R>R_f$, the peak of the pulse is
rounder and the peak flux is higher compared to an abrupt turn-off of
the emission with radius, which is approached in the opposite limit of
$q \gg q_{\rm crit} \sim 1$. This can nicely be seen in
the right panel of fig.~\ref{fig_comp_ecoff}, which shows the shape of a pulse for different
values of $q$, including the limiting case of $q \to \infty$, which corresponds
to an abrupt turn-off of the emission at $R_f$.

Such an exponential turn-off could therefore be useful when fitting
our our model with data, in order to reproduce round-peaks pulses. Of
course, one should be aware that this adds a free parameter ($q$ or
$Q$), and might thus increases the degeneracy between the different
fit parameters. Therefore, adding this extra model parameter should be
done only when it is required by the data.

$ $
\label{lastpage}

\end{document}